\documentclass[fleqn,10pt]{wlscirep}
\usepackage[utf8]{inputenc}
\usepackage[T1]{fontenc}
\usepackage{bm}

\usepackage{booktabs}
\usepackage{adjustbox}
\usepackage{longtable}
\usepackage{caption}
\usepackage{mathtools}
\usepackage{amssymb}
\usepackage{amsmath}
\usepackage{etoolbox}
\usepackage{amsthm}
\usepackage{tocloft}
\usepackage{float}

\usepackage{lineno}

\addtocontents{toc}{\protect\setcounter{tocdepth}{-1}}

\usepackage{titlesec}
\titleformat{\section}
  {\normalfont\Large\bfseries}
  {\thesection}
  {0em} 
  {#1}
  
\allowdisplaybreaks

\usepackage{upgreek}
\renewcommand{\pi}{\uppi}
\renewcommand{\delta}{\updelta}
\renewcommand{\tau}{\uptau}
\renewcommand{\rho}{\uprho}
\renewcommand{\phi}{\upphi}
\renewcommand{\gamma}{\upgamma}

\title{Public goods games on any population structure}

\author[1]{Chaoqian Wang}
\author[2,3,4,*]{Qi Su}
\affil[1]{School of Mathematics and Statistics, Nanjing University of Science and Technology, Nanjing 210094, China}
\affil[2]{School of Automation and Intelligent Sensing, Shanghai Jiao Tong University, Shanghai, China}
\affil[3]{Key Laboratory of System Control and Information Processing, Ministry of Education of China, Shanghai, China}
\affil[4]{Shanghai Engineering Research Center of Intelligent Control and Management, Shanghai, China}
\affil[*]{Corresponding to: qisu@sjtu.edu.cn}

\begin{abstract}
Understanding the emergence of cooperation in social networks has advanced through pairwise interactions, but the corresponding theory for group-based public goods games (PGGs) remains less explored. Here, we provide theoretical conditions under which cooperation thrives in PGGs on arbitrary population structures, which are accurate under weak selection. We find that a class of networks that would otherwise fail to produce cooperation, such as star graphs, are particularly conducive to cooperation in PGGs. More generally, PGGs can support cooperation on almost all networks, which is robust across all kinds of model details. This fundamental advantage of PGGs derives from self-reciprocity realized by group separations and from clustering through second-order interactions. We also apply PGGs to empirical networks, which shows that PGGs could be a promising interaction mode for the emergence of cooperation in real-world systems.
\end{abstract}
\begin{document}

\flushbottom
\maketitle

\maketitle

\section*{Short title}
PGGs on any population structure

\section*{Teaser}
A mathematical theory shows when cooperation emerges in public goods games on any network structure.

\section*{Introduction}
Cooperation is essential for the emergence of higher-level complexity in biological and social systems~\cite{nowak2006evolutionary,sigmund2010calculus,perc2017statistical}. To answer the question of how individual selfishness leads to selfless cooperation in evolution, many theories have been developed, such as evolutionary games on networks (also known as graphs). When individuals interact locally and adopt strategies with higher payoffs, cooperation may emerge through spatial reciprocity~\cite{nowak1992evolutionary}. Over the past three decades, agent-based simulations on networks have revealed numerous mechanisms that promote cooperation~\cite{hauert2004spatial,perc2010coevolutionary}. In parallel, the analytical theory of pairwise games, initially on the regular graphs~\cite{ohtsuki2006simple,taylor2007evolution,allen2014games}, has developed to identify the conditions for the success of cooperation on any population structure~\cite{ allen2017evolutionary,allen2019mathematical,mcavoy2021fixation}---some networks are more conducive to cooperation than others~\cite{fotouhi2018conjoining,mcavoy2020social}.

However, pairwise games are limited in their ability to capture diverse natural and social phenomena, including nonlinear effects~\cite{hauert2006synergy,allen2024nonlinear}. Real-world interactions may involve more than two individuals. The evolution of cooperation in these group interactions is naturally described by multiplayer games~\cite{nash1950equilibrium}. Applying the framework of evolutionary dynamics on graphs, a natural approach is to let individuals form groups with their neighbors and play multiplayer games within these groups. In this way, individuals participate not only in the groups they initiate but also in those initiated by their neighbors. Individuals then use the average or accumulated payoffs from these groups as the basis for strategy updates~\cite{perc2013evolutionary}. This principle can be applied to any multiplayer game~\cite{wang2024evolutionary}, including the public goods game~\cite{szabo2002phase} and others~\cite{wang2022reversed, wang2022modeling}.

The public goods game (PGG) is among the most widely studied multiplayer games, originating from the tragedy of the commons~\cite{hardin1968tragedy}. Players choose whether to contribute to the common pool. All contributions are multiplied by a synergy factor and then evenly distributed among all players. From a group perspective, this amplification makes contributions collectively beneficial. However, from an individual perspective, one can still receive an equal share without contributing. This results in higher payoffs for non-contributors and incentivizes individuals not to contribute. Apart from human experiments~\cite{semmann2003volunteering,hauser2019social,shi2020freedom,otten2022human}, previous research on PGGs has primarily relies on agent-based simulations~\cite{szabo2002phase,santos2008social}. Although agent-based simulations allow for great flexibility in studying new mechanisms, they require intensive computational resources and make it difficult to identify the underlying principles behind the phenomena. Therefore, recent work has attempted to analyze these mechanisms at a theoretical level~\cite{wang2023evolution,wang2023conflict,wang2023inertia,wang2023greediness,wang2023imitation,meng2024dynamics}. The feasibility of such efforts depends on the extent to which mathematical theory can support them.

The most advanced analytical theory for PGGs remains limited to regular graphs. There is no efficient algorithm at general selection strengths due to computational complexity~\cite{ibsen2015computational}, but a feasible alternative is to analyze the weak selection limit~\cite{allen2013spatial}. Building on this framework, Li~et~al.~\cite{li2014cooperation,li2016evolutionary} developed a theory for PGGs in infinite structured populations using pair approximation, although this method cannot capture clustering effects or higher-order interactions. Su~et~al.~\cite{su2018understanding,su2019spatial} further addressed this limitation and derived a corresponding theory for PGGs in finite structured populations. Both of these theories were confined to homogeneous networks, where all individuals have the same number of neighbors. General results for heterogeneous networks remains lacking.

Here, we develop the analytical theory for the evolution of cooperation in PGGs on any population structure. We identify the critical synergy factor for the success of cooperation on different networks, which is accurate under weak selection. Using this theory, we examine a large number of networks, finding that a class of structures, especially the star graph, can strongly promote cooperation in PGGs. We also explore PGGs on a series of random networks, including random, small-world, to scale-free, and verify the general effects of local structures, such as clustering, on cooperation. Finally, we analyze all small networks and four empirical networks, and find that cooperation consistently emerge across various model details in PGGs, whereas in pairwise games it cannot. These results imply that PGGs may represent a more plausible and effective interaction mechanism in the real world than pairwise games.

\section*{Results}\label{sec_model}

\subsection*{PGGs on any population structure}

\begin{figure}
	\centering
	\includegraphics[width=.9\textwidth]{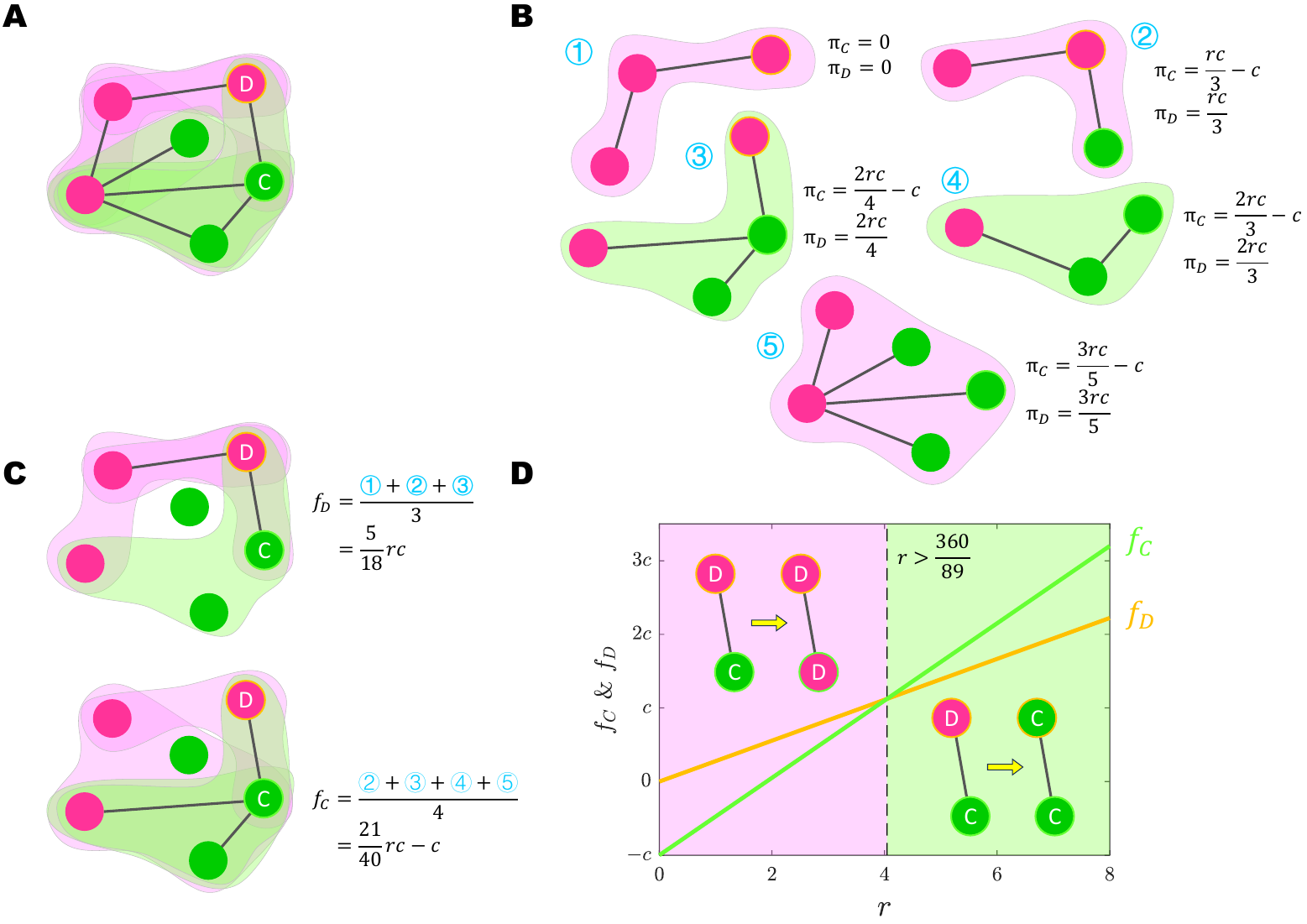}
	\caption{\textbf{The public goods game (PGG) on a general network.}
		(\textbf{A}) Each agent organizes a PGG in the group of itself and its neighbors. (\textbf{B}) In each PGG, payoffs for cooperation and defection are calculated by Eq.~(\ref{eq_singlepgg}). Obviously, $\pi_D>\pi_C$, defectors always have higher payoffs than cooperators, seemingly driving cooperation towards defection. (\textbf{C}) Since each agent organizes a PGG, an agent plays not only the PGG organized by itself, but also the PGGs organized by its neighbors. We average the payoffs from these PGGs as an agent's actual payoff. (\textbf{D}) The emergence of spatial reciprocity. For the presented two agents, their actual payoffs $f_C>f_D$ when $r>360/89$, driving defection towards cooperation. } \label{fig_demo}
\end{figure}

In a public goods game (PGG) of $G$ players, each player chooses either cooperation ($C$) or defection ($D$). A cooperator pays a cost $c$ and contributes to the common pool, whereas a defector contributes nothing. If there are $g_C$ ($0\leq g_C\leq G$) cooperators, the total contribution to the pool is $g_C c$. These contributions are multiplied by a synergy factor $r$ ($r>1$) to produce the public goods $r g_C c$. This amount is evenly redistributed among all $G$ players, so each receives $r g_C c/G$. Therefore, the payoffs for cooperation and defection, $\pi_C$ and $\pi_D$, are 
\begin{subequations}\label{eq_singlepgg}
	\begin{align}
		\pi_C&=\frac{rg_C c}{G}-c, \\
		\pi_D&=\frac{rg_C c}{G}.
	\end{align}
\end{subequations}
At the group level, the synergy factor $r>1$ makes cooperation collectively beneficial. On the other hand, defectors also receive the public goods produced by cooperators, resulting in higher payoffs for defectors than for cooperators. The social dilemma thus emerges.

In this work, we study the PGG on general unweighted networks. We consider a population of size $N$, whose node set is denoted by $\mathcal{N}=\{1,2,\dots,N\}$. Each node represents an agent in the population. The adjacency between agents $i$ and $j$ is represented by $k_{ij}$: if they are neighbors, $k_{ij}=1$; otherwise, $k_{ij}=0$. The number of neighbors of agent $i$ is thus $k_i=\sum_{j\in\mathcal{N}}k_{ij}$. For convenience, we denote the neighbor set of agent $i$ as $\mathcal{N}_i$: if $j$ is $i$'s neighbor ($k_{ij}=1$), then $j\in\mathcal{N}_i$.

On this network, each agent $i$ organizes a group of size $G_i=k_i+1$, consisting of its neighbors and itself  (Fig.~\ref{fig_demo}\textbf{A}). A PGG is played within this group, and payoffs are calculated according to Eq.~(\ref{eq_singlepgg}) (Fig.~\ref{fig_demo}\textbf{B}). Also, agent $i$ participates in the $k_i$ PGGs organized by its neighbors. Thus, agent $i$ plays $1+k_i=G_i$ PGGs organized by itself and its neighbors. We take the average payoffs that agent $i$ receives from these $G_i$ PGGs as the actual payoff, $f_i=(1/G_i)\sum_{j\in\mathcal{G}_i}\pi_i^j$ (Fig.~\ref{fig_demo}\textbf{C}), where $\pi_i^j$ denotes the payoff that agent $i$ receives in the PGG organized by agent $j$ and $\mathcal{G}_i=\{i\}\cup\mathcal{N}_i$ denotes the group organized by agent $i$. These actual payoffs $f_i$ determine strategy evolution (Fig.~\ref{fig_demo}\textbf{D}).

In each elementary step, a random focal agent $i$ is chosen to update its strategy. The actual payoff $f_i$ is calculated and mapped to fitness, $F_i=\exp{(\delta f_i)}$~\cite{mcavoy2020social,wang2024evolutionary}. Here, $0<\delta\ll 1$ is a weak selection strength, meaning that game payoffs contribute only a small perturbation to baseline fitness~\cite{ohtsuki2006simple,allen2017evolutionary}. The weak selection assumption reflects the idea that many other factors dominate the overall fitness of individuals. The actual payoffs and fitnesses of agent $i$’s neighbors are also calculated for comparison. Commonly used update rules, such as pairwise comparison (PC)~\cite{szabo1998evolutionary}, death-birth (DB)~\cite{ohtsuki2006simple}, and birth-death (BD)~\cite{lieberman2005evolutionary}, vary in details but follow the same principle that strategies with higher payoffs have an advantage to propagate. We present the PC rule here as an example and study other update rules in \ref{sec_theory}. The focal agent $i$ selects a random neighbor $j\in\mathcal{N}_i$ and adopts the strategy of agent $j$ with probability
\begin{equation}\label{eq_Wpc}
	W_{i\gets j}=\frac{1}{1+\exp{(-\delta(f_j-f_i))}}.
\end{equation}
Otherwise, agent $i$ keeps the current strategy. The probability in Eq.~(\ref{eq_Wpc}) can also be interpreted as $W_{i\gets j}=F_j/(F_i+F_j)$, with the probability of keeping the strategy understood as $F_i/(F_i+F_j)$.

The process then proceeds to the next elementary step, in which a new random focal agent $i$ is chosen to update its strategy, and this process iterates (Methods). We track the fraction of cooperators $\rho_C$ in the population, which changes over time and may reach steady states after sufficiently many steps.

\subsection*{Conditions for the evolution of cooperation}
Although the cooperation fraction could fluctuate in non-equilibrium states for a very long time~\cite{perc2013evolutionary}, it is expected to reach a fixation state eventually under weak selection, where all individuals choose to use the same strategies. There are only two fixation states: full cooperation ($\rho_C=1$) and full defection ($\rho_C=0$). 

In a fully defective population, a mutant cooperator introduced at a particular node has a certain probability of spreading and eventually taking over the entire population. On heterogeneous networks, this probability generally depends on the node at which the mutant first appears. We define the fixation probability of a cooperator as the average takeover probability when the initial cooperator is placed uniformly at random. In other words, it is the likelihood that a single randomly positioned cooperator can convert a population of defectors into one of cooperators.
Under neutral selection, i.e., when $\delta = 0$ so that payoffs from game interactions have no influence on the evolutionary process, the fixation probability of a cooperator is $1/N$~\cite{allen2019mathematical}. Natural selection is said to favor the evolution of cooperation if the fixation probability of a cooperator exceeds this neutral benchmark.

We find that evolution favors cooperation in PGGs once the synergy factor $r$ exceeds a critical threshold $r^\star$. The critical synergy factor $r^\star$ depends on the network and model details. Under the PC update rule, the condition for the success of cooperation on any network is 
\begin{equation}\label{eq_PCcondi}
	r>\frac{\tau^{(1)}}{\Upsilon^{(1)}}.
\end{equation}
Here, $\tau^{(n)}=\sum_{i,j\in\mathcal{N}} k_i p_{ij}^{(n)}\tau_{ij}$ and $\Upsilon^{(n)}=\sum_{i,j\in\mathcal{N}} k_i p_{ij}^{(n)}\Upsilon_{ij}$ can be obtained for a given network. First, $p^{(n)}_{ij}$ is the probability of arriving at node $j$ after an $n$-step random walks starting from node $i$. Second, $\tau_{ij}$ (the coalescence times of ancestral random walks~\cite{kingman1982coalescent,wakeley2009coalescent,cox1989coalescing,durrett1994importance,allen2017evolutionary}) between nodes $i$ and $j$ are obtained by solving the following linear equations, with $\tau_{ji}\equiv \tau_{ij}$:
\begin{align}\label{eq_tauij}
	\begin{cases}
		\displaystyle{
			\tau_{ij}=1+\frac{1}{2k_i}\sum_{l\in\mathcal{N}_i}\tau_{jl}+\frac{1}{2k_j}\sum_{l\in\mathcal{N}_j}\tau_{il}}, & \mbox{if $j\neq i$},
		\\[1em]
		\displaystyle{
			\tau_{ij}=0}, & \mbox{if $j=i$}.
	\end{cases}
\end{align}
Third, $\Upsilon_{ij}$ can be calculated from the coalescence times $\tau_{ij}$:
\begin{equation}\label{eq_Upsilonij}
	\Upsilon_{ij}=
	\frac{1}{G_i}\left(
	\frac{\tau_{ij}+\sum_{l\in\mathcal{N}_i}(\tau_{jl}-\tau_{il})}{G_i}+
	\sum_{l\in\mathcal{N}_i}\frac{(\tau_{jl}-\tau_{il})+ \sum_{\ell\in\mathcal{N}_l}(\tau_{j\ell}-\tau_{i\ell})}{G_l}
	\right).
\end{equation}
Note that $\Upsilon_{ii}=0$ since $\tau_{ii}=0$. However, $\Upsilon_{ij}\neq \Upsilon_{ji}$ in general because of differences in group sizes on heterogeneous networks.

We are thus able to determine the condition for cooperation in PGGs on all networks. Similar to Eq.~(\ref{eq_PCcondi}), we also identify the condition under the DB rule (at each elementary step, a random individual $i$ dies, and neighbors $j\in\mathcal{N}_i$ compete for the vacant site proportionally to their fitness $F_j/\sum_{l\in\mathcal{N}_i}F_l$), which is $r>\tau^{(2)}/\Upsilon^{(2)}$. See \ref{sec_theory} for the conditions under the BD rule and full theoretical derivations for other model details.

\subsection*{Synthetic networks}

\begin{figure}[!ht]
	\centering
	\includegraphics[width=.8\textwidth]{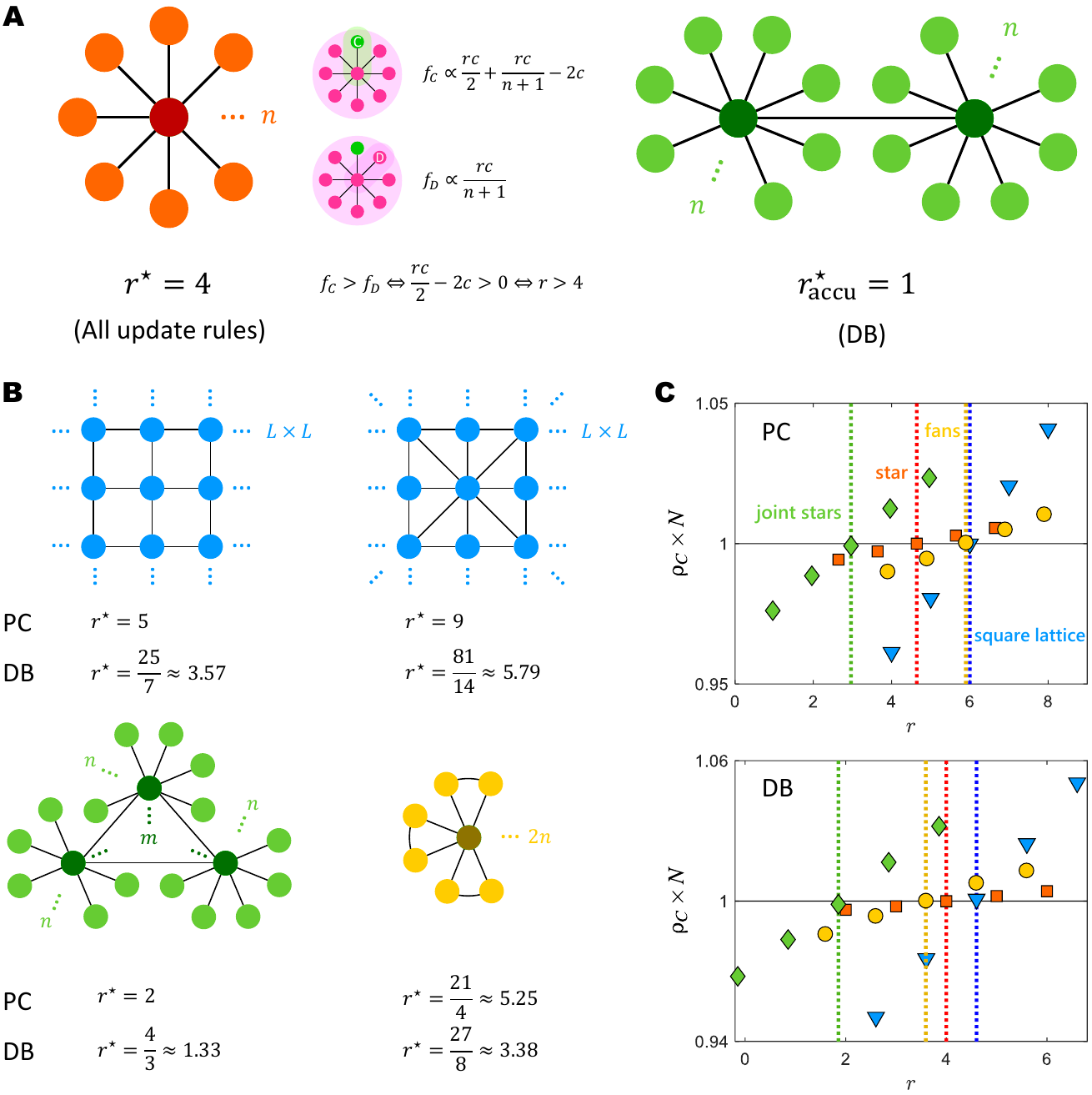}
	\caption{\textbf{Cooperation conditions for PGGs on homogeneous and heterogeneous networks.} (\textbf{A}) Star graphs support cooperation in PGGs as simple as when $r>4$, and the essence of which is $2\times 2$: there are 2 groups for a leaf, with 2 players in its own group. When using accumulated payoffs, connecting the hubs of two or more star graphs leads to $r^\star_\text{accu}=1$ under the DB rule, which maximally supports cooperation. (\textbf{B}) Comparison with other networks,including square lattices with von Neumann (left) and Moore (right) neighborhoods, joint stars with any number of hubs, and ceiling fans. The $r^\star$ values reported here are for infinite population size; see \ref{sec_appl} for finite size. (\textbf{C}) Agent-based simulations confirm the theoretical predictions on $L=5$ square lattice with von Neumann neighborhood, $n=9$ star, $m=3$ \& $n=9$ joint stars, and $n=9$ ceiling fans. The dots represent the average cooperation fraction in the steady states (Methods), while the dashed lines are theoretical $r^\star$ over which $\rho_C>1/N$. }
	\label{fig_specialgraph}
\end{figure}

We start the discussion from synthetic networks, which are uniquely determined by their network parameters. The conditions for the success of cooperation in PGGs can be expressed in terms of these network parameters.

A simple example is the star graph, composed of one hub and $n$ leaves (Fig.~\ref{fig_specialgraph}\textbf{A}). We find that star graphs consistently promote cooperation in PGGs for $r>4$ under all update rules (and in the infinite population limit $n\to \infty$). This differs from the previous conclusion in pairwise donation games (DGs), where cooperation cannot emerge on star graphs~\cite{allen2017evolutionary}. In this case, the so-called graph surgery, such as connecting the hubs of two stars, was a way to rescue cooperation in pairwise games. Here, if we further connect two hubs, we get a super structure to support cooperation in PGGs: with a variation of model details (accumulated instead of average payoffs), the condition for the success of cooperation is $r>1$, that is, cooperation is maximally favored.

The intuition of $r^\star=4$ for star graphs can be interpreted as $2\times 2$, two groups for a leaf times two players in the group organized by the leaf. We explain this under the DB update rule, for which $r^\star\equiv 4$ holds independently of the population size. Since the focal agent's payoff does not influence strategy updates (only neighbors compete for the focal vacant position), a focal leaf always takes the strategy of the hub neighbor. On the other hand, when the hub updates, the competition happens among all leaves and is independent of the hub. Therefore, the hub's payoff does no work, and we only need to discuss the competition among all leaves. A leaf participates in 2 PGGs, organized by itself and the hub neighbor. In the PGG organized by itself, the payoff is $r(1+x_H)c/2-c$ if cooperating or $rx_H c/2$ if defecting ($x_H=1$ if the hub cooperates and $x_H=0$ if it defects), where the overlapping term $rx_H c/2$ can be eliminated. In the PGG organized by the hub, the payoff is $rg_C c/(n+1)-c$ or $rg_C c/(n+1)$, where the overlapping term $rg_C c/(n+1)$ can be eliminated. In this way, a cooperator leaf has a higher payoff than a defector leaf if and only if $(rc/2-2c)/2>0$ or $r/2-2>0$. That is, $r>r^\star=2\times 2=4$.

On the $L\times L$ square lattices with periodic boundary conditions of different neighborhood sizes ($G=5$ and $G=9$), our results agree with previous findings on regular networks~\cite{su2019spatial,wang2023inertia} (Fig.~\ref{fig_specialgraph}\textbf{B}). Observing these results, we notice that star graphs can be even more conducive to cooperation than regular graphs, as the critical synergy factor rises with group size on regular graphs but remains constant on star graphs. For example, the square lattice of $G=9$ has $r^\star\approx 5.79$ under death-birth, which is worse than the constant $r^\star=4$ of the star graph. 

Fig.~\ref{fig_specialgraph}\textbf{B} also shows results on more heterogeneous networks. For $m$ fully connected hubs with $n$ leaves on each, we have $r^\star=4m/(2m-1)$ (PC update) and $r^\star=(12m-4)/(9m-7)$ (DB update) for large $n$, which are $r^\star=2$ and $r^\star=4/3$ for large $m$. As another simple extension, for ceiling fans with $n$ fans (each has 2 leaves), we have $r^\star=21/4$ for PC update and $r^\star=27/8$ for DB update. The DB update promotes cooperation on ceiling fans, while the PC rule inhibits it (compared to the original star graph). This is because the DB rule can utilize the clustering coefficient to promote cooperation in PGGs, which will be explained in the next section. Our predictions for these synthetic networks are also validated by agent-based simulations under different update rules (Fig.~\ref{fig_specialgraph}\textbf{C}).

\subsection*{General roles of local structures}

\begin{figure}[!ht]
	\centering
	\includegraphics[width=\textwidth]{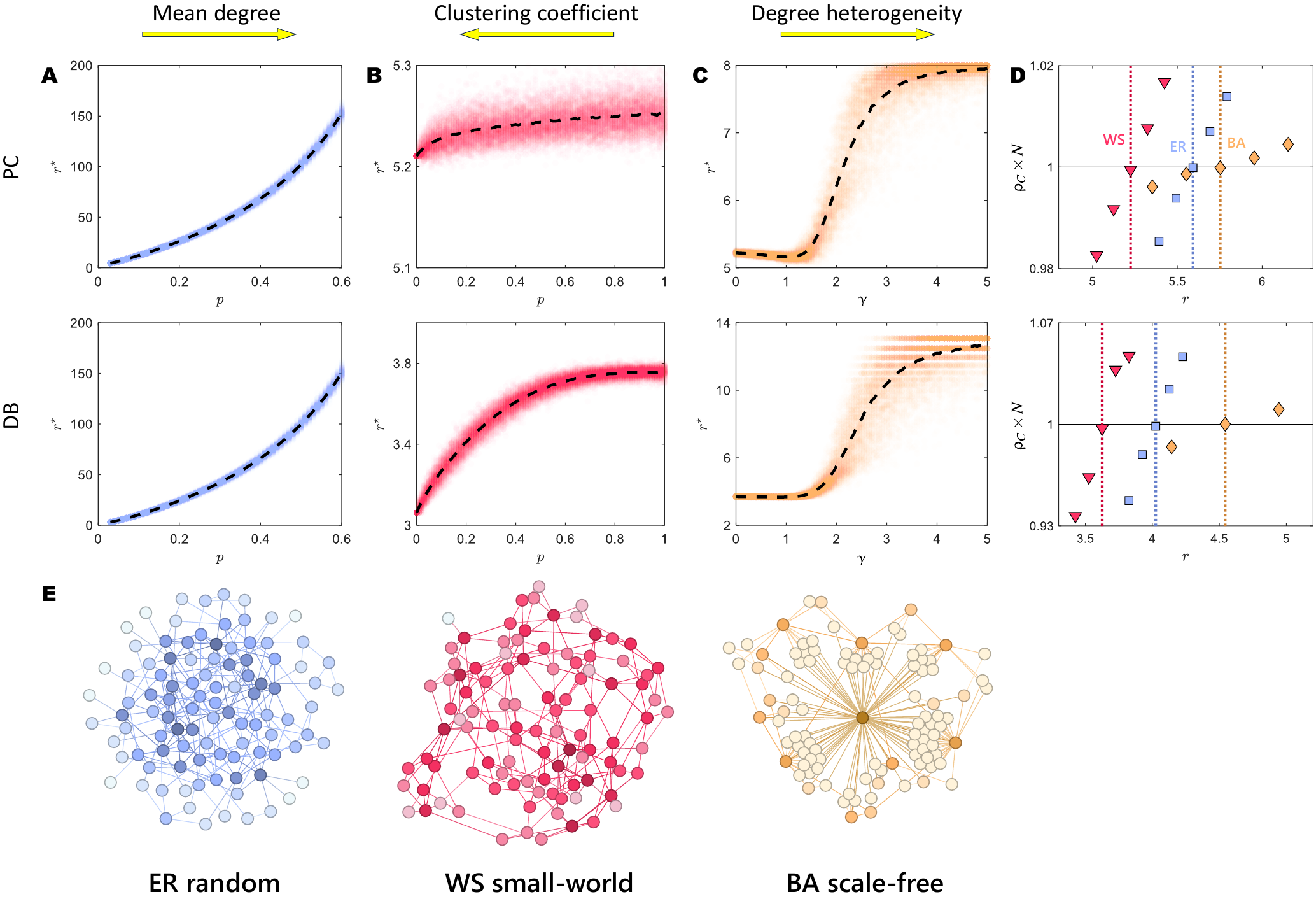}
	\caption{\textbf{Effects of local structures on cooperation in PGGs.} (\textbf{A}) Critical synergy factor $r^\star$ as a function of connecting probability $p$ on random graphs (ER)~\cite{erdos1959random}. The increasing average degree inhibits cooperation. (\textbf{B}) $r^\star$ as a function of rewiring probability $p$ on small-world networks (WS)~\cite{watts1998collective} with $d=2$. The increasing clustering coefficient promotes cooperation. (\textbf{C}) $r^\star$ as a function of generated degree distribution heterogeneity $\gamma$ on scale-free networks (BA)~\cite{barabasi1999emergence,krapivsky2000connectivity} with $m=2$. The increasing degree heterogeneity initially promotes but ultimately inhibits cooperation. In \textbf{A}--\textbf{C}, each point on dashed lines is the average result of 1,000 randomly generated networks (totaling 58,000 in \textbf{A}, 101,000 in \textbf{B}, 101,000 in \textbf{C}). All networks are of size $N=100$ and are connected. (\textbf{D}) Agent-based simulations on selected sample networks confirm theoretical predictions. For ER, $p=4/99$; for WS, $d=2$, $p=0.5$; for BA, $m=2$, $\gamma=2$. (\textbf{E}) Visualization of the sample networks in agent-based simulations. }
	\label{fig_randomgraph}
\end{figure}

Next, we investigate the general roles of local structures on classic random networks, including random (ER), small-world (WS), and scale-free (BA) as shown in Fig.~\ref{fig_randomgraph}\textbf{A}--\textbf{C}. These networks are randomly generated by given parameters.

Random networks reflect the effect of average degree on cooperation when heterogeneity is present (Fig.~\ref{fig_randomgraph}\textbf{A}). We use the Erd{\H{o}}s--R{\'e}nyi (ER) algorithm~\cite{erdos1959random} to generate random networks, which leads to a Poisson degree distribution. Among $N=100$ nodes, there can be at most $N(N-1)/2$ edges. Let the probability of an edge existing be $p$, then the average degree of the network is approximately $\langle k\rangle\approx p(N-1)$, which increases with $p$. We find that the critical synergy factor in PGGs increases with $p$, which indicates that an average degree increase is detrimental to cooperation. This is consistent with previous findings for pairwise games~\cite{ohtsuki2006simple, allen2017evolutionary}.

Small-world networks (WS) reflect the role of clustering coefficients (Fig.~\ref{fig_randomgraph}\textbf{B}). We use the Watts--Strogatz algorithm~\cite{watts1998collective} to generate small-world networks, starting from a ring network of $N=100$ where each node has $2d$ neighbors within distance $d=2$ on both sides. Then, each node rewires the other end of each edge with probability $p$ (the same edge cannot rewire twice; no self-loops or duplicate edges). The clustering coefficient, measuring the abundance of triangles, is approximately $\frac{3(d-1)}{2(2d-1)}(1-p)^3$ (accurate in large populations)~\cite{barrat2000properties}, which decreases as $p$ increases. The critical synergy factor $r^\star$ increases with $p$, which indicates that high clustering promotes cooperation. This result is consistent with previous findings for regular graphs~\cite{su2019spatial} (\ref{sec_appl}), which can be understood as an impact of ``higher-order interactions''~\cite{sheng2024strategy}. Notably, high clustering only notably promotes cooperation under the DB rule. On the one hand, agents interact with second-order neighbors in PGGs (while pairwise games only involve first-order neighbors). On the other hand, the essence of DB update is competition with second-order neighbors, while the PC rule is with first-order neighbors~\cite{allen2014games}. The combination of second-order game interactions and strategy competitions provides enough reach to be influenced by triangles.

Scale-free networks reflect the impact of degree heterogeneity on cooperation (Fig.~\ref{fig_randomgraph}\textbf{C}). We use the algorithm proposed by Krapivsky~et~al.~\cite{krapivsky2000connectivity}, which is an extension of the Barab{\'a}si--Albert (BA)~\cite{barabasi1999emergence}, to generate scale-free networks. Compared to the original BA model, this algorithm allows for adjusting degree heterogeneity directly. We start with $m=2$ isolated initial nodes. The remaining $N-m=98$ nodes then join the existing network one by one. Each new node attaches to $m$ existing nodes (hence the approximate average degree $\langle k\rangle=2m(1-m/N)$). The probability of selecting node $i$ is proportional to ${k_i}^\gamma$, where $\gamma$ is the strength of preferential attachment, determining the degree heterogeneity. When $\gamma=1$, we reduce to the standard BA scale-free network. We find that increasing degree heterogeneity $\gamma$ initially slightly promotes cooperation but ultimately hinders cooperation. In other words, a moderate network heterogeneity is most conducive to cooperation~\cite{pinheiro2017intermediate}. Such a non-monotonic transition in cooperation is due to a complex interplay of multiple factors. For example, increasing degree heterogeneity may also change the clustering coefficient. This is different from some previous studies on pairwise games~\cite{santos2006evolutionary,maciejewski2014evolutionary}. 

The conclusions are more subtle across other model details (Fig.~\ref{fig_randomgraph_sup}). The theoretical results on these random networks are also supported by agent-based simulations (Fig.~\ref{fig_randomgraph}\textbf{D}) on selected structures (Fig.~\ref{fig_randomgraph}\textbf{E}). According to these simulations with consistent average degrees, the small-world network ($\langle k\rangle=4$) is most conducive to cooperation, the random ($\langle k\rangle\approx 4$) is secondary, and the scale-free ($\langle k\rangle=3.84$) is least conducive.

\subsection*{Fundamental advantages of PGGs}

\begin{figure}[!ht]
	\centering
	\includegraphics[width=.95\textwidth]{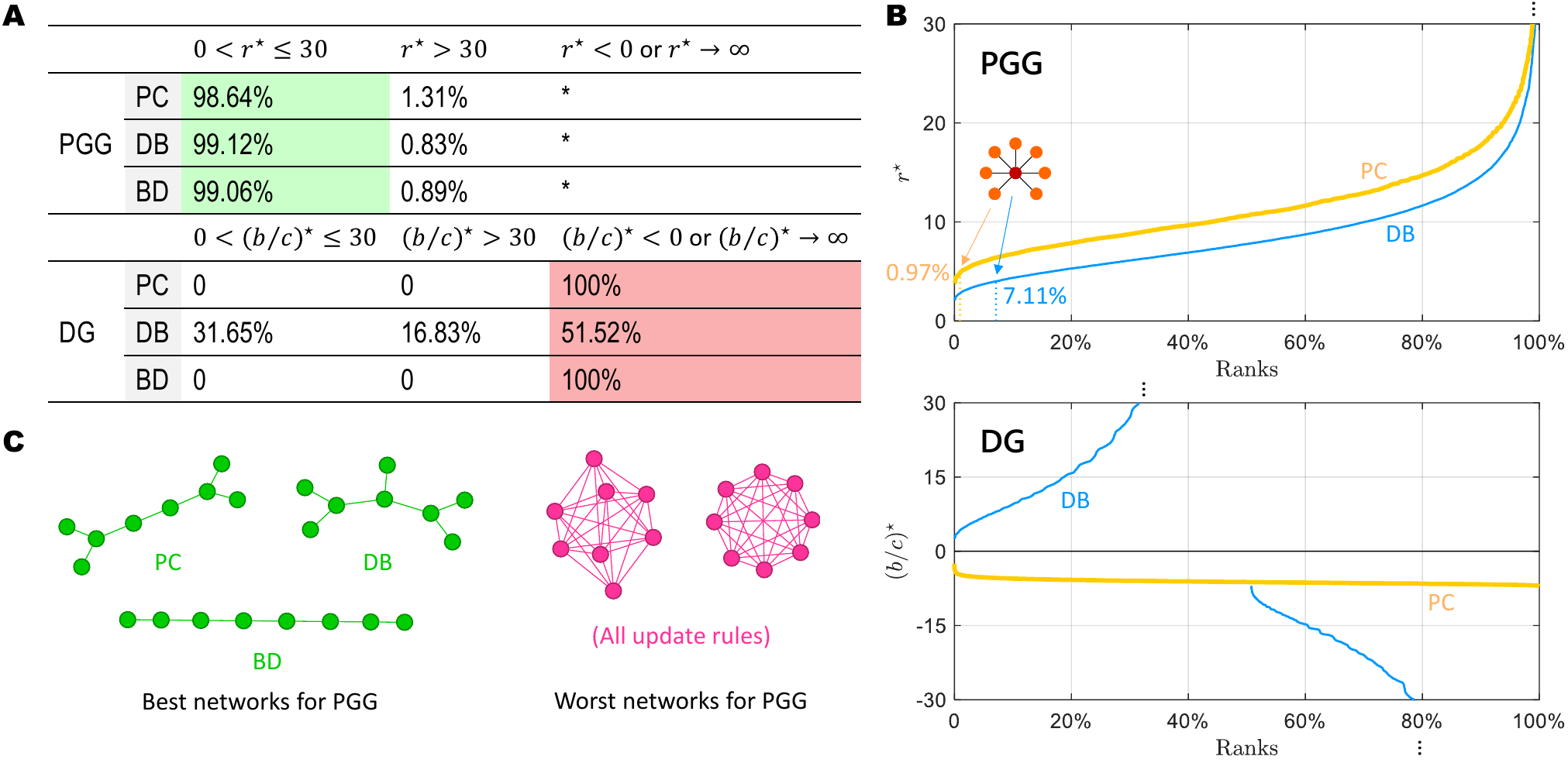}
	\caption{\textbf{Cooperation emerges consistently in PGGs on almost all networks under different update rules, outperforming pairwise DGs.} (\textbf{A}) Among all 12,111 networks of sizes $3\leq N\leq 8$, the fraction of networks classified by their critical synergy factors. Almost all networks have $0<r^\star \leq 30$ in supporting cooperation. The symbol * means that the only structure that does not support cooperation is the fully connected network. In contrast, for pairwise DGs, cooperation is only possible under DB update, with more than half of networks not supporting cooperation. (\textbf{B}) The ranks of all 11,117 networks of size $N=8$ in supporting cooperation for PGGs and DGs under PC and DB update rules. The star graph ranks the top 0.97\% (PC) and 7.11\% (DB). The results in PGGs are consistent under different update rules, while in pairwise DGs they are quite different. (\textbf{C}) The best and worst networks of size $N=8$ for cooperation in PGGs. The best networks differ with update rules. The worst two networks are consistent under all update rules, which are the fully connected (right) and a similar network (left).}
	\label{fig_allnetwork}
\end{figure}

We further investigate the conditions for cooperation in PGGs across all small networks. For population sizes $3\leq N\leq 8$, there are 2 ($N=3$), 6 ($N=4$), 21 ($N=5$), 112 ($N=6$), 853 ($N=7$), and 11,117 ($N=8$) possible networks, respectively. The critical synergy factors on these 12,111 networks are summarized in Fig.~\ref{fig_allnetwork}\textbf{A}. There are 98.64\% (PC), 99.12\% (DB), and 99.06\% (BD) networks where the critical synergy factors are $0<r^\star\leq 30$. The conditions for the success of cooperation are relaxed on almost all networks under various update rules. There are only a few networks (1.31\% for PC, 0.83\% for DB, and 0.89\% for BD) with strict cooperation conditions $r^\star>30$. The last category is $r^\star<0$ or $r^\star\to\infty$ (we also numerically categorize $r>10^3$ as $r\to\infty$ here). When $r^\star<0$, the cooperation condition becomes $r<r^\star$ and cooperation is impossible for meaningful $r>1$. The symbol * means that the only network that does not support cooperation in PGGs is the fully connected network. 

In comparison, cooperation cannot thrive well in the donation game (DG)~\cite{allen2017evolutionary}, where a cooperator donates $b$ to the other player by paying $c$ ($b>c$). The studied quantity here is the benefit-to-cost ratio, $b/c$, which has a critical value $b/c>(b/c)^\star$ over which cooperation is favored. We find the critical values $(b/c)^\star$ fall within unfavorable intervals. Under PC and BD updates, no network can support cooperation ($(b/c)^\star<0$ or $(b/c)^\star\to\infty$ for 100\%). Under the DB update, more than half networks (51.52\%) cannot support cooperation, with only 31.65\% of structures having relaxed cooperation conditions. These conclusions on group PGGs and pairwise DGs remain  valid if we use accumulated payoffs (Figs.~\ref{fig_allnetwork_sup} and \ref{fig_allnetwork_accu_sup}). 

From these insights, we see a fundamental advantage of PGGs, that cooperation can emerge on all networks (except fully connected) and under various update rules, which outperforms pairwise DGs. In PGGs, self-reciprocity plays a key role in the emergence of cooperation: a cooperator can generate personal benefits $rc/G$. Once the synergy factor exceeds the group size, $r>G$, a single cooperator can generate a net benefit independently ($rc/G-c>0$). To clarify, such self-reciprocity cannot induce cooperation alone. One example is the fully connect graph (rightmost of Fig.~\ref{fig_allnetwork}\textbf{C}), where the critical synergy factor is $r\to\infty$, and individuals do not choose cooperation even when $r>N$. This is because we did not assume individual intelligence that perceives the net benefit of cooperation. Instead, individuals follow the update rule (PC, DB, or others), in which individuals compare their payoff to their neighbors. Obviously, on the fully connected graph, the payoff of defectors is always higher than that of cooperators. Only a certain group separation on the network can make the net benefits from cooperation comparable, thus making individuals choose to cooperate. Therefore, we attribute the fundamental advantage of PGGs to self-reciprocity through group separation on networks, where the role of networks is indispensable.

We show the ranks of all 11,117 networks of size $N=8$ under PC and DB rules in Fig.~\ref{fig_allnetwork}\textbf{B}. Networks with smaller critical synergy factors $r^\star$ are more advantageous for cooperation and appear earlier in the ranking. PC and DB updates show consistency in their rank trends, with the DB rule slightly more favorable for cooperation. In contrast, the ranks in pairwise DGs are quite different under PC and DB rules. This is curious, since there is no qualitative difference between these update rules, which consistently assume the advantages of high payoffs in strategy evolution. Here, PGGs show another fundamental advantage, that they perform consistently across various details in update rules.

Additionally, it is worth mentioning that the star graph ranks in the top 0.97\% and 7.11\% for the PC and DB updates, which agrees with our previous conclusion that star graphs are among the most conducive networks for cooperation in PGGs. More generally, we present the best and worst networks of size $N=8$ as shown in Fig.~\ref{fig_allnetwork}\textbf{C}. The best networks vary in different update rules, but they all have low average degrees. From BD, PC, to DB rules, the best networks shift from linear to star-like in shape. The worst networks are consistent across all update rules, which have high degrees, from fully connected to similar structures.

\subsection*{Robustness of PGGs on empirical networks}

\begin{figure}[!ht]
	\centering
	\includegraphics[width=\textwidth]{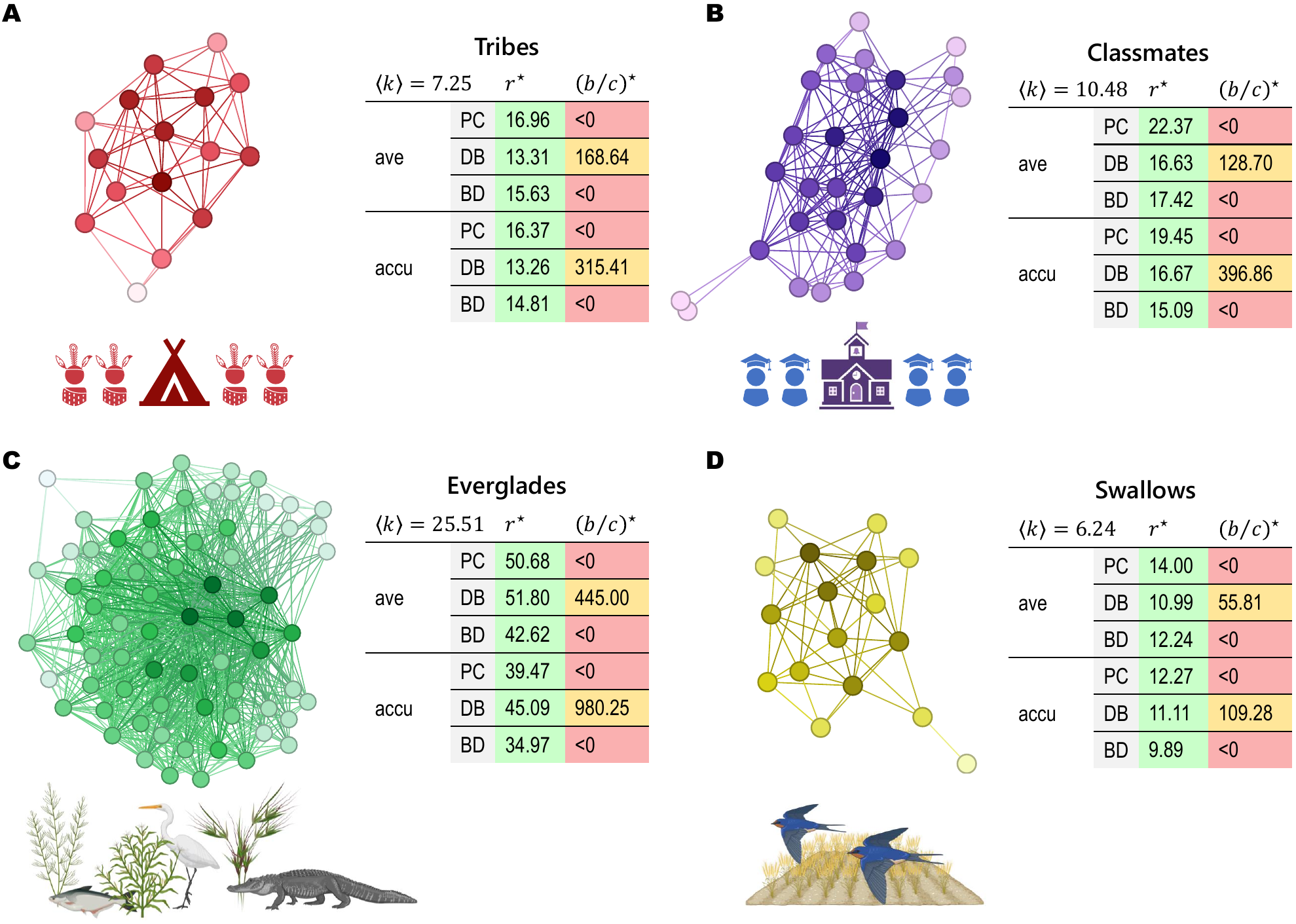}
	\caption{\textbf{PGGs could be a promising interaction mode for cooperation on empirical networks.} We analyze PGGs on four empirical networks, where the cooperation conditions are consistently relaxed across various model details in PGGs. In contrast, the conditions for cooperation are strict and inconsistent across different model details in pairwise DGs. The examined empirical systems include: (\textbf{A}) 16 tribes of the Gahuku--Gama alliance structure of the Eastern Central Highlands of New Guinea~\cite{nr,read1954cultures}. (\textbf{B}) 29 seventh-grade students in Victoria, Australia, connected by whom they would prefer to work with~\cite{vickers1981representing}. (\textbf{C}) The networks of trophic interactions that occur among the 69 major taxonomic groups of everglades habitats in South Florida ecosystems~\cite{nr,ulanowicz1998network,melian2004food}. (\textbf{D}) The network of body contact interactions among 17 North American barn swallows~\cite{nr,levin2016stress}. }
	\label{fig_empirical}
\end{figure}

The fundamental advantages of PGGs were presented on small networks of sizes $N\leq 8$, but we are also interested in the robustness of these advantages in large real-world systems. Here, we analyze four empirical networks and calculate the critical synergy factors for the success of cooperation on each of them (Fig.~\ref{fig_empirical}). The results are presented across various model details, including three update rules (PC, DB, and BD) and two payoff calculations (average (ave) and accumulated (accu)). 

The first two cases are human societies, including a pre-modern and a modern social structure. The pre-modern example is 16 tribes of the Gahuku--Gama alliance structure of the Eastern Central Highlands of New Guinea~\cite{nr,read1954cultures} (Fig.~\ref{fig_empirical}\textbf{A}), with normalized critical synergy factors $r^\star/\langle k\rangle=1.83\sim 2.33$ in PGGs and benefit-to-cost ratios $(b/c)^\star/\langle k\rangle<0$ \& $(b/c)^\star/\langle k\rangle\geq 23.26$ in DGs. The modern social structure is 29 seventh-grade students in Australia based on their preferred working partners~\cite{vickers1981representing} (Fig.~\ref{fig_empirical}\textbf{B}), with normalized critical synergy factors $r^\star/\langle k\rangle=1.44\sim 2.13$ and benefit-to-cost ratios $(b/c)^\star/\langle k\rangle<0$ \& $(b/c)^\star/\langle k\rangle\geq 12.28$. The third case is an ecological structure, the trophic interactions among 69 major taxonomic groups of the various everglades habitats in South Florida ecosystems~\cite{nr,ulanowicz1998network,melian2004food} (Fig.~\ref{fig_empirical}\textbf{C}), which has $r^\star/\langle k\rangle=1.37\sim 2.03$ and $(b/c)^\star/\langle k\rangle<0$ \& $(b/c)^\star/\langle k\rangle\geq 17.44$. The last case is an animal social network, constructed with edges of body contact interactions among 17 North American barn swallows~\cite{nr,levin2016stress} (Fig.~\ref{fig_empirical}\textbf{D}), with $r^\star/\langle k\rangle=1.58\sim 2.24$ and $(b/c)^\star/\langle k\rangle<0$ \& $(b/c)^\star/\langle k\rangle\geq 8.94$. Since the cooperation conditions increase with average degrees, we mentioned the corresponding normalized conditions above for comparison. According to these real data, we verify that PGGs can support cooperation on these real-world population structures with relaxed conditions and are robust across all model details. From this perspective, PGGs could be a promising interaction mode for cooperation in real-world systems.

On the other hand, we can see that the required critical synergy factors are always greater than the average degree (e.g., for primitive tribes, $r^\star>7.25$), which means that these real networks are not as conducive to cooperation as regular graphs. Therefore, even with an interaction mode like PGGs that are conducive to cooperation, networks alone cannot fully explain the emergence of cooperation. There could be other factors that come together to support cooperation in these social structures.

We also see that accumulated payoffs are slightly more conducive to cooperation in PGGs (the only exceptions are the DB update in ``classmates'' and ``swallows''). This is also consistent with our intuition that payoffs from different groups are physically accumulated in the real world. In contrast, the average payoffs serve normalization and theoretical analysis, which are less conducive to cooperation on the studied real-world networks.

\section*{Discussion}
As a group-based interaction, PGGs capture the empirical fact that individuals not only interact with their direct friends, but also with their second-order neighbors in the circle of the direct friends~\cite{granovetter1973strength,newman2001structure,hao2024proper}. Here, we develop an analytical theory to predict the evolution of cooperation in PGGs on arbitrary population structures. Given a network as input, the theory yields the cooperation condition for PGGs on that network, which is accurate under weak selection and can remain qualitatively consistent when selection becomes numerically stronger (Fig.~\ref{fig_largeselection}). While prior work focused on pairwise games~\cite{lieberman2005evolutionary,allen2017evolutionary}, our theory advances this line of research to group-based PGGs.

Using our theory, we find that a class of networks, such as the star graph, which were considered unfavorable for cooperation in pairwise games~\cite{allen2017evolutionary,fotouhi2018conjoining}, can in fact notably promote cooperation in PGGs. An intuitive explanation may be self-reciprocity embedded in PGGs, i.e., cooperation can yield a net benefit $rc/G$. Accordingly, when the net benefit exceeds the cost, $rc/G>c\Leftrightarrow r>G$, cooperation becomes individually favorable and the situation may not even constitute a social dilemma~\cite{hauert2002volunteering}. However, we emphasize that self-reciprocity alone is insufficient to fully explain the emergence of cooperation in PGGs. One counterexample is fully connected networks, in which cooperation cannot emerge even when $r>N$. This is because update rules (e.g., PC) in evolutionary dynamics rely on local fitness-based imitation rather than rational decision-making. In other words, only the network creates certain group separations can individuals ``perceive'' the net benefit of cooperation. Our analytical framework captures such structural effects, showing that a non-dilemma condition~\cite{hauert2002volunteering} must be complemented by network-driven group structure for cooperation to emerge in evolution.

These insights also raise questions about the role of evolution in cooperation. Evolution can not only promote cooperation beyond the reach of individual rationality, but also hinder cooperation that rational individuals might otherwise pursue. A similar phenomenon was observed in the social goods dilemma~\cite{mcavoy2020social}, where evolution promotes prosocial behavior even when such behavior is not collectively optimal. Compared to the social goods dilemma---where interactions are limited to first-order neighbors and second-order edges---our PGG model incorporates second-order neighbors, capturing a broader scope of interaction. An even broader interaction structure was found in the diffusible public goods model~\cite{allen2013spatial}, where public goods can spread to and influence more distant nodes. However, our work provides analytical results for the most widely studied setting~\cite{szabo2002phase,perc2013evolutionary}, where the diffusion is minimal and thus more conducive to cooperation~\cite{allen2013spatial}.

Given the typically high clustering coefficients in real-world networks~\cite{watts1998collective,albert2002statistical}, we highlight the significance of clustering effects arising from second-order interactions~\cite{battiston2021physics} in PGGs---an effect absent in pairwise games. Previous studies have also revealed clustering effects in PGGs, but these were confined to regular graphs~\cite{su2019spatial,wang2023inertia}. We demonstrate the positive role of clustering in promoting cooperation on general networks. Similar phenomena have also been observed in studies of higher-order networks~\cite{alvarez2021evolutionary,sheng2024strategy,civilini2024explosive}. However, the notion of clustering in higher-order frameworks differs from that in traditional network models, because they require a separately defined higher-order interaction graph. In addition, higher-order studies~\cite{sheng2024strategy} primarily investigate the effects of nonlinear payoff functions~\cite{hauert2006synergy,allen2024nonlinear}, and their scope is limited to interactions involving no more than three individuals. By contrast, our framework supports standard PGGs with arbitrary group sizes on conventional networks.

More generally, we test all possible networks with sizes $3\leq N\leq 8$ and find that PGGs consistently promote cooperation (except on fully connected networks). In contrast, pairwise games fail to produce cooperation on most networks~\cite{allen2017evolutionary} (Figs.~\ref{fig_allnetwork}, \ref{fig_allnetwork_sup}, and \ref{fig_allnetwork_accu_sup}). This difference can be attributed to how benefits and costs are accounted, or the self-reciprocity embedded in PGGs. As previously noted, this built-in feature relies on the network to manifest its effect. Our main contribution is to establish a general theory of PGGs on networks, which is equally important as the foundational theory of pairwise donation games.

We examine four empirical networks, including primitive tribes, junior students, everglades, and barn swallows. The PGG can produce consistent cooperation across all kinds of evolutionary details on these real networks. The results thus imply that PGGs could be a promising interaction mode for the emergence of cooperation in real-world systems. However, these real-world networks do not maximize cooperation. Their cooperation thresholds are stricter than those of regular graphs with the same average degree. In the real world, cooperation is shaped by the coexistence of multiple behavioral and social mechanisms~\cite{nowak2006five}, such as memory of past interactions, conformity to social norms, and concerns about punishment or reward. By excluding these mechanisms, our analysis isolates the effect of network structure, and under such minimal assumptions it is expected that the critical synergy factors $r^\star$ appear relatively large.

Although real-world interactions are complex to empirically disentangle the influence of network from other behavioral factors, controlled human experiments provide a means to do so. Rand~et~al.~\cite{rand2014static} showed that theoretical predictions about cooperation thresholds qualitatively match evolutionary outcomes in human groups and that networks with smaller average degree have lower cooperation thresholds and exhibit higher cooperation levels. Similar consistencies of theoretical analysis and realistic observations have been evidenced by many experimental studies~\cite{santos2006cooperation,rand2011dynamic}. By analogy, we expect that network structures with lower critical synergy factors $r^\star$ in our framework should likewise be more conducive to the evolution of cooperation.

For applications to larger population sizes, one challenge is the computational complexity associated with Eqs.~(\ref{eq_PCcondi})--(\ref{eq_Upsilonij}). The main linear equations~(\ref{eq_tauij}) of coalescent time $\tau_{ij}$ are the same as those in Ref.~\cite{allen2017evolutionary}, which can be solved in polynomial time. The additional quantities $\Upsilon_{ij}$ (Eq.~(\ref{eq_Upsilonij})) required for PGGs are directly derived from the $\tau_{ij}$ values and do not lead to a higher computational complexity. We examine six larger empirical networks to demonstrate the computational feasibility of our theory (Fig.~\ref{fig_empirical_sup}).

Our theory lays the foundation for exploring a large number of extended mechanisms~\cite{perc2010coevolutionary} in PGGs, such as inertia~\cite{wang2023inertia,wang2023imitation}, on general population structures. One can investigate the additional effects of different networks on these mechanisms. One can also study the evolutionary dynamics of PGGs on multi-layer~\cite{su2022evolution} and dynamic networks~\cite{su2023strategy}, which were only studied in pairwise games previously. Moreover, one can generalize our results to weighted networks and also study the outcomes with arbitrary initial conditions~\cite{mcavoy2021fixation}. The algorithm for group interactions based on second-order neighbors can also be used to study other multiplayer games, including the nonlinear PGGs~\cite{archetti2011coexistence,archetti2012game,pena2012group,pena2015evolutionary}, which bear unknown complexity~\cite{wu2013dynamic,mcavoy2016structure}.

\section*{Methods}
%

\subsection*{Theoretical conditions for the success of cooperation}
Here, we briefly summarize the mathematical derivations of the cooperation condition in PGGs. The system state is denoted by $\mathbf{x}=(x_1,x_2,\dots,x_N)$, where $x_i=1$ if agent $i$ cooperates and $x_i=0$ if it defects. In this way, we can formalize the payoff calculation on networks. The actual payoff of agent $i$ at system state $\mathbf{x}$, denoted by $f_i(\mathbf{x})$, is expressed as 
\begin{align}\label{eq_payoff_mth}
	f_i(\mathbf{x})
	&=\frac{1}{G_i}\sum_{l\in\mathcal{G}_i}\left(\frac{r\sum_{\ell\in\mathcal{G}_l} x_\ell c}{G_l}-x_i c\right)
	\nonumber\\
	&=\frac{1}{1+k_i}\left[
	\left(\frac{r(x_i+\sum_{l\in\mathcal{N}_i} x_l)c}{k_i+1}-x_i c\right)
	+\sum_{l\in\mathcal{N}_i} \left(\frac{r(x_l+\sum_{\ell\in\mathcal{N}_l} x_\ell)c}{k_l+1}-x_i c\right)
	\right].
\end{align}
Agent $i$ plays $G_i=1+k_i$ games, organized by itself and its $k_i$ neighbors $l\in\mathcal{N}_i$, which form the group $\mathcal{G}_i=\{i\}\cup\mathcal{N}_i$. In the game organized by agent $l$, there are $G_l$ players in the group.

In the weak selection limit ($0<\delta\ll 1$), the dynamics of strategy evolution almost reduces to the Voter model~\cite{clifford1973model}, where the marginal role of games does not influence strategy distributions. Therefore, in previous literature~\cite{allen2017evolutionary,allen2019mathematical} (or \ref{sec_theory}), the conditions for the success of cooperation were obtained under neutral drift ($\delta=0$) and remain unchanged for different payoff calculations. 

In other words, the key difference in analyzing PGGs is the payoff calculation. We only need to substitute the payoffs of PGGs, i.e., Eq.~(\ref{eq_payoff_mth}), into the previously obtained cooperation conditions, which (under the PC rule) is
\begin{equation}\label{eq_coopcondi_PC_mth}
	\frac{1}{4N^2\langle k\rangle} \sum_{i,j\in\mathcal{N}}k_i p_{ij}\mathbb{E}_\text{RMC}^\circ [(x_i-x_j)(f_i(\mathbf{x})-f_j(\mathbf{x}))]>0.
\end{equation}
Applying $f_i(\mathbf{x})$ ($f_j(\mathbf{x})$) of Eq.~(\ref{eq_payoff_mth}) and considering 
\begin{equation}\label{eq_tauijERMC_PC_mth}
	\tau_{ij}=\frac{\dfrac{1}{2}-\mathbb{E}_\text{RMC}^\circ[x_i x_j]}{K/4},
\end{equation}
we can calculate the condition of Eq.~(\ref{eq_coopcondi_PC_mth}) as
\begin{align}\label{eq_coopcondi_2_PC_mth}
	&~\sum_{i,j\in\mathcal{N}}k_i p_{ij}\Bigg\{\left(\frac{rc}{(k_i+1)^2}-c\right) \tau_{ij}
	+\frac{rc}{k_i+1}\sum_{l\in\mathcal{N}_i}\left(\frac{1}{k_i+1}+\frac{1}{k_l+1}\right) \left(-\tau_{il}+\tau_{jl}\right)
	\nonumber\\
	&+\frac{rc}{k_i+1}\sum_{l\in\mathcal{N}_i}\frac{1}{k_l+1}\sum_{\ell\in\mathcal{N}_l}\left(-\tau_{i\ell}+\tau_{j\ell}\right)
	-\left(\frac{rc}{(k_j+1)^2}-c\right)\left(-\tau_{ij}\right)
	\nonumber\\
	&-\frac{rc}{k_j+1}\sum_{l\in\mathcal{N}_j}\left(\frac{1}{k_j+1}+\frac{1}{k_l+1}\right)\left(-\tau_{il}+\tau_{jl}\right)
	-\frac{rc}{k_j+1}\sum_{l\in\mathcal{N}_j}\frac{1}{k_l+1}\sum_{\ell\in\mathcal{N}_l}\left(-\tau_{i\ell}+\tau_{j\ell}\right)\Bigg\}
	>0 \nonumber\\
	\Leftrightarrow
	&~r>\frac{2\sum_{i,j\in\mathcal{N}}k_i p_{ij}\tau_{ij}}{\sum_{i,j\in\mathcal{N}}k_i p_{ij}(\Upsilon_{ij}+\Upsilon_{ji})},
\end{align}
which is equivalent to Eq.~(\ref{eq_PCcondi}) in the main text, with $\tau_{ij}$ and $\Upsilon_{ij}$ obtained through Eqs.~(\ref{eq_tauij}) and (\ref{eq_Upsilonij}), respectively. See \ref{sec_theory} for the meaning of mathematical symbols in Eqs.~(\ref{eq_coopcondi_PC_mth})--(\ref{eq_coopcondi_2_PC_mth}) and their detailed deductions. 

The results under the DB and BD update rules, including those with accumulated payoffs, follow the same idea. We have the cooperation conditions (Eq.~(\ref{eq_coopcondi_DB}) for DB and Eq.~(\ref{eq_coopcondi_BD}) for BD) which are independent of payoff calculations. Applying $f_i(\mathbf{x})$ ($f_j(\mathbf{x})$) of Eq.~(\ref{eq_payoff_mth}) (or the ones for accumulated payoffs) and their respect $\tau_{ij}$ values leads to the resultant cooperation conditions, as detailed in \ref{sec_theory}.

\subsection*{Agent-based simulations}
We conduct the agent-based simulations using the standard Monte Carlo method. The selection strength $\delta$ is between 0.01~\cite{su2019spatial,wang2023inertia} and 0.025~\cite{gore2009snowdrift,allen2017evolutionary}, which is considered numerically weak. The cost of cooperation $c$ is set to 1~\cite{szabo2002phase}. In the initial state, there is one random cooperator and $N-1$ defectors. Each full Monte Carlo step (MCS) contains $N$ elementary time steps where a random focal agent is selected to update the strategy, so that every agent is updated once on average. We allow for up to $4\times 10^5$ full MCS, which is theoretically infinite~\cite{allen2017evolutionary}. If the fraction of cooperators hits a fixation state ($\rho_C=1$ or $\rho_C=0$), we end the current run and record the result. If a fixation state is not reached within the maximally allowed MCS, we take the actual $\rho_C$ at the last step as the result. We repeat the simulations $10^6$--$10^9$ times independently under the given game parameters and network, averaging the final fraction of cooperators $\rho_C$ in these runs as the actual result of $\rho_C$. If the average cooperation fraction $\rho_C>1/N$, then evolution favors cooperation.


\newpage
\subsection*{Acknowledgements}
We thank the referees for constructive feedback.

\subsection*{Funding}
Q.S. acknowledges support from the National Natural Science Foundation of China (No. 62473252, 24Z990200677) and from the State Key Laboratory of Autonomous Intelligent Unmanned Systems (No. ZZKF2025-1-4).

\subsection*{Author contributions}
C.W.: Writing - original draft, Conceptualization, Investigation, Writing - review \& editing, Methodology, Validation, Supervision, Formal analysis, Software, Project administration, Visualization.

Q.S.: Writing - original draft, Conceptualization, Writing - review \& editing, Funding acquisition, Supervision, Formal analysis, Project administration.

\subsection*{Competing interests}
The authors declare no competing interests.

\subsection*{Data, code, and materials availability}
All data needed to evaluate the conclusions in the paper are present in the paper and/or the Supplementary Materials. All numerical solutions for the theoretical predictions were performed in MATLAB R2022a. All agent-based simulations were performed in Python 3.9. Codes are publicly available on GitHub at \url{https://github.com/ChaoqianSCI/Spatial-PGGs-any-network}. To ensure long-term accessibility and reproducibility, the version of the code used for this study has been archived in Zenodo at \url{https://doi.org/10.5281/zenodo.18060514}. The paper did not generate new materials.

\newpage
\begin{center}
  \Huge Supplementary Information for
\end{center}

\noindent
\textbf{\Huge Public goods games on any population structure}
\vspace{2em}

\noindent
{\Large Chaoqian Wang, Qi Su}
\vspace{1em} 

\noindent
e-mail: CqWang814921147@outlook.com~(C.~Wang); qisu@sjtu.edu.cn (Q.~Su)

\addtocontents{toc}{\protect\setcounter{tocdepth}{3}}
\tableofcontents

\renewcommand{\theequation}{S\arabic{equation}}
\setcounter{equation}{0}

\let\oldsection\thesection
\renewcommand{\thesection}{Supplementary Note~\arabic{section}}
\renewcommand{\thesubsection}{\oldsection.\arabic{subsection}}

\renewcommand{\thefigure}{S\arabic{figure}}
\setcounter{figure}{0}

\newpage

\section{: Conditions for the success of cooperation in PGGs}\label{sec_theory}

\subsection{Payoff calculation for PGGs on any network}
According to the description in the main text, agents participate in the PGGs organized by themselves and their neighbors, taking the average payoffs obtained in these games as the actual payoff. To formally express this algorithm, we denote the system state (i.e., the strategies of all agents) by $\mathbf{x}=(x_1,x_2,\dots,x_N)$, in a population of size $N$. If agent $i$ cooperates, $x_i=1$. If agent $i$ defects, $x_i=0$. The full cooperation state can be written by $\mathbf{C}=\mathbf{1}=(1,1,\dots,1)$ and the full defection state is $\mathbf{D}=\mathbf{0}=(0,0,\dots,0)$. The set consisting of all possible system states is denoted by $\mathbf{X}$. 

We denote $f_i(\mathbf{x})$ as the average payoff that agent $i$ obtains from the PGGs organized by itself and its neighbors at system state $\mathbf{x}$. Literally, $f_i(\mathbf{x})$ follows the calculation in Eq.~(\ref{eq_payoff}):
\begin{align}\label{eq_payoff}
	f_i(\mathbf{x})
	&=\frac{1}{G_i}\sum_{l\in\mathcal{G}_i}\left(\frac{r\sum_{\ell\in\mathcal{G}_l} x_\ell c}{G_l}-x_i c\right)
	\nonumber\\
	&=\frac{1}{1+k_i}\left[
	\left(\frac{r(x_i+\sum_{l\in\mathcal{N}_i} x_l)c}{k_i+1}-x_i c\right)
	+\sum_{l\in\mathcal{N}_i} \left(\frac{r(x_l+\sum_{\ell\in\mathcal{N}_l} x_\ell)c}{k_l+1}-x_i c\right)
	\right]
	\nonumber\\
	&=\left(\frac{rc}{(k_i+1)^2}-c\right)x_i
	+\frac{rc}{k_i+1}\sum_{l\in\mathcal{N}_i}\left(\frac{1}{k_i+1}+\frac{1}{k_l+1}\right)x_l
	+\frac{rc}{k_i+1}\sum_{l\in\mathcal{N}_i}\frac{1}{k_l+1}\sum_{\ell\in\mathcal{N}_l}x_\ell.
\end{align}
In the first line, $\mathcal{G}_i=\{i\}\cup\mathcal{N}_i$ is the group containing oneself and its neighbors. In the second line, the former item is the PGG organized by agent $i$, and the latter item is the PGGs organized by its neighbors $l\in\mathcal{N}_i$. The third line is a simplification of the second line for later use.

\subsection{General conditions for the success of cooperation}\label{sec_coopcondi}
According to Ref.~\cite{allen2019mathematical}, the general condition for the success of cooperation on arbitrary networks is
\begin{equation}\label{eq_coopcondi}
	\mathbb{E}_\text{RMC}^\circ [\hat{\Delta}'_\text{sel}(\mathbf{x})]>0,
\end{equation}
where, the upper-right corner label $\circ$ of a quantity is to take the value of this quantity at $\delta=0$ (neutral drift). The upper-right corner label $'$ of a quantity is to calculate the first-order derivative of this quantity with respect to $\delta$ at $\delta=0$.

$\hat{\Delta}_\text{sel}(\mathbf{x})$ is the change in the cooperation fraction within an elementary Monte Carlo step (MCS) due to strategy learning (i.e., selection, abbreviated as ``sel''), weighted by reproduction numbers, at system state $\mathbf{x}$. The expression of $\hat{\Delta}_\text{sel}(\mathbf{x})$ is~\cite{nowak2010evolution}
\begin{equation}\label{eq_RVDeltasel}
	\hat{\Delta}_\text{sel}(\mathbf{x})=\frac{1}{N}\sum_{i\in\mathcal{N}}x_i (\hat{b}_i(\mathbf{x})-\hat{d}_i(\mathbf{x})),
\end{equation}
where $\hat{b}_i(\mathbf{x})$ and $\hat{d}_i(\mathbf{x})$ represent the birth and death probabilities of agent $i$, weighted by reproduction numbers, respectively.

Before weighted by reproduction numbers, $\hat{b}_i(\mathbf{x})$ and $\hat{d}_i(\mathbf{x})$ are given by $b_i(\mathbf{x})=\sum_{j\in\mathcal{N}}e_{ij}(\mathbf{x})$ and $d_i(\mathbf{x})=\sum_{j\in\mathcal{N}}e_{ji}(\mathbf{x})$. An agent's strategy reproduces if learned by other agents, or dies if the agent adopts the strategy of others. Here, $e_{ij}(\mathbf{x})$ is the probability that agent $i$ transmits its strategy to agent $j$, determined by specific strategy update rules.

The birth and death probabilities of agent $i$ weighted by reproduction numbers, $\hat{b}_i(\mathbf{x})$ and $\hat{d}_i(\mathbf{x})$, take the following form:
\begin{subequations}\label{eq_RVbidi}
	\begin{align}
		\hat{b}_i(\mathbf{x})&=\sum_{j\in\mathcal{N}}e_{ij}(\mathbf{x})v_j, \\
		\hat{d}_i(\mathbf{x})&=\sum_{j\in\mathcal{N}}e_{ji}(\mathbf{x})v_i,
	\end{align}
\end{subequations}
where, $v_i$ and $v_j$ represent the reproduction numbers of agents $i$ and $j$, respectively. The reproduction numbers $\{v_i\}_{i\in\mathcal{N}}$ of all agents in the population are defined based on the fact that under neutral drift, natural selection does not influence the change in strategy proportions ($\hat{\Delta}^\circ_\text{sel}(\mathbf{x})=0$). Additionally, considering normalization, the average reproduction number for each agent is set to 1. Therefore, the following equation holds~\cite{allen2019mathematical}:
\begin{subequations}\label{eq_veqs}
	\begin{align}
		\hat{d}^\circ_i(\mathbf{x})&=\hat{b}^\circ_i(\mathbf{x})\Leftrightarrow \sum_{j\in\mathcal{N}}e^\circ_{ji}(\mathbf{x})v_i=\sum_{j\in\mathcal{N}}e^\circ_{ij}(\mathbf{x})v_j, \\
		\sum_{i\in\mathcal{N}}v_i&=N.
	\end{align}
\end{subequations}
The first line results from $\hat{\Delta}^\circ_\text{sel}(\mathbf{x})=0$, while the second line is for normalization. From Eqs.~(\ref{eq_veqs}), the reproduction numbers of all agents can be solved given the network and update rules.

Since there is no expected change in $\hat{\Delta}_\text{sel}(\mathbf{x})$ caused by natural selection under neutral drift, we can study the first derivative of $\hat{\Delta}_\text{sel}(\mathbf{x})$ with respect to $\delta$ at $\delta=0$ in order to quantitatively analyze $\hat{\Delta}_\text{sel}(\mathbf{x})$. Substituting Eqs.~(\ref{eq_RVbidi}) into Eq.~(\ref{eq_RVDeltasel}) and taking the derivative, we obtain
\begin{align}\label{eq_RVDeltasel'}
	\hat{\Delta}'_\text{sel}(\mathbf{x})
	&=\frac{1}{N}\sum_{i,j\in\mathcal{N}}x_i (e'_{ij}(\mathbf{x})v_j-e'_{ji}(\mathbf{x})v_i) \nonumber\\
	&=\frac{1}{2N}\sum_{i,j\in\mathcal{N}}(x_i-x_j) (e'_{ij}(\mathbf{x})v_j-e'_{ji}(\mathbf{x})v_i).
\end{align}
Taking the equivalent form in the second line (based on the symmetry between $i$ and $j$) can facilitate subsequent calculations. At this point, by calculating the strategy reproduction probabilities $\{e_{ij}(\mathbf{x})\}_{i,j\in\mathcal{N},\mathbf{x}\in\mathbf{X}}$ and reproduction numbers $\{v_i\}_{i\in\mathcal{N}}$ under the given update rule, we can obtain the corresponding value of $\hat{\Delta}'_\text{sel}(\mathbf{x})$.

Another concept that appears in Eq.~(\ref{eq_coopcondi}) is $\mathbb{E}_\text{RMC}[\cdot]$, where RMC stands for the rare-mutation conditional distribution. Before defining RMC, it is necessary to define MSS: the mutation-selection stationary distribution. This distribution describes the system's state when a mutation mechanism is present. The introduction of the mutation mechanism is intended to construct a mathematically tractable, complete Markov chain. Suppose that in each elementary Monte Carlo step, the focal agent mutates with probability $u$ (i.e., mutation): it switches to either cooperation or defection with probability $1/2$ respectively. With the remaining probability $1-u$, the agent update the strategy (i.e., selection) according to the given update rule. The weak mutation limit $u\to 0$ leads to the model we study in the main text, where strategy updates depend solely on the strategy update rule.

$\Pi_\text{MSS}(\mathbf{x})$ represents the probability that the system stablizes at state $\mathbf{x}$ under the mutation-selection stationary distribution. The sum of probabilities for all possible stationary states is 1, i.e., $\sum_{\mathbf{x}\in\mathbf{X}}\Pi_\text{MSS}(\mathbf{x})=1$. Obviously, in the weak mutation limit $u\to 0$, the system has only two stable states: full cooperation ($\mathbf{x}=\mathbf{C}$) and full defection ($\mathbf{x}=\mathbf{D}$): $\Pi_\text{MSS}(\mathbf{C})\to \rho_C/(\rho_C+\rho_D)$, $\Pi_\text{MSS}(\mathbf{D})\to \rho_D/(\rho_C+\rho_D)$~\cite{allen2014measures,van2015social}. For $\mathbf{x}\notin\{\mathbf{C},\mathbf{D}\}$, $\Pi_\text{MSS}(\mathbf{x})\to 0$.

On this basis, the rare-mutation conditional (RMC) distribution describes the distribution of system states among possible states other than full cooperation and full defection as $u\to 0$. $\Pi_\text{RMC}(\mathbf{x})$ represents the probability that the system is in state $\mathbf{x}$ under the RMC distribution, satisfying the normalization condition $\sum_{\mathbf{x}\in\mathbf{X}\setminus\{\mathbf{C},\mathbf{D}\}}\Pi_\text{RMC}(\mathbf{x})=1$ (note that here $\mathbf{x}$ is restricted to $\mathbf{x}\in\mathbf{X}\setminus\{\mathbf{C},\mathbf{D}\}$, i.e., $\mathbf{x}\notin\{\mathbf{C},\mathbf{D}\}$). By this definition, $\Pi_\text{RMC}(\mathbf{x})$ can be derived from $\Pi_\text{MSS}(\mathbf{x})$, 
\begin{equation}
	\Pi_\text{RMC}(\mathbf{x})=\lim_{u\to 0}\frac{\Pi_\text{MSS}(\mathbf{x})}{1-\Pi_\text{MSS}(\mathbf{C})-\Pi_\text{MSS}(\mathbf{D})}.
\end{equation}

$\mathbb{E}_\text{MSS}[\cdot]$ and $\mathbb{E}_\text{RMC}[\cdot]$ represent the expected value under the corresponding distribution, i.e., the sum of the products of all possible state variables in $[\cdot]$ and their respective probabilities. For example, given a function $f(\mathbf{x})$ of the system state $\mathbf{x}$, we have 
\begin{align}
	&\mathbb{E}_\text{MSS}[f(\mathbf{x})]=\sum_{\mathbf{x}\in\mathbf{X}} \Pi_\text{MSS}(\mathbf{x})f(\mathbf{x}), \\
	&\mathbb{E}_\text{RMC}[f(\mathbf{x})]=\sum_{\mathbf{x}\in\mathbf{X}\setminus\{\mathbf{C},\mathbf{D}\}} \Pi_\text{RMC}(\mathbf{x})f(\mathbf{x}).
\end{align}

Later, we will need the following property: in the limit $u\to 0$, for any agent $i\in\mathcal{N}$, we have $\mathbb{E}^\circ_\text{MSS}[x_i]=1/2$ and $\mathbb{E}^\circ_\text{RMC}[x_i]=1/2$. To prove, under neutral drift, strategies $C$ and $D$ are indistinguishable and thus interchangeable. From $\rho_C=\rho_D$ and $\rho_C+\rho_D=1$, we can solve for $\rho_C=\rho_D=1/2$. Therefore, based on $\Pi_\text{MSS}(\mathbf{C})\to \rho_C/(\rho_C+\rho_D)$ and $\Pi_\text{MSS}(\mathbf{D})\to \rho_D/(\rho_C+\rho_D)$, we have $\Pi_\text{MSS}(\mathbf{C}),\Pi_\text{MSS}(\mathbf{D})\to 1/2$. Consequently, $\mathbb{E}^\circ_\text{MSS}[x_i]=1/2\times 1+1/2\times 0=1/2$. Similarly, due to the interchangeability of $C$ and $D$, we have $\Pi_\text{RMC}(\mathbf{x})=\Pi_\text{RMC}(\mathbf{1}-\mathbf{x})$, and we can calculate $\mathbb{E}^\circ_\text{RMC}[x_i]=(1/2)\sum_{\mathbf{x}\notin\{\mathbf{C},\mathbf{D}\}} (\Pi_\text{RMC}(\mathbf{x})x_i+\Pi_\text{RMC}(\mathbf{1}-\mathbf{x})(1-x_i))=(1/2)\sum_{\mathbf{x}\notin\{\mathbf{C},\mathbf{D}\}} \Pi_\text{RMC}(\mathbf{x})(x_i+1-x_i)=1/2$.

We define a quantity $K$,
\begin{equation}
	K=\lim_{u\to 0}\frac{u}{1-\Pi_\text{MSS}(\mathbf{C})-\Pi_\text{MSS}(\mathbf{D})}.
\end{equation}
Ref.~\cite{allen2019mathematical} has shown that $K$ exists and is positive.

Later, we will need another property: let $\phi(\mathbf{x})$ be a function that satisfies $\phi(\mathbf{C})=\phi(\mathbf{D})=0$, then we can calculate and find that $\mathbb{E}_\text{RMC}[\phi(\mathbf{x})]$ and $\mathbb{E}_\text{MSS}[\phi(\mathbf{x})]$ are related by Eq.~(\ref{eq_ERMCfromEMSS}):
\begin{align}\label{eq_ERMCfromEMSS}
	\mathbb{E}_\text{RMC}[\phi(\mathbf{x})]
	&=\sum_{\mathbf{x}\in\mathbf{X}\setminus\{\mathbf{C},\mathbf{D}\}} \Pi_\text{RMC}(\mathbf{x})\phi(\mathbf{x}) \nonumber\\
	&=\sum_{\mathbf{x}\in\mathbf{X}\setminus\{\mathbf{C},\mathbf{D}\}} \lim_{u\to 0}\frac{\Pi_\text{MSS}(\mathbf{x})}{1-\Pi_\text{MSS}(\mathbf{C})-\Pi_\text{MSS}(\mathbf{D})}\phi(\mathbf{x}) \nonumber\\
	&=\lim_{u\to 0}\frac{\sum_{\mathbf{x}\in\mathbf{X}\setminus\{\mathbf{C},\mathbf{D}\}}\Pi_\text{MSS}(\mathbf{x})\phi(\mathbf{x})}{1-\Pi_\text{MSS}(\mathbf{C})-\Pi_\text{MSS}(\mathbf{D})} \nonumber\\
	&=\lim_{u\to 0}\frac{\sum_{\mathbf{x}\in\mathbf{X}}\Pi_\text{MSS}(\mathbf{x})\phi(\mathbf{x})}{1-\Pi_\text{MSS}(\mathbf{C})-\Pi_\text{MSS}(\mathbf{D})} \nonumber\\
	&=\lim_{u\to 0}\frac{\mathbb{E}_\text{MSS}[\phi(\mathbf{x})]}{1-\Pi_\text{MSS}(\mathbf{C})-\Pi_\text{MSS}(\mathbf{D})} \nonumber\\
	&=\left(\lim_{u\to 0}\frac{u}{1-\Pi_\text{MSS}(\mathbf{C})-\Pi_\text{MSS}(\mathbf{D})}\right)
	\left(\lim_{u\to 0}\frac{\mathbb{E}_\text{MSS}[\phi(\mathbf{x})]}{u}\right) \nonumber\\
	&=K\left(\lim_{u\to 0}\frac{\mathbb{E}_\text{MSS}[\phi(\mathbf{x})]}{u}\right) \nonumber\\
	&=K\left.\frac{\mathrm{d}\mathbb{E}_\text{MSS}[\phi(\mathbf{x})]}{\mathrm{d}u}\right|_{u=0}.
\end{align}
The final step in Eq.~(\ref{eq_ERMCfromEMSS}) employs L'H\^{o}pital's Rule, which allows for the calculation of the limit when both the numerator and denominator approach zero by taking the derivative of the numerator and denominator separately.

By utilizing the relationship between $\mathbb{E}_\text{MSS}[\cdot]$ and $\mathbb{E}_\text{RMC}[\cdot]$, we can start from the MSS distribution and, using stability, derive a recurrence relation by strategy updates within an elementary MCS under the given update rule. Then, by taking the weak mutation limit, we can obtain the results needed for the evolutionary dynamics (RMC distribution) as discussed in the main text (see details below).

\subsection{Pairwise comparison (PC)}
For the PC rule, the probability $e_{ij}(\mathbf{x})$ that agent $i$ transmits its strategy to agent $j$ can be calculated as follows. In each elementary MCS, agent $j$ is selected as the focal agent with probability $1/N$ to update the strategy. Agent $i$ is chosen as the reference by agent $j$ with probability $k_{ji}/k_j=p_{ji}$ (i.e., $k_{ji}/k_j=1/k_j$ if $i$ neighbors $j$; otherwise this probability is zero), and agent $i$'s strategy is learned by agent $j$ with the learning probability $W_{j\gets i}(\mathbf{x})$ (defined by Eq.~(\ref{eq_Wpc}) in the main text). That is,
\begin{equation}\label{eq_eij_PC}
	e_{ij}(\mathbf{x})=\frac{p_{ji}}{N}\times W_{j\gets i}(\mathbf{x})=\frac{p_{ji}}{N}\times \frac{1}{1+\exp{(-\delta(f_i(\mathbf{x})-f_j(\mathbf{x})))}}.
\end{equation}
Taking $\delta=0$ in Eq.~(\ref{eq_eij_PC}), we have
\begin{equation}\label{eq_eijo_PC}
	e^\circ_{ij}(\mathbf{x})=\frac{p_{ji}}{2N}.
\end{equation}
Taking the derivative of Eq.~(\ref{eq_eij_PC}) with respect to $\delta$ at $\delta=0$, we have
\begin{equation}\label{eq_eij'_PC}
	e'_{ij}(\mathbf{x})=\frac{p_{ji}}{4N}(f_i(\mathbf{x})-f_j(\mathbf{x})).
\end{equation}

Substituting Eq.~(\ref{eq_eijo_PC}) into Eqs.~(\ref{eq_veqs}), we obtain 
\begin{subequations}\label{eq_veqs_PC}
	\begin{align}
		\sum_{j\in\mathcal{N}}e^\circ_{ji}(\mathbf{x})v_i&=\sum_{j\in\mathcal{N}}e^\circ_{ij}(\mathbf{x})v_j 
		\Leftrightarrow \frac{v_i}{2N}=\sum_{j\in\mathcal{N}}\frac{p_{ji}}{2N} v_j,
		\\
		\sum_{i\in\mathcal{N}}v_i&=N.
	\end{align}
\end{subequations}
Given that $p_{ji}=k_{ji}/k_j$, the solution to Eqs.~(\ref{eq_veqs_PC}) is $v_i=k_i/\langle k\rangle$ for $i\in\mathcal{N}$, where $\langle k\rangle=(\sum_{j\in\mathcal{N}}k_j)/N$ represents the average degree of all nodes on the network.

Substituting $v_i=k_i/\langle k\rangle$ and Eq.~(\ref{eq_eij'_PC}) into Eq.~(\ref{eq_RVDeltasel'}), we can calculate $\hat{\Delta}'_\text{sel}(\mathbf{x})$ under the PC rule:
\begin{align}\label{eq_RVDeltasel'_PC}
	\hat{\Delta}'_\text{sel}(\mathbf{x})
	&=\frac{1}{2N}\sum_{i,j\in\mathcal{N}}(x_i-x_j) (e'_{ij}(\mathbf{x})v_j-e'_{ji}(\mathbf{x})v_i) \nonumber\\
	&=\frac{1}{2N}\sum_{i,j\in\mathcal{N}}(x_i-x_j) \left(\frac{p_{ji}}{4N}(f_i(\mathbf{x})-f_j(\mathbf{x}))\frac{k_j}{\langle k\rangle}-\frac{p_{ij}}{4N}(f_j(\mathbf{x})-f_i(\mathbf{x}))\frac{k_i}{\langle k\rangle}\right) \nonumber\\
	&=\frac{1}{2N}\sum_{i,j\in\mathcal{N}}(x_i-x_j) \frac{k_{ij}}{4N\langle k\rangle}(f_i(\mathbf{x})-f_j(\mathbf{x})-f_j(\mathbf{x})+f_i(\mathbf{x})) \nonumber\\
	&=\frac{1}{4N^2\langle k\rangle}\sum_{i,j\in\mathcal{N}}(x_i-x_j) k_i p_{ij}(f_i(\mathbf{x})-f_j(\mathbf{x})).
\end{align}

Substituting Eq.~(\ref{eq_RVDeltasel'_PC}) into Eq.~(\ref{eq_coopcondi}), we obtain the condition for the success of cooperation under the PC rule:
\begin{equation}\label{eq_coopcondi_PC}
	\mathbb{E}_\text{RMC}^\circ [\hat{\Delta}'_\text{sel}(\mathbf{x})]>0 \Leftrightarrow 
	\frac{1}{4N^2\langle k\rangle} \sum_{i,j\in\mathcal{N}}k_i p_{ij}\mathbb{E}_\text{RMC}^\circ [(x_i-x_j)(f_i(\mathbf{x})-f_j(\mathbf{x}))]>0.
\end{equation}
Note that quantities such as $N$, $k_i$, and $p_{ij}$ are input parameters with constant expected values. Therefore, they can be factored out of $\mathbb{E}_\text{RMC}^\circ[\cdot]$.

We first calculate $\mathbb{E}_\text{RMC}^\circ [(x_i-x_j) (f_i(\mathbf{x})-f_j(\mathbf{x}))]$ in Eq.~(\ref{eq_coopcondi_PC}). Inserting the payoffs in PGGs, $f_i(\mathbf{x})$ and $f_j(\mathbf{x})$, by using Eq.~(\ref{eq_payoff}) and notice that $r$ and $c$ are also input parameters that remain constant, we have 
\begin{align}\label{eq_coopcondi_1_PC}
	&~\mathbb{E}_\text{RMC}^\circ [(x_i-x_j)(f_i(\mathbf{x})-f_j(\mathbf{x}))]
	\nonumber\\
	=&~\mathbb{E}_\text{RMC}^\circ \Bigg[
	\left(\frac{rc}{(k_i+1)^2}-c\right)(x_i^2-x_i x_j)
	+\frac{rc}{k_i+1}\sum_{l\in\mathcal{N}_i}\left(\frac{1}{k_i+1}+\frac{1}{k_l+1}\right)(x_i x_l-x_j x_l)
	\nonumber\\
	&+\frac{rc}{k_i+1}\sum_{l\in\mathcal{N}_i}\frac{1}{k_l+1}\sum_{\ell\in\mathcal{N}_l}(x_i x_\ell-x_j x_\ell)
	-\left(\frac{rc}{(k_j+1)^2}-c\right)(x_i x_j-x_j^2)
	\nonumber\\
	&-\frac{rc}{k_j+1}\sum_{l\in\mathcal{N}_j}\left(\frac{1}{k_j+1}+\frac{1}{k_l+1}\right)(x_i x_l-x_j x_l)-\frac{rc}{k_j+1}\sum_{l\in\mathcal{N}_j}\frac{1}{k_l+1}\sum_{\ell\in\mathcal{N}_l}(x_i x_\ell-x_j x_\ell)
	\Bigg]
	\nonumber\\
	=&~\left(\frac{rc}{(k_i+1)^2}-c\right) \left(\mathbb{E}_\text{RMC}^\circ[x_i^2]-\mathbb{E}_\text{RMC}^\circ[x_i x_j]\right)
	+\frac{rc}{k_i+1}\sum_{l\in\mathcal{N}_i}\left(\frac{1}{k_i+1}+\frac{1}{k_l+1}\right) \left(\mathbb{E}_\text{RMC}^\circ[x_i x_l]-\mathbb{E}_\text{RMC}^\circ[x_j x_l]\right)
	\nonumber\\
	&+\frac{rc}{k_i+1}\sum_{l\in\mathcal{N}_i}\frac{1}{k_l+1}\sum_{\ell\in\mathcal{N}_l}\left(\mathbb{E}_\text{RMC}^\circ[x_i x_\ell]-\mathbb{E}_\text{RMC}^\circ[x_j x_\ell]\right)
	-\left(\frac{rc}{(k_j+1)^2}-c\right)\left(\mathbb{E}_\text{RMC}^\circ[x_i x_j]-\mathbb{E}_\text{RMC}^\circ[x_j^2]\right)
	\nonumber\\
	&-\frac{rc}{k_j+1}\sum_{l\in\mathcal{N}_j}\left(\frac{1}{k_j+1}+\frac{1}{k_l+1}\right)\left(\mathbb{E}_\text{RMC}^\circ[x_i x_l]-\mathbb{E}_\text{RMC}^\circ[x_j x_l]\right)
	\nonumber\\
	&-\frac{rc}{k_j+1}\sum_{l\in\mathcal{N}_j}\frac{1}{k_l+1}\sum_{\ell\in\mathcal{N}_l}\left(\mathbb{E}_\text{RMC}^\circ[x_i x_\ell]-\mathbb{E}_\text{RMC}^\circ[x_j x_\ell]\right).
\end{align}
Since $x_i\in\{0,1\}$, we have $x_i^2=x_i$, and thus $\mathbb{E}_\text{RMC}^\circ[x_i^2]=\mathbb{E}_\text{RMC}^\circ[x_i]=1/2$ for $i\in\mathcal{N}$. The remaining work is to calculate $\mathbb{E}_\text{RMC}^\circ[x_i x_j]$ for all $i,j\in\mathcal{N}$.

We begin with the MSS distribution. Since the MSS distribution is stationary, the expected system's state remains unchanged after strategy updates. We aim to derive a recurrence relation by working through the strategy update within an elementary MCS. For convenience of calculation, we study $\mathbb{E}_\text{MSS}^\circ[(x_i-1/2)(x_j-1/2)]$, which has a useful property $\mathbb{E}_\text{MSS}^\circ[x_i-1/2]=0$ due to $\mathbb{E}_\text{MSS}^\circ[x_i]=1/2$.

Integrating the mutation mechanism described in Section~\ref{sec_coopcondi}, the possible events that happen within an elementary MCS can be classified into the following categories based on their impact on $x_i$ or $x_j$.

\begin{itemize}
	\item Agent $i$ is selected as the focal agent with probability $1/N$ to update its strategy:
	\subitem (\romannumeral 1) The focal agent $i$ mutates with probability $u$, becoming cooperation ($x_i\gets 1$) with probability $1/2$, or defection ($x_i\gets 0$) with probability $1/2$;
	\subitem (\romannumeral 2) Agent $i$ updates its strategy under the PC rule with probability $1-u$. With probability $p_{il}$, agent $i$ chooses reference agent $l$ ($l\in\mathcal{N}$), learning $l$'s strategy, $x_i\gets x_l$, with probability $W^\circ_{i\gets l}(\mathbf{x})=1/2$ (note that we are discussing neutral drift now), or keeping the current strategy $x_i$ unchanged with probability $1-W^\circ_{i\gets l}(\mathbf{x})=1/2$. The probabilities summarized here are consistent with the strategy transmission probability $e^\circ_{li}=p_{il}/(2N)$ in Eq.~(\ref{eq_eijo_PC}), but the probability $1/N$ to select focal agent $i$ is not repeatedly considered.
	
	\item Similarly, agent $j$ is selected as the focal agent with probability $1/N$ to update its strategy:
	\subitem (\romannumeral 1) The focal agent $j$ mutates with probability $u$, becoming cooperation ($x_j\gets 1$) with probability $1/2$, or defection ($x_j\gets 0$) with probability $1/2$;
	\subitem (\romannumeral 2) Agent $j$ updates its strategy under the PC rule with probability $1-u$. With probability $p_{jl}$, agent $j$ chooses reference agent $l$ ($l\in\mathcal{N}$), learning $l$'s strategy, $x_j\gets x_l$, with probability $W^\circ_{j\gets l}(\mathbf{x})=1/2$, or keeping the current strategy $x_j$ unchanged with probability $1-W^\circ_{j\gets l}(\mathbf{x})=1/2$. 
	
	\item The focal agent is one of the remaining $N-2$ agents other than $i$ and $j$, with probability $1/N$. Since only the focal agent's strategy may update, both $x_i$ and $x_j$ remain unchanged.
\end{itemize}

Combining all the above possibilities of an elementary MCS, we can obtain the following recurrence relation under the MSS distribution:
\begin{align}
	&~\mathbb{E}_\text{MSS}^\circ[(x_i-1/2)(x_j-1/2)] \nonumber\\
	=&~\frac{1}{N}\Bigg\{
	u\left(\frac{1}{2}\mathbb{E}_\text{MSS}^\circ[(1-1/2)(x_j-1/2)]+\frac{1}{2}\mathbb{E}_\text{MSS}^\circ[(0-1/2)(x_j-1/2)]\right) \nonumber\\
	&+(1-u)\sum_{l\in\mathcal{N}}p_{il}\Big(W^\circ_{i\gets l}(\mathbf{x})\mathbb{E}_\text{MSS}^\circ[(x_l-1/2)(x_j-1/2)]+(1-W^\circ_{i\gets l}(\mathbf{x}))\mathbb{E}_\text{MSS}^\circ[(x_i-1/2)(x_j-1/2)]\Big)\Bigg\} \nonumber\\
	&+\frac{1}{N}\Bigg\{
	u\left(\frac{1}{2}\mathbb{E}_\text{MSS}^\circ[(x_i-1/2)(1-1/2)]+\frac{1}{2}\mathbb{E}_\text{MSS}^\circ[(x_i-1/2)(0-1/2)]\right) \nonumber\\
	&+(1-u)\sum_{l\in\mathcal{N}}p_{jl}\Big(W^\circ_{j\gets l}(\mathbf{x})\mathbb{E}_\text{MSS}^\circ[(x_i-1/2)(x_l-1/2)]+(1-W^\circ_{j\gets l}(\mathbf{x}))\mathbb{E}_\text{MSS}^\circ[(x_i-1/2)(x_j-1/2)]\Big)\Bigg\} \nonumber\\
	&+(N-2)\frac{1}{N}\mathbb{E}_\text{MSS}^\circ[(x_i-1/2)(x_j-1/2)] \nonumber\\
	=&~\frac{1}{N}\Bigg\{
	0+(1-u)\sum_{l\in\mathcal{N}}\frac{p_{il}}{2}\Big( \mathbb{E}_\text{MSS}^\circ[(x_l-1/2)(x_j-1/2)]+\mathbb{E}_\text{MSS}^\circ[(x_i-1/2)(x_j-1/2)]\Big)\Bigg\} \nonumber\\
	&+\frac{1}{N}\Bigg\{
	0+(1-u)\sum_{l\in\mathcal{N}}\frac{p_{jl}}{2}\Big( \mathbb{E}_\text{MSS}^\circ[(x_i-1/2)(x_l-1/2)]+\mathbb{E}_\text{MSS}^\circ[(x_i-1/2)(x_j-1/2)]\Big)\Bigg\} \nonumber\\
	&+(N-2)\frac{1}{N}\mathbb{E}_\text{MSS}^\circ[(x_i-1/2)(x_j-1/2)].
\end{align}
We integrate $\mathbb{E}_\text{MSS}^\circ[(x_i-1/2)(x_j-1/2)]$ into the left-hand side and denote $\underline{x}_i=x_i-1/2$ for convenience. Then, we have
\begin{equation}\label{eq_EMSSxixj_PC}
	\mathbb{E}_\text{MSS}^\circ[\underline{x}_i \underline{x}_j]=\frac{1-u}{2}
	\left(\sum_{l\in\mathcal{N}}p_{il}\mathbb{E}_\text{MSS}^\circ[\underline{x}_l \underline{x}_j]+\sum_{l\in\mathcal{N}}p_{jl}\mathbb{E}_\text{MSS}^\circ[\underline{x}_i \underline{x}_l]\right).
\end{equation}

We define variables $\phi_{ij}(\mathbf{x})$, 
\begin{equation}\label{eq_phi_PC}
	\phi_{ij}(\mathbf{x})=\underline{x}_i \underline{x}_j-\frac{1}{2}\left(\sum_{l\in\mathcal{N}}p_{il}\underline{x}_l \underline{x}_j+\sum_{l\in\mathcal{N}}p_{jl}\underline{x}_i \underline{x}_l\right),
\end{equation}
which satisfy the properties $\phi_{ij}(\mathbf{C})=1/4-1/2\times (1/4+1/4)=0$ and similarly, $\phi_{ij}(\mathbf{D})=0$. Therefore, $\phi_{ij}(\mathbf{x})$ can be used to relate the MSS and RMC distributions through Eq.~(\ref{eq_ERMCfromEMSS}) as $u\to 0$.

We first calculate $\mathbb{E}_\text{MSS}^\circ[\phi_{ij}(\mathbf{x})]$. Writing down the expected value of Eq.~(\ref{eq_phi_PC}) and using Eq.~(\ref{eq_EMSSxixj_PC}), we have 
\begin{align}\label{eq_EMSSphi_PC}
	\mathbb{E}_\text{MSS}^\circ[\phi_{ij}(\mathbf{x})]
	&=\mathbb{E}_\text{MSS}^\circ[\underline{x}_i \underline{x}_j]-\frac{1}{2}\left(\sum_{l\in\mathcal{N}}p_{il}\mathbb{E}_\text{MSS}^\circ[\underline{x}_l \underline{x}_j]+\sum_{l\in\mathcal{N}}p_{jl}\mathbb{E}_\text{MSS}^\circ[\underline{x}_i \underline{x}_l]\right) \nonumber\\
	&=\mathbb{E}_\text{MSS}^\circ[\underline{x}_i \underline{x}_j]-\frac{1}{1-u}\mathbb{E}_\text{MSS}^\circ[\underline{x}_i \underline{x}_j] \nonumber\\
	&=-\frac{u}{1-u}\mathbb{E}_\text{MSS}^\circ[\underline{x}_i \underline{x}_j].
\end{align}

According to Eq.~(\ref{eq_ERMCfromEMSS}), we can calculate $\mathbb{E}_\text{RMC}^\circ[\phi_{ij}(\mathbf{x})]$ from $\mathbb{E}_\text{MSS}^\circ[\phi_{ij}(\mathbf{x})]$,
\begin{align}\label{eq_ERMCfromEMSS_PC}
	\mathbb{E}_\text{RMC}^\circ[\phi_{ij}(\mathbf{x})]
	&=K\left.\frac{\mathrm{d}\mathbb{E}_\text{MSS}^\circ[\phi_{ij}(\mathbf{x})]}{\mathrm{d}u}\right|_{u=0} \nonumber\\
	&=K\left.\frac{\mathrm{d}}{\mathrm{d}u}\right|_{u=0} \left(-\frac{u}{1-u}\mathbb{E}_\text{MSS}^\circ[\underline{x}_i \underline{x}_j]\right) \nonumber\\
	&=K\left(-\left.\mathbb{E}_\text{MSS}^\circ[\underline{x}_i \underline{x}_j]\right|_{u=0}+0\right) \nonumber\\
	&=-K\left(\frac{1}{2}\times(1-\frac{1}{2})(1-\frac{1}{2})+\frac{1}{2}\times(0-\frac{1}{2})(0-\frac{1}{2})\right) \nonumber\\
	&=-\frac{K}{4}.
\end{align}
In the second-to-last step, we recalled that as $u\to 0$, there are only two stationary states under MSS, $\mathbf{x}=\mathbf{1}$ or $\mathbf{x}=\mathbf{0}$, each with probability $1/2$ under neutral drift.

On the other hand, we write down the expected value of Eq.~(\ref{eq_phi_PC}) under the RMC distribution and obtain another expression of $\mathbb{E}_\text{RMC}^\circ[\phi_{ij}(\mathbf{x})]$:
\begin{equation}\label{eq_ERMCphi_PC}
	\mathbb{E}_\text{RMC}^\circ[\phi_{ij}(\mathbf{x})]
	=\mathbb{E}_\text{RMC}^\circ[\underline{x}_i \underline{x}_j]-\frac{1}{2}\left(\sum_{l\in\mathcal{N}}p_{il}\mathbb{E}_\text{RMC}^\circ[\underline{x}_l \underline{x}_j]+\sum_{l\in\mathcal{N}}p_{jl}\mathbb{E}_\text{RMC}^\circ[\underline{x}_i \underline{x}_l]\right).
\end{equation}
Substituting $\mathbb{E}_\text{RMC}^\circ[\phi_{ij}(\mathbf{x})]=-K/4$ (Eq.~(\ref{eq_ERMCfromEMSS_PC})) into Eq.~(\ref{eq_ERMCphi_PC}), we have 
\begin{equation}\label{eq_ERMCxixj_PC}
	\mathbb{E}_\text{RMC}^\circ[\underline{x}_i \underline{x}_j]
	=\frac{1}{2}\left(\sum_{l\in\mathcal{N}}p_{il}\mathbb{E}_\text{RMC}^\circ[\underline{x}_l \underline{x}_j]+\sum_{l\in\mathcal{N}}p_{jl}\mathbb{E}_\text{RMC}^\circ[\underline{x}_i \underline{x}_l]\right)-\frac{K}{4}.
\end{equation}

We define variables $\tau_{ij}$ for $i,j\in\mathcal{N}$, 
\begin{equation}\label{eq_tauijERMC_PC}
	\tau_{ij}=\frac{\dfrac{1}{2}-\mathbb{E}_\text{RMC}^\circ[x_i x_j]}{K/4}.
\end{equation}
Obviously, $\tau_{ii}=0$ when $i=j$, because $\mathbb{E}_\text{RMC}^\circ[x_i^2]=1/2$. Also, $\tau_{ij}=\tau_{ji}$, because $\mathbb{E}_\text{RMC}^\circ[x_i x_j]=\mathbb{E}_\text{RMC}^\circ[x_j x_i]$.

When $i\neq j$, we can solve for the values of $\tau_{ij}$ by the recurrence relation. We know that $\mathbb{E}_\text{RMC}^\circ[x_i x_j]$ and $\mathbb{E}_\text{RMC}^\circ[\underline{x}_i \underline{x}_j]$ have the following relation:
\begin{align}\label{eq_relation_RMC}
	\mathbb{E}_\text{RMC}^\circ[\underline{x}_i \underline{x}_j]
	&=\mathbb{E}_\text{RMC}^\circ[(x_i-1/2)(x_j-1/2)] \nonumber\\
	&=\mathbb{E}_\text{RMC}^\circ[x_i x_j]-\frac{1}{2}\mathbb{E}_\text{RMC}^\circ[x_i]-\frac{1}{2}\mathbb{E}_\text{RMC}^\circ[x_j]+\frac{1}{4} \nonumber\\
	&=\mathbb{E}_\text{RMC}^\circ[x_i x_j]-\frac{1}{4},
\end{align}
and therefore, Eq.~(\ref{eq_tauijERMC_PC}) can be written as 
\begin{equation}
	\tau_{ij}
	=\frac{\dfrac{1}{2}-\left(\mathbb{E}_\text{RMC}^\circ[\underline{x}_i \underline{x}_j]+\dfrac{1}{4}\right)}{K/4}
	=\frac{1-4\mathbb{E}_\text{RMC}^\circ[\underline{x}_i \underline{x}_j]}{K}
\end{equation}
or 
\begin{equation}\label{eq_ERMCtauij_PC}
	\mathbb{E}_\text{RMC}^\circ[\underline{x}_i \underline{x}_j]
	=\frac{1-K\tau_{ij}}{4}.
\end{equation}

Substituting Eq.~(\ref{eq_ERMCtauij_PC}) into Eq.~(\ref{eq_ERMCxixj_PC}), we obtain the recurrence relation of $\tau_{ij}$:
\begin{align}\label{eq_tauij_PC}
	&~\frac{1-K\tau_{ij}}{4}
	=\frac{1}{2}\left(\sum_{l\in\mathcal{N}}p_{il}\frac{1-K\tau_{lj}}{4}+\sum_{l\in\mathcal{N}}p_{jl}\frac{1-K\tau_{il}}{4}\right)-\frac{K}{4} \nonumber\\
	\Leftrightarrow 
	&~\tau_{ij}=1+\frac{1}{2}\left(\sum_{l\in\mathcal{N}}p_{il}\tau_{lj}+\sum_{l\in\mathcal{N}}p_{jl}\tau_{il}\right).
\end{align}
The recurrence relation Eq.~(\ref{eq_tauij_PC}), together with $\tau_{ii}=0$, form a system of linear equations, through which all $\tau_{ij}$ values ($i,j\in\mathcal{N}$) can be determined on a given network.

Back to the halfway calculation in Eq.~(\ref{eq_coopcondi_1_PC}) of the cooperation success condition. Substituting all $\mathbb{E}_\text{RMC}^\circ[x_i x_j]$ into Eq.~(\ref{eq_coopcondi_1_PC}) with computable $\tau_{ij}$ using Eq.~(\ref{eq_tauijERMC_PC}), and then substituting the result into Eq.~(\ref{eq_coopcondi_PC}), where the positive factors $K/4$ and $1/(4N^2\langle k\rangle)$ can be canceled out, we arrive at the following condition for the success of cooperation:
\begin{align}\label{eq_coopcondi_2_PC}
	&~\mathbb{E}_\text{RMC}^\circ [\hat{\Delta}'_\text{sel}(\mathbf{x})]>0 \nonumber\\
	\Leftrightarrow
	&~\sum_{i,j\in\mathcal{N}}k_i p_{ij}\Bigg\{\left(\frac{rc}{(k_i+1)^2}-c\right) \tau_{ij}
	+\frac{rc}{k_i+1}\sum_{l\in\mathcal{N}_i}\left(\frac{1}{k_i+1}+\frac{1}{k_l+1}\right) \left(-\tau_{il}+\tau_{jl}\right)
	\nonumber\\
	&+\frac{rc}{k_i+1}\sum_{l\in\mathcal{N}_i}\frac{1}{k_l+1}\sum_{\ell\in\mathcal{N}_l}\left(-\tau_{i\ell}+\tau_{j\ell}\right)
	-\left(\frac{rc}{(k_j+1)^2}-c\right)\left(-\tau_{ij}\right)
	\nonumber\\
	&-\frac{rc}{k_j+1}\sum_{l\in\mathcal{N}_j}\left(\frac{1}{k_j+1}+\frac{1}{k_l+1}\right)\left(-\tau_{il}+\tau_{jl}\right)
	-\frac{rc}{k_j+1}\sum_{l\in\mathcal{N}_j}\frac{1}{k_l+1}\sum_{\ell\in\mathcal{N}_l}\left(-\tau_{i\ell}+\tau_{j\ell}\right)\Bigg\}
	>0 \nonumber\\
	\Leftrightarrow
	&~r>\frac{2\sum_{i,j\in\mathcal{N}}k_i p_{ij}\tau_{ij}}{\sum_{i,j\in\mathcal{N}}k_i p_{ij}(\Upsilon_{ij}+\Upsilon_{ji})},
\end{align}
where $\Upsilon_{ij}$ are defined by (equivalent to Eq.~(\ref{eq_Upsilonij}) in the main text)
\begin{equation}\label{eq_Upsilonij_PC}
	\Upsilon_{ij}=
	\frac{1}{k_i+1}\left(
	\frac{\tau_{ij}+k_i \sum_{l\in\mathcal{N}}p_{il} (\tau_{jl}-\tau_{il})}{k_i+1}+
	k_i \sum_{l\in\mathcal{N}}p_{il}\frac{(\tau_{jl}-\tau_{il})+k_l \sum_{\ell\in\mathcal{N}}p_{l\ell}(\tau_{j\ell}-\tau_{i\ell})}{k_l+1}
	\right).
\end{equation}
And according to the previous discussion, $\tau_{ij}$ can be solved by the following system of linear equations (equivalent to Eq.~(\ref{eq_tauij}) in the main text):
\begin{align}\label{eq_eqtauij_PC}
	\begin{cases}
		\displaystyle{
			\tau_{ij}=1+\frac{1}{2}\sum_{l\in\mathcal{N}}
			(p_{il}\tau_{jl}+p_{jl}\tau_{il})}, & \mbox{if $j\neq i$},
		\\[1em]
		\displaystyle{
			\tau_{ij}=0}, & \mbox{if $j=i$}.
	\end{cases}
\end{align}

Finally, although usually $\Upsilon_{ij}\neq \Upsilon_{ji}$, we can still infer that $\sum_{i,j\in\mathcal{N}}k_i p_{ij}\Upsilon_{ji}=\sum_{i,j\in\mathcal{N}}k_{ij}\Upsilon_{ji}=\sum_{j,i\in\mathcal{N}}k_{ji}\Upsilon_{ij}=\sum_{j,i\in\mathcal{N}}k_{ij}\Upsilon_{ij}=\sum_{j,i\in\mathcal{N}}k_i p_{ij}\Upsilon_{ij}$, such that $\sum_{i,j\in\mathcal{N}}k_i p_{ij}(\Upsilon_{ij}+\Upsilon_{ji})=2\sum_{i,j\in\mathcal{N}}k_i p_{ij}\Upsilon_{ij}$. Therefore, $\sum_{i,j\in\mathcal{N}}k_i p_{ij}(\Upsilon_{ij}+\Upsilon_{ji})=2\sum_{i,j\in\mathcal{N}}k_i p_{ij}\Upsilon_{ij}$. As a result, Eq.~(\ref{eq_coopcondi_2_PC}) can be further simplified as 
\begin{equation}\label{eq_coopcondi_3_PC}
	r>\frac{\sum_{i,j\in\mathcal{N}}k_i p_{ij}\tau_{ij}}{\sum_{i,j\in\mathcal{N}}k_i p_{ij}\Upsilon_{ij}}.
\end{equation}
The right-hand side is the $r^\star$ value under the PC rule, as mentioned in the main text.

\subsection{Death-birth (DB)}
For the DB rule, the probability $e_{ij}(\mathbf{x})$ that agent $i$ transmits its strategy to agent $j$ can be calculated as follows. In each elementary MCS, agent $j$ is selected as the focal agent with probability $1/N$ to update the strategy. Agent $j$'s strategy ``dies'', and agent $i$'s strategy occupies the vacant position with probability $W_{j\gets i}(\mathbf{x})$, which is proportional to its fitness among $j$'s neighbors (see Eq.~(\ref{eq_eij_DB})). That is,
\begin{equation}\label{eq_eij_DB}
	e_{ij}(\mathbf{x})=\frac{1}{N}\times W_{j\gets i}(\mathbf{x})=\frac{1}{N}\times \frac{k_{ji}F_i(\mathbf{x})}{\sum_{l\in \mathcal{N}}k_{jl}F_l(\mathbf{x})}.
\end{equation}
Taking $\delta=0$ in Eq.~(\ref{eq_eij_DB}), we have 
\begin{equation}\label{eq_eijo_DB}
	e^\circ_{ij}(\mathbf{x})=\frac{k_{ji}}{Nk_j}=\frac{p_{ji}}{N}.
\end{equation}
Taking the derivative of Eq.~(\ref{eq_eij_DB}) with respect to $\delta$ at $\delta=0$, we have
\begin{equation}\label{eq_eij'_DB}
	e'_{ij}(\mathbf{x})
	=\frac{1}{N}\frac{k_{ji}f_i(\mathbf{x})k_j-k_{ji}\sum_{l\in\mathcal{N}}k_{jl}f_l(\mathbf{x})}{k_j^2}
	=\frac{p_{ji}}{N}\left(f_i(\mathbf{x})-\sum_{l\in\mathcal{N}}p_{jl}f_l(\mathbf{x})\right).
\end{equation}

Substituting Eq.~(\ref{eq_eijo_DB}) into Eqs.~(\ref{eq_veqs}), we obtain 
\begin{subequations}\label{eq_veqs_DB}
	\begin{align}
		\sum_{j\in\mathcal{N}}e^\circ_{ji}(\mathbf{x})v_i&=\sum_{j\in\mathcal{N}}e^\circ_{ij}(\mathbf{x})v_j 
		\Leftrightarrow \frac{v_i}{N}=\sum_{j\in\mathcal{N}}\frac{p_{ji}}{N} v_j,
		\\
		\sum_{i\in\mathcal{N}}v_i&=N.
	\end{align}
\end{subequations}
Similar to the PC rule, the solution to Eqs.~(\ref{eq_veqs_DB}) is also $v_i=k_i/\langle k\rangle$ for $i\in\mathcal{N}$, where $\langle k\rangle=(\sum_{j\in\mathcal{N}}k_j)/N$ represents the average degree of all nodes.

Substituting $v_i=k_i/\langle k\rangle$ and Eq.~(\ref{eq_eij'_DB}) into Eq.~(\ref{eq_RVDeltasel'}), we can calculate $\hat{\Delta}'_\text{sel}(\mathbf{x})$ under the DB rule. We start from the first line in Eq.~(\ref{eq_RVDeltasel'}), 
\begin{align}\label{eq_RVDeltasel'1_DB}
	\frac{1}{N}\sum_{i,j\in\mathcal{N}}x_i (e'_{ij}(\mathbf{x})v_j-e'_{ji}(\mathbf{x})v_i)
	&=\frac{1}{N}\sum_{i,j\in\mathcal{N}}x_i \left(\frac{p_{ji}}{N}\left(f_i(\mathbf{x})-\sum_{l\in\mathcal{N}}p_{jl}f_l(\mathbf{x})\right)\frac{k_j}{\langle k\rangle}-\frac{p_{ij}}{N}\left(f_j(\mathbf{x})-\sum_{l\in\mathcal{N}}p_{il}f_l(\mathbf{x})\right)\frac{k_i}{\langle k\rangle}\right) \nonumber\\
	&=\frac{1}{N}\sum_{i,j\in\mathcal{N}}x_i \frac{k_i p_{ij}}{N\langle k\rangle}\left(f_i(\mathbf{x})-\sum_{l\in\mathcal{N}}p_{jl}f_l(\mathbf{x})-f_j(\mathbf{x})+\sum_{l\in\mathcal{N}}p_{il}f_l(\mathbf{x})\right) \nonumber\\
	&=\frac{1}{N}\sum_{i\in\mathcal{N}}x_i \frac{k_i}{N\langle k\rangle}\left(f_i(\mathbf{x})-\sum_{j,l\in\mathcal{N}}p_{ij}p_{jl}f_l(\mathbf{x})-\sum_{j\in\mathcal{N}}p_{ij}f_j(\mathbf{x})+\sum_{l\in\mathcal{N}}p_{il}f_l(\mathbf{x})\right) \nonumber\\
	&=\frac{1}{N}\sum_{i\in\mathcal{N}}x_i \frac{k_i}{N\langle k\rangle}\left(f_i(\mathbf{x})-\sum_{l\in\mathcal{N}}p^{(2)}_{il}f_l(\mathbf{x})\right) \nonumber\\
	&=\frac{1}{N}\sum_{i,j\in\mathcal{N}}x_i \frac{k_i p^{(2)}_{ij}}{N\langle k\rangle}\left(f_i(\mathbf{x})-f_j(\mathbf{x})\right),
\end{align}
where $p^{(2)}_{ij}$ is defined as $\sum_{l\in\mathcal{N}}p_{il}p_{lj}$. Then, by comparing Eq.~(\ref{eq_RVDeltasel'1_DB}) and Eq.~(\ref{eq_RVDeltasel'}), we know that 
\begin{align}\label{eq_RVDeltasel'_DB}
	\hat{\Delta}'_\text{sel}(\mathbf{x})
	&=\frac{1}{N}\sum_{i,j\in\mathcal{N}}x_i (e'_{ij}(\mathbf{x})v_j-e'_{ji}(\mathbf{x})v_i) \nonumber\\
	&=\frac{1}{2N}\sum_{i,j\in\mathcal{N}}(x_i-x_j) (e'_{ij}(\mathbf{x})v_j-e'_{ji}(\mathbf{x})v_i) \nonumber\\
	&=\frac{1}{2N^2\langle k\rangle}\sum_{i,j\in\mathcal{N}}(x_i-x_j) k_i p^{(2)}_{ij}(f_i(\mathbf{x})-f_j(\mathbf{x})).
\end{align}

Substituting Eq.~(\ref{eq_RVDeltasel'_DB}) into Eq.~(\ref{eq_coopcondi}), we obtain the condition for the success of cooperation under the DB rule:
\begin{equation}\label{eq_coopcondi_DB}
	\mathbb{E}_\text{RMC}^\circ [\hat{\Delta}'_\text{sel}(\mathbf{x})]>0 \Leftrightarrow 
	\frac{1}{2N^2\langle k\rangle} \sum_{i,j\in\mathcal{N}}k_i p^{(2)}_{ij}\mathbb{E}_\text{RMC}^\circ [(x_i-x_j)(f_i(\mathbf{x})-f_j(\mathbf{x}))]>0.
\end{equation}

Similarly, we first calculate $\mathbb{E}_\text{RMC}^\circ [(x_i-x_j) (f_i(\mathbf{x})-f_j(\mathbf{x}))]$ in Eq.~(\ref{eq_coopcondi_DB}), which is completely the same as the one under the PC rule (see Eq.~(\ref{eq_coopcondi_1_PC}) for the same result).

The remaining work is to calculate $\mathbb{E}_\text{RMC}^\circ[x_i x_j]$ for all $i,j\in\mathcal{N}$ under the DB rule. Similarly, we begin with the MSS distribution and aim to derive a recurrence relation by working through the strategy update within an elementary MCS. Under the DB rule, the possible events that happen within an elementary MCS can be classified into the following categories based on their impact on $x_i$ or $x_j$.

\begin{itemize}
	\item Agent $i$ is selected as the focal agent with probability $1/N$ to update its strategy:
	\subitem (\romannumeral 1) The focal agent $i$ mutates with probability $u$, becoming cooperation ($x_i\gets 1$) with probability $1/2$, or defection ($x_i\gets 0$) with probability $1/2$;
	\subitem (\romannumeral 2) Agent $i$ learns the strategy of a neighbor under the DB rule with probability $1-u$. With probability $W^\circ_{i\gets l}(\mathbf{x})=p_{il}$, agent $i$ learns the strategy of agent $l$, $x_i\gets x_l$. Note that under the DB rule, $\sum_{l\in\mathcal{N}}W^\circ_{i\gets l}(\mathbf{x})=1$; the focal agent cannot keep its own strategy.
	
	\item Similarly, agent $j$ is selected as the focal agent with probability $1/N$ to update its strategy:
	\subitem (\romannumeral 1) The focal agent $j$ mutates with probability $u$, becoming cooperation ($x_j\gets 1$) with probability $1/2$, or defection ($x_j\gets 0$) with probability $1/2$;
	\subitem (\romannumeral 2) Agent $j$ learns the strategy of a neighbor under the DB rule with probability $1-u$. With probability $W^\circ_{j\gets l}(\mathbf{x})=p_{jl}$, agent $j$ learns the strategy of agent $l$, $x_j\gets x_l$.
	
	\item The focal agent is one of the remaining $N-2$ agents other than $i$ and $j$, with probability $1/N$. Since only the focal agent's strategy can update, both $x_i$ and $x_j$ remain unchanged.
\end{itemize}

Combining all the above possibilities of an elementary MCS, we can obtain the following recurrence relation under the MSS distribution:
\begin{align}
	&~\mathbb{E}_\text{MSS}^\circ[(x_i-1/2)(x_j-1/2)] \nonumber\\
	=&~\frac{1}{N}\Bigg\{
	u\left(\frac{1}{2}\mathbb{E}_\text{MSS}^\circ[(1-1/2)(x_j-1/2)]+\frac{1}{2}\mathbb{E}_\text{MSS}^\circ[(0-1/2)(x_j-1/2)]\right) \nonumber\\
	&+(1-u)\sum_{l\in\mathcal{N}}W^\circ_{i\gets l}(\mathbf{x})\mathbb{E}_\text{MSS}^\circ[(x_l-1/2)(x_j-1/2)]\Bigg\} \nonumber\\
	&+\frac{1}{N}\Bigg\{
	u\left(\frac{1}{2}\mathbb{E}_\text{MSS}^\circ[(x_i-1/2)(1-1/2)]+\frac{1}{2}\mathbb{E}_\text{MSS}^\circ[(x_i-1/2)(0-1/2)]\right) \nonumber\\
	&+(1-u)\sum_{l\in\mathcal{N}}W^\circ_{j\gets l}(\mathbf{x})\mathbb{E}_\text{MSS}^\circ[(x_i-1/2)(x_l-1/2)]\Bigg\} \nonumber\\
	&+(N-2)\frac{1}{N}\mathbb{E}_\text{MSS}^\circ[(x_i-1/2)(x_j-1/2)] \nonumber\\
	=&~\frac{1}{N}\Bigg\{
	0+(1-u)\sum_{l\in\mathcal{N}}p_{il}\mathbb{E}_\text{MSS}^\circ[(x_l-1/2)(x_j-1/2)]\Bigg\} \nonumber\\
	&+\frac{1}{N}\Bigg\{
	0+(1-u)\sum_{l\in\mathcal{N}}p_{jl}\mathbb{E}_\text{MSS}^\circ[(x_i-1/2)(x_l-1/2)])\Bigg\} \nonumber\\
	&+(N-2)\frac{1}{N}\mathbb{E}_\text{MSS}^\circ[(x_i-1/2)(x_j-1/2)].
\end{align}
Integrating $\mathbb{E}_\text{MSS}^\circ[(x_i-1/2)(x_j-1/2)]$ into the left-hand side and denoting $\underline{x}_i=x_i-1/2$, we have 
\begin{equation}\label{eq_EMSSxixj_DB}
	\mathbb{E}_\text{MSS}^\circ[\underline{x}_i \underline{x}_j]=\frac{1-u}{2}
	\left(\sum_{l\in\mathcal{N}}p_{il}\mathbb{E}_\text{MSS}^\circ[\underline{x}_l \underline{x}_j]+\sum_{l\in\mathcal{N}}p_{jl}\mathbb{E}_\text{MSS}^\circ[\underline{x}_i \underline{x}_l]\right),
\end{equation}
which is completely the same as the one under the PC rule (see Eq.~(\ref{eq_EMSSxixj_PC}) for the result). Therefore, the subsequent steps are also the same and are not repeated here. We can ultimately use the defined variables $\tau_{ij}$,
\begin{equation}\label{eq_tauijERMC_DB}
	\tau_{ij}=\frac{\dfrac{1}{2}-\mathbb{E}_\text{RMC}^\circ[x_i x_j]}{K/4},
\end{equation}
to replace all $\mathbb{E}_\text{RMC}^\circ[x_i x_j]$.

Substituting all $\mathbb{E}_\text{RMC}^\circ[x_i x_j]$ into $\mathbb{E}_\text{RMC}^\circ [(x_i-x_j) (f_i(\mathbf{x})-f_j(\mathbf{x}))]$ (presented in Eq.~(\ref{eq_coopcondi_1_PC})) with computable $\tau_{ij}$ using Eq.~(\ref{eq_tauijERMC_DB}), and then substituting the result into Eq.~(\ref{eq_coopcondi_PC}), where the positive factors $K/4$ and $1/(2N^2\langle k\rangle)$ can be canceled out, we can organize and obtain the following condition for the success of cooperation:
\begin{align}\label{eq_coopcondi_2_DB}
	&~\mathbb{E}_\text{RMC}^\circ [\hat{\Delta}'_\text{sel}(\mathbf{x})]>0 \nonumber\\
	\Leftrightarrow
	&~\sum_{i,j\in\mathcal{N}}k_i p^{(2)}_{ij}\Bigg\{\left(\frac{rc}{(k_i+1)^2}-c\right) \tau_{ij}
	+\frac{rc}{k_i+1}\sum_{l\in\mathcal{N}_i}\left(\frac{1}{k_i+1}+\frac{1}{k_l+1}\right) \left(-\tau_{il}+\tau_{jl}\right)
	\nonumber\\
	&+\frac{rc}{k_i+1}\sum_{l\in\mathcal{N}_i}\frac{1}{k_l+1}\sum_{\ell\in\mathcal{N}_l}\left(-\tau_{i\ell}+\tau_{j\ell}\right)
	-\left(\frac{rc}{(k_j+1)^2}-c\right)\left(-\tau_{ij}\right)
	\nonumber\\
	&-\frac{rc}{k_j+1}\sum_{l\in\mathcal{N}_j}\left(\frac{1}{k_j+1}+\frac{1}{k_l+1}\right)\left(-\tau_{il}+\tau_{jl}\right)
	-\frac{rc}{k_j+1}\sum_{l\in\mathcal{N}_j}\frac{1}{k_l+1}\sum_{\ell\in\mathcal{N}_l}\left(-\tau_{i\ell}+\tau_{j\ell}\right)\Bigg\}
	>0 \nonumber\\
	\Leftrightarrow
	&~r>\frac{2\sum_{i,j\in\mathcal{N}}k_i p^{(2)}_{ij}\tau_{ij}}{\sum_{i,j\in\mathcal{N}}k_i p^{(2)}_{ij}(\Upsilon_{ij}+\Upsilon_{ji})},
\end{align}
where $\Upsilon_{ij}$ are also defined by Eq.~(\ref{eq_Upsilonij_PC}) (equivalent to Eq.~(\ref{eq_Upsilonij}) in the main text) and $\tau_{ij}$ can also be obtained by solving the system of Eqs.~(\ref{eq_eqtauij_PC}) (or Eq.~(\ref{eq_tauij}) in the main text).

According to Eq.~(\ref{eq_kipijreverse}), we have $k_i p^{(2)}_{ij}=k_j p^{(2)}_{ji}$. Therefore, we can infer that $\sum_{i,j\in\mathcal{N}}k_i p^{(2)}_{ij}\Upsilon_{ji}=\sum_{i,j\in\mathcal{N}}k_j p^{(2)}_{ji}\Upsilon_{ji}=\sum_{j,i\in\mathcal{N}}k_i p^{(2)}_{ij}\Upsilon_{ij}$, such that $\sum_{i,j\in\mathcal{N}}k_i p^{(2)}_{ij}(\Upsilon_{ij}+\Upsilon_{ji})=2\sum_{i,j\in\mathcal{N}}k_i p^{(2)}_{ij}\Upsilon_{ij}$. Therefore, Eq.~(\ref{eq_coopcondi_2_DB}) can be further simplified as 
\begin{equation}\label{eq_coopcondi_3_DB}
	r>\frac{\sum_{i,j\in\mathcal{N}}k_i p^{(2)}_{ij}\tau_{ij}}{\sum_{i,j\in\mathcal{N}}k_i p^{(2)}_{ij}\Upsilon_{ij}},
\end{equation}
which gives the $r^\star$ value under the DB rule.
\subsection{Beath-dirth (BD)}
For the BD rule, the probability $e_{ij}(\mathbf{x})$ that agent $i$ transmits its strategy to agent $j$ can be calculated as follows. In each elementary MCS, agent $i$ is selected as the focal agent with a probability $W_i(\mathbf{x})$ proportional to its fitness in the population,
\begin{equation}
	W_i(\mathbf{x})=\frac{F_i(\mathbf{x})}{\sum_{l\in\mathcal{N}}F_l(\mathbf{x})},
\end{equation}
and transmits its strategy $x_i$ to a random neighbor. That is, 
\begin{equation}\label{eq_eij_BD}
	e_{ij}(\mathbf{x})=W_i(\mathbf{x})\times p_{ij}=\frac{F_i(\mathbf{x})}{\sum_{l\in\mathcal{N}}F_l(\mathbf{x})}\times p_{ij}.
\end{equation}
Taking $\delta=0$ in Eq.~(\ref{eq_eij_BD}), we have
\begin{equation}\label{eq_eijo_BD}
	e^\circ_{ij}(\mathbf{x})=\frac{p_{ij}}{N}.
\end{equation}
Taking the derivative of Eq.~(\ref{eq_eij_BD}) with respect to $\delta$ at $\delta=0$, we have 
\begin{equation}\label{eq_eij'_BD}
	e'_{ij}(\mathbf{x})
	=\frac{p_{ij}}{N}\left(f_i(\mathbf{x})-\frac{1}{N}\sum_{l\in\mathcal{N}}f_l(\mathbf{x})\right).
\end{equation}

Substituting Eq.~(\ref{eq_eijo_BD}) into Eqs.~(\ref{eq_veqs}), we obtain 
\begin{subequations}\label{eq_veqs_BD}
	\begin{align}
		\sum_{j\in\mathcal{N}}e^\circ_{ji}(\mathbf{x})v_i&=\sum_{j\in\mathcal{N}}e^\circ_{ij}(\mathbf{x})v_j 
		\Leftrightarrow \sum_{j\in\mathcal{N}}\frac{p_{ji}}{N}v_i=\sum_{j\in\mathcal{N}}\frac{p_{ij}}{N} v_j,
		\\
		\sum_{i\in\mathcal{N}}v_i&=N.
	\end{align}
\end{subequations}
The solution to Eqs.~(\ref{eq_veqs_BD}) is $v_i=k_i^{-1}/\langle k^{-1}\rangle$ for $i\in\mathcal{N}$, where $k_i^{-1}=1/k_i$, and $\langle k^{-1}\rangle=(\sum_{i\in\mathcal{N}}k_i^{-1})/N$ represents the average of the reciprocals of the degree of all nodes.

Substituting $v_i=k_i^{-1}/\langle k^{-1}\rangle$ and Eq.~(\ref{eq_eij'_BD}) into Eq.~(\ref{eq_RVDeltasel'}), we can calculate $\hat{\Delta}'_\text{sel}(\mathbf{x})$ under the BD rule:
\begin{align}\label{eq_RVDeltasel'_BD}
	\hat{\Delta}'_\text{sel}(\mathbf{x})
	&=\frac{1}{2N}\sum_{i,j\in\mathcal{N}}(x_i-x_j) (e'_{ij}(\mathbf{x})v_j-e'_{ji}(\mathbf{x})v_i) \nonumber\\
	&=\frac{1}{2N}\sum_{i,j\in\mathcal{N}}(x_i-x_j) \left(\frac{p_{ij}}{N}\Big(f_i(\mathbf{x})-\frac{1}{N}\sum_{l\in\mathcal{N}}f_l(\mathbf{x})\Big)\frac{k_j^{-1}}{\langle k^{-1}\rangle}-\frac{p_{ji}}{N}\Big(f_j(\mathbf{x})-\frac{1}{N}\sum_{l\in\mathcal{N}}f_l(\mathbf{x})\Big)\frac{k_i^{-1}}{\langle k^{-1}\rangle}\right) \nonumber\\
	&=\frac{1}{2N}\sum_{i,j\in\mathcal{N}}(x_i-x_j) \left(\frac{k_{ij}}{N}\Big(f_i(\mathbf{x})-\frac{1}{N}\sum_{l\in\mathcal{N}}f_l(\mathbf{x})\Big)\frac{k_i^{-1} k_j^{-1}}{\langle k^{-1}\rangle}-\frac{k_{ji}}{N}\Big(f_j(\mathbf{x})-\frac{1}{N}\sum_{l\in\mathcal{N}}f_l(\mathbf{x})\Big)\frac{k_i^{-1} k_j^{-1}}{\langle k^{-1}\rangle}\right) \nonumber\\
	&=\frac{1}{2N^2\langle k^{-1}\rangle}\sum_{i,j\in\mathcal{N}}(x_i-x_j) \frac{k_{ij}}{k_i k_j}(f_i(\mathbf{x})-f_j(\mathbf{x})).
\end{align}

Substituting Eq.~(\ref{eq_RVDeltasel'_BD}) into Eq.~(\ref{eq_coopcondi}), we obtain the condition for the success of cooperation under the BD rule:
\begin{equation}\label{eq_coopcondi_BD}
	\mathbb{E}_\text{RMC}^\circ [\hat{\Delta}'_\text{sel}(\mathbf{x})]>0 \Leftrightarrow 
	\frac{1}{2N^2\langle k^{-1}\rangle} \sum_{i,j\in\mathcal{N}}\frac{k_{ij}}{k_i k_j}\mathbb{E}_\text{RMC}^\circ [(x_i-x_j)(f_i(\mathbf{x})-f_j(\mathbf{x}))]>0.
\end{equation}

We first calculate $\mathbb{E}_\text{RMC}^\circ [(x_i-x_j) (f_i(\mathbf{x})-f_j(\mathbf{x}))]$ in Eq.~(\ref{eq_coopcondi_BD}), which is the same as the one calculated under the PC rule (see Eq.~(\ref{eq_coopcondi_1_PC}) for the same result). The remaining work is to calculate $\mathbb{E}_\text{RMC}^\circ[x_i x_j]$ for all $i,j\in\mathcal{N}$ under the BD rule.

Similarly, we begin with the MSS distribution and derive the recurrence relation by working through an elementary MCS. For the mutation mechanism under the BD rule, we specify that the event of a mutation is assigned to the focal agent rather than being transmitted to a random neighbor by the BD rule. This ensures the probability of the initial mutation on each node is $1/N$ in a fixation state, consistent with the PC and DB rules.

Therefore, under the BD rule, the possible events that happen within an elementary MCS can be classified into the following categories.

\begin{itemize}
	\item Agent $i$ is selected as the focal agent with probability $W_i^\circ(\mathbf{x})=1/N$ to propagate its strategy: 
	\subitem (\romannumeral 1) The focal agent $i$ mutates with probability $u$, becoming cooperation ($x_i\gets 1$) with probability $1/2$, or defection ($x_i\gets 0$) with probability $1/2$;
	\subitem (\romannumeral 2) Agent $i$ transmits the strategy to a random neighbor with probability $1-u$. With probability $p_{i\ell}$, agent $i$ transmits its strategy to agent $\ell$ ($\ell\in\mathcal{N}$), $x_\ell \gets x_i$. If $\ell=j$, this influences the quantity $\mathbb{E}_\text{MSS}^\circ[(x_i-1/2)(x_j-1/2)]$; otherwise the quantity keeps unchanged.
	
	\item Similarly, agent $j$ is selected as the focal agent with probability $W_j^\circ(\mathbf{x})=1/N$ to propagate its strategy: 
	\subitem (\romannumeral 1) The focal agent $j$ mutates with probability $u$, becoming cooperation ($x_j\gets 1$) with probability $1/2$, or defection ($x_j\gets 0$) with probability $1/2$; 
	\subitem (\romannumeral 2) Agent $j$ transmits the strategy to a random neighbor with probability $1-u$. With probability $p_{j\ell}$, agent $j$ transmits its strategy to agent $\ell$ ($\ell\in\mathcal{N}$), $x_\ell \gets x_j$. Similarly, if $\ell=i$, this influences the quantity $\mathbb{E}_\text{MSS}^\circ[(x_i-1/2)(x_j-1/2)]$; otherwise the quantity keeps unchanged.
	
	\item The focal agent, denoted by $l\in\mathcal{N}\setminus\{i,j\}$, is one of the remaining $N-2$ agents other than $i$ and $j$, with probability $W_l^\circ(\mathbf{x})=1/N$:
	\subitem (\romannumeral 1) The focal agent $l$ mutates with probability $u$, which can only change $x_l$ and has nothing to do with $x_i$ or $x_j$;
	\subitem (\romannumeral 2) Agent $l$ transmits the strategy to a random neighbor with probability $1-u$. With probability $p_{l\ell}$, agent $l$ transmits its strategy to agent $\ell$ ($\ell\in\mathcal{N}$), $x_\ell \gets x_l$. If $\ell=i$ or $\ell=j$, this influences the quantity $\mathbb{E}_\text{MSS}^\circ[(x_i-1/2)(x_j-1/2)]$; otherwise the quantity keeps unchanged.
\end{itemize}

Combining all the above possibilities of an elementary MCS, we obtain the following recurrence relation under the MSS distribution:
\begin{align}
	&~\mathbb{E}_\text{MSS}^\circ[(x_i-1/2)(x_j-1/2)] \nonumber\\
	=&~W_i^\circ(\mathbf{x})\Bigg\{
	u\left(\frac{1}{2}\mathbb{E}_\text{MSS}^\circ[(1-1/2)(x_j-1/2)]+\frac{1}{2}\mathbb{E}_\text{MSS}^\circ[(0-1/2)(x_j-1/2)]\right) \nonumber\\
	&+(1-u)\Bigg(\sum_{\ell\in\mathcal{N}\setminus\{j\}}p_{i\ell}\mathbb{E}_\text{MSS}^\circ[(x_i-1/2)(x_j-1/2)]+p_{ij}\mathbb{E}_\text{MSS}^\circ[(x_i-1/2)(x_i-1/2)]\Bigg)\Bigg\} \nonumber\\
	&+W_j^\circ(\mathbf{x})\Bigg\{
	u\left(\frac{1}{2}\mathbb{E}_\text{MSS}^\circ[(x_i-1/2)(1-1/2)]+\frac{1}{2}\mathbb{E}_\text{MSS}^\circ[(x_i-1/2)(0-1/2)]\right) \nonumber\\
	&+(1-u)\Bigg(\sum_{\ell\in\mathcal{N}\setminus\{i\}}p_{j\ell}\mathbb{E}_\text{MSS}^\circ[(x_i-1/2)(x_j-1/2)]+p_{ji}\mathbb{E}_\text{MSS}^\circ[(x_j-1/2)(x_j-1/2)]\Bigg)\Bigg\} \nonumber\\
	&+\sum_{l\in\mathcal{N}\setminus\{i,j\}}W_l^\circ(\mathbf{x})\Bigg\{
	u\mathbb{E}_\text{MSS}^\circ[(x_i-1/2)(x_j-1/2)] +(1-u)\Bigg(\sum_{\ell\in\mathcal{N}\setminus\{i,j\}}p_{l\ell}\mathbb{E}_\text{MSS}^\circ[(x_i-1/2)(x_j-1/2)] \nonumber\\
	&+p_{li}\mathbb{E}_\text{MSS}^\circ[(x_l-1/2)(x_j-1/2)]+p_{lj}\mathbb{E}_\text{MSS}^\circ[(x_i-1/2)(x_l-1/2)] \Bigg)\Bigg\} \nonumber\\
	=&~\frac{1}{N}\Bigg\{
	0+(1-u)\Bigg(\sum_{\ell\in\mathcal{N}\setminus\{j\}}p_{i\ell}\mathbb{E}_\text{MSS}^\circ[(x_i-1/2)(x_j-1/2)]+p_{ij}\mathbb{E}_\text{MSS}^\circ[(x_i-1/2)(x_i-1/2)]\Bigg)\Bigg\} \nonumber\\
	&+\frac{1}{N}\Bigg\{
	0+(1-u)\Bigg(\sum_{\ell\in\mathcal{N}\setminus\{i\}}p_{j\ell}\mathbb{E}_\text{MSS}^\circ[(x_i-1/2)(x_j-1/2)]+p_{ji}\mathbb{E}_\text{MSS}^\circ[(x_j-1/2)(x_j-1/2)]\Bigg)\Bigg\} \nonumber\\
	&+\frac{1}{N}\sum_{l\in\mathcal{N}\setminus\{i,j\}}\Bigg\{
	u\mathbb{E}_\text{MSS}^\circ[(x_i-1/2)(x_j-1/2)] +(1-u)\Bigg(\sum_{\ell\in\mathcal{N}\setminus\{i,j\}}p_{l\ell}\mathbb{E}_\text{MSS}^\circ[(x_i-1/2)(x_j-1/2)] \nonumber\\
	&+p_{li}\mathbb{E}_\text{MSS}^\circ[(x_l-1/2)(x_j-1/2)]+p_{lj}\mathbb{E}_\text{MSS}^\circ[(x_i-1/2)(x_l-1/2)] \Bigg)\Bigg\} \nonumber\\
	=&~\frac{1-u}{N}\sum_{l\in\mathcal{N}}\Big(p_{li}\mathbb{E}_\text{MSS}^\circ[(x_l-1/2)(x_j-1/2)]+p_{lj}\mathbb{E}_\text{MSS}^\circ[(x_i-1/2)(x_l-1/2)] \Big) \nonumber\\
	&+\frac{u}{N}\sum_{l\in\mathcal{N}\setminus\{i,j\}}\mathbb{E}_\text{MSS}^\circ[(x_i-1/2)(x_j-1/2)]+\frac{1-u}{N}\sum_{l\in\mathcal{N}}\sum_{\ell\in\mathcal{N}\setminus\{i,j\}}p_{l\ell}\mathbb{E}_\text{MSS}^\circ[(x_i-1/2)(x_j-1/2)] \nonumber\\
	=&~\frac{1-u}{N}\sum_{l\in\mathcal{N}}\Big(p_{li}\mathbb{E}_\text{MSS}^\circ[(x_l-1/2)(x_j-1/2)]+p_{lj}\mathbb{E}_\text{MSS}^\circ[(x_i-1/2)(x_l-1/2)] \Big) \nonumber\\
	&+\frac{u}{N}(N-2)\mathbb{E}_\text{MSS}^\circ[(x_i-1/2)(x_j-1/2)]+\frac{1-u}{N}\sum_{l\in\mathcal{N}}(1-p_{li}-p_{lj})\mathbb{E}_\text{MSS}^\circ[(x_i-1/2)(x_j-1/2)] \nonumber\\
	=&~\frac{1-u}{N}\sum_{l\in\mathcal{N}}\Big(p_{li}\mathbb{E}_\text{MSS}^\circ[(x_l-1/2)(x_j-1/2)]+p_{lj}\mathbb{E}_\text{MSS}^\circ[(x_i-1/2)(x_l-1/2)] \Big) \nonumber\\
	&+\left(1-\frac{2u}{N}-\frac{1-u}{N}\sum_{l\in\mathcal{N}}(p_{li}+p_{lj})\right)\mathbb{E}_\text{MSS}^\circ[(x_i-1/2)(x_j-1/2)].
\end{align}
Integrating $\mathbb{E}_\text{MSS}^\circ[(x_i-1/2)(x_j-1/2)]$ into the left-hand side and denoting $\underline{x}_i=x_i-1/2$, we obtain
\begin{align}\label{eq_EMSSxixj_BD}
	&~\mathbb{E}_\text{MSS}^\circ[\underline{x}_i \underline{x}_j]=\frac{1-u}{2u+(1-u)\sum_{l\in\mathcal{N}}(p_{li}+p_{lj})}
	\left(\sum_{l\in\mathcal{N}}p_{li}\mathbb{E}_\text{MSS}^\circ[\underline{x}_l \underline{x}_j]+\sum_{l\in\mathcal{N}}p_{lj}\mathbb{E}_\text{MSS}^\circ[\underline{x}_i \underline{x}_l]\right) \nonumber\\
	\Leftrightarrow 
	&~\left(\frac{2u}{(1-u)\sum_{l\in\mathcal{N}}(p_{li}+p_{lj})}+1\right)\mathbb{E}_\text{MSS}^\circ[\underline{x}_i \underline{x}_j]=\frac{1}{\sum_{l\in\mathcal{N}}(p_{li}+p_{lj})}
	\left(\sum_{l\in\mathcal{N}}p_{li}\mathbb{E}_\text{MSS}^\circ[\underline{x}_l \underline{x}_j]+\sum_{l\in\mathcal{N}}p_{lj}\mathbb{E}_\text{MSS}^\circ[\underline{x}_i \underline{x}_l]\right).
\end{align}

We define variables $\tilde{\phi}_{ij}(\mathbf{x})$, 
\begin{equation}\label{eq_phi_BD}
	\tilde{\phi}_{ij}(\mathbf{x})=\underline{x}_i \underline{x}_j-\frac{1}{\sum_{l\in\mathcal{N}}(p_{li}+p_{lj})}\left(\sum_{l\in\mathcal{N}}p_{li}\underline{x}_l \underline{x}_j+\sum_{l\in\mathcal{N}}p_{lj}\underline{x}_i \underline{x}_l\right),
\end{equation}
which satisfy the properties: $\tilde{\phi}_{ij}(\mathbf{C})=\tilde{\phi}_{ij}(\mathbf{D})=0$. Therefore, $\tilde{\phi}_{ij}(\mathbf{x})$ can be used to relate the MSS and RMC distributions through Eq.~(\ref{eq_ERMCfromEMSS}) as $u\to 0$.

We first calculate $\mathbb{E}_\text{MSS}^\circ[\tilde{\phi}_{ij}(\mathbf{x})]$. Writing down the expected value of Eq.~(\ref{eq_phi_BD}) and using Eq.~(\ref{eq_EMSSxixj_BD}), we have 
\begin{align}\label{eq_EMSSphi_BD}
	\mathbb{E}_\text{MSS}^\circ[\tilde{\phi}_{ij}(\mathbf{x})]
	&=\mathbb{E}_\text{MSS}^\circ[\underline{x}_i \underline{x}_j]-\frac{1}{\sum_{l\in\mathcal{N}}(p_{li}+p_{lj})}\left(\sum_{l\in\mathcal{N}}p_{li}\mathbb{E}_\text{MSS}^\circ[\underline{x}_l \underline{x}_j]+\sum_{l\in\mathcal{N}}p_{lj}\mathbb{E}_\text{MSS}^\circ[\underline{x}_i \underline{x}_l]\right) \nonumber\\
	&=\mathbb{E}_\text{MSS}^\circ[\underline{x}_i \underline{x}_j]-\left(\frac{2u}{(1-u)\sum_{l\in\mathcal{N}}(p_{li}+p_{lj})}+1\right)\mathbb{E}_\text{MSS}^\circ[\underline{x}_i \underline{x}_j] \nonumber\\
	&=-\frac{2u}{(1-u)\sum_{l\in\mathcal{N}}(p_{li}+p_{lj})}\mathbb{E}_\text{MSS}^\circ[\underline{x}_i \underline{x}_j].
\end{align}

According to Eq.~(\ref{eq_ERMCfromEMSS}), we can calculate $\mathbb{E}_\text{RMC}^\circ[\tilde{\phi}_{ij}(\mathbf{x})]$ from $\mathbb{E}_\text{MSS}^\circ[\tilde{\phi}_{ij}(\mathbf{x})]$, 
\begin{align}\label{eq_ERMCfromEMSS_BD}
	\mathbb{E}_\text{RMC}^\circ[\tilde{\phi}_{ij}(\mathbf{x})]
	&=K\left.\frac{\mathrm{d}\mathbb{E}_\text{MSS}^\circ[\tilde{\phi}_{ij}(\mathbf{x})]}{\mathrm{d}u}\right|_{u=0} \nonumber\\
	&=K\left.\frac{\mathrm{d}}{\mathrm{d}u}\right|_{u=0} \left(-\frac{2u}{(1-u)\sum_{l\in\mathcal{N}}(p_{li}+p_{lj})}\mathbb{E}_\text{MSS}^\circ[\underline{x}_i \underline{x}_j]\right) \nonumber\\
	&=K\left(-\frac{2}{\sum_{l\in\mathcal{N}}(p_{li}+p_{lj})}\left.\mathbb{E}_\text{MSS}^\circ[\underline{x}_i \underline{x}_j]\right|_{u=0}+0\right) \nonumber\\
	&=-\frac{K}{2\sum_{l\in\mathcal{N}}(p_{li}+p_{lj})}.
\end{align}

On the other hand, writing down the expected value of Eq.~(\ref{eq_phi_BD}) under the RMC distribution leads to another expression of $\mathbb{E}_\text{RMC}^\circ[\tilde{\phi}_{ij}(\mathbf{x})]$: 
\begin{equation}\label{eq_ERMCphi_BD}
	\mathbb{E}_\text{RMC}^\circ[\tilde{\phi}_{ij}(\mathbf{x})]
	=\mathbb{E}_\text{RMC}^\circ[\underline{x}_i \underline{x}_j]-\frac{1}{\sum_{l\in\mathcal{N}}(p_{li}+p_{lj})}\left(\sum_{l\in\mathcal{N}}p_{li}\mathbb{E}_\text{RMC}^\circ[\underline{x}_l \underline{x}_j]+\sum_{l\in\mathcal{N}}p_{lj}\mathbb{E}_\text{RMC}^\circ[\underline{x}_i \underline{x}_l]\right).
\end{equation}
Substituting the result of Eq.~(\ref{eq_ERMCfromEMSS_BD}) into Eq.~(\ref{eq_ERMCphi_BD}), we have 
\begin{equation}\label{eq_ERMCxixj_BD}
	\mathbb{E}_\text{RMC}^\circ[\underline{x}_i \underline{x}_j]
	=\frac{1}{\sum_{l\in\mathcal{N}}(p_{li}+p_{lj})}\left(\sum_{l\in\mathcal{N}}p_{li}\mathbb{E}_\text{RMC}^\circ[\underline{x}_l \underline{x}_j]+\sum_{l\in\mathcal{N}}p_{lj}\mathbb{E}_\text{RMC}^\circ[\underline{x}_i \underline{x}_l]\right)
	-\frac{K}{2\sum_{l\in\mathcal{N}}(p_{li}+p_{lj})}.
\end{equation}

We define variables $\tilde{\tau}_{ij}$ for $i,j\in\mathcal{N}$, 
\begin{equation}\label{eq_tauijERMC_BD}
	\tilde{\tau}_{ij}=\frac{\dfrac{1}{2}-\mathbb{E}_\text{RMC}^\circ[x_i x_j]}{K/2}.
\end{equation}
Obviously, $\tilde{\tau}_{ii}=0$ when $i=j$, because $\mathbb{E}_\text{RMC}^\circ[x_i^2]=1/2$. Also, $\tilde{\tau}_{ij}=\tilde{\tau}_{ji}$, because $\mathbb{E}_\text{RMC}^\circ[x_i x_j]=\mathbb{E}_\text{RMC}^\circ[x_j x_i]$.

When $i\neq j$, we can solve for the values of $\tilde{\tau}_{ij}$ by the recurrence relation. According to Eq.~(\ref{eq_relation_RMC}), we know that Eq.~(\ref{eq_tauijERMC_BD}) can be written as 
\begin{equation}
	\tilde{\tau}_{ij}
	=\frac{\dfrac{1}{2}-\left(\mathbb{E}_\text{RMC}^\circ[\underline{x}_i \underline{x}_j]+\dfrac{1}{4}\right)}{K/2}
	=\frac{\dfrac{1}{2}-2\mathbb{E}_\text{RMC}^\circ[\underline{x}_i \underline{x}_j]}{K}
\end{equation}
or
\begin{equation}\label{eq_ERMCtauij_BD}
	\mathbb{E}_\text{RMC}^\circ[\underline{x}_i \underline{x}_j]
	=\frac{\dfrac{1}{2}-K\tilde{\tau}_{ij}}{2}.
\end{equation}

Substituting Eq.~(\ref{eq_ERMCtauij_BD}) into Eq.~(\ref{eq_ERMCxixj_BD}), we obtain the recurrence relation of $\tilde{\tau}_{ij}$: 
\begin{align}\label{eq_tauij_BD}
	&~\frac{\dfrac{1}{2}-K\tilde{\tau}_{ij}}{2}
	=\frac{1}{\sum_{l\in\mathcal{N}}(p_{li}+p_{lj})}\left(\sum_{l\in\mathcal{N}}p_{li}\frac{\dfrac{1}{2}-K\tilde{\tau}_{jl}}{2}+\sum_{l\in\mathcal{N}}p_{lj}\frac{\dfrac{1}{2}-K\tilde{\tau}_{il}}{2}\right)
	-\frac{K}{2\sum_{l\in\mathcal{N}}(p_{li}+p_{lj})} \nonumber\\
	\Leftrightarrow 
	&~\tilde{\tau}_{ij}=\frac{1}{\sum_{l\in\mathcal{N}}(p_{li}+p_{lj})}\left(1+\sum_{l\in\mathcal{N}}p_{li}\tilde{\tau}_{jl}+\sum_{l\in\mathcal{N}}p_{lj}\tilde{\tau}_{il}\right).
\end{align}
The recurrence relation Eq.~(\ref{eq_tauij_BD}), together with $\tilde{\tau}_{ii}=0$, form a system of linear equations, through which all $\tilde{\tau}_{ij}$ values ($i,j\in\mathcal{N}$) can be determined on a given network.

Substituting all $\mathbb{E}_\text{RMC}^\circ[x_i x_j]$ into Eq.~(\ref{eq_coopcondi_1_PC}) with computable $\tilde{\tau}_{ij}$ using Eq.~(\ref{eq_tauijERMC_BD}), and then substituting the result into the cooperation success condition Eq.~(\ref{eq_coopcondi_BD}), where the positive factors $K/2$ and $1/(2N^2\langle k^{-1}\rangle)$ can be canceled out, we arrive at the following condition for the success of cooperation:
\begin{align}\label{eq_coopcondi_2_BD}
	&~\mathbb{E}_\text{RMC}^\circ [\hat{\Delta}'_\text{sel}(\mathbf{x})]>0 \nonumber\\
	\Leftrightarrow
	&~\sum_{i,j\in\mathcal{N}}\frac{k_{ij}}{k_i k_j}\Bigg\{\left(\frac{rc}{(k_i+1)^2}-c\right) \tilde{\tau}_{ij}
	+\frac{rc}{k_i+1}\sum_{l\in\mathcal{N}_i}\left(\frac{1}{k_i+1}+\frac{1}{k_l+1}\right) \left(-\tilde{\tau}_{il}+\tilde{\tau}_{jl}\right)
	\nonumber\\
	&+\frac{rc}{k_i+1}\sum_{l\in\mathcal{N}_i}\frac{1}{k_l+1}\sum_{\ell\in\mathcal{N}_l}\left(-\tilde{\tau}_{i\ell}+\tilde{\tau}_{j\ell}\right)
	-\left(\frac{rc}{(k_j+1)^2}-c\right)\left(-\tilde{\tau}_{ij}\right)
	\nonumber\\
	&-\frac{rc}{k_j+1}\sum_{l\in\mathcal{N}_j}\left(\frac{1}{k_j+1}+\frac{1}{k_l+1}\right)\left(-\tilde{\tau}_{il}+\tilde{\tau}_{jl}\right)
	-\frac{rc}{k_j+1}\sum_{l\in\mathcal{N}_j}\frac{1}{k_l+1}\sum_{\ell\in\mathcal{N}_l}\left(-\tilde{\tau}_{i\ell}+\tilde{\tau}_{j\ell}\right)\Bigg\}
	>0 \nonumber\\
	\Leftrightarrow
	&~r>\frac{2\sum_{i,j\in\mathcal{N}}\dfrac{k_{ij}}{k_i k_j}\tilde{\tau}_{ij}}{\sum_{i,j\in\mathcal{N}}\dfrac{k_{ij}}{k_i k_j}(\tilde{\Upsilon}_{ij}+\tilde{\Upsilon}_{ji})},
\end{align}
where $\tilde{\Upsilon}_{ij}$ are defined by
\begin{equation}\label{eq_Upsilonij_BD}
	\tilde{\Upsilon}_{ij}=
	\frac{1}{k_i+1}\left(
	\frac{\tilde{\tau}_{ij}+k_i \sum_{l\in\mathcal{N}}p_{il} (\tilde{\tau}_{jl}-\tilde{\tau}_{il})}{k_i+1}+
	k_i \sum_{l\in\mathcal{N}}p_{il}\frac{(\tilde{\tau}_{jl}-\tilde{\tau}_{il})+k_l \sum_{\ell\in\mathcal{N}}p_{l\ell}(\tilde{\tau}_{j\ell}-\tilde{\tau}_{i\ell})}{k_l+1}
	\right).
\end{equation}
And according to the previous discussion, $\tilde{\tau}_{ij}$ can be solved by the following system of linear equations: 
\begin{align}\label{eq_eqtauij_BD}
	\begin{cases}
		\displaystyle{
			\tilde{\tau}_{ij}=\frac{1}{\sum_{l\in\mathcal{N}}(p_{li}+p_{lj})}\left(1+\sum_{l\in\mathcal{N}}
			(p_{li}\tilde{\tau}_{jl}+p_{lj}\tilde{\tau}_{il})\right)}, & \mbox{if $j\neq i$},
		\\[1em]
		\displaystyle{
			\tilde{\tau}_{ij}=0}, & \mbox{if $j=i$}.
	\end{cases}
\end{align}
In applications, we can use the following equivalent form, which is more intuitive for calculation:
\begin{align}\label{eq_tauij_BD_2}
	\begin{cases}
		\displaystyle{
			\tilde{\tau}_{ij}=\frac{1}{\sum_{l\in\mathcal{N}_i}k_l^{-1}+\sum_{l\in\mathcal{N}_j}k_l^{-1}}\left(1+\sum_{l\in\mathcal{N}_i}
			k_l^{-1}\tilde{\tau}_{jl}+\sum_{l\in\mathcal{N}_j}
			k_l^{-1}\tilde{\tau}_{il}\right)}, & \mbox{if $j\neq i$},
		\\[1em]
		\displaystyle{
			\tilde{\tau}_{ij}=0}, & \mbox{if $j=i$}.
	\end{cases}
\end{align}

Finally, $\sum_{i,j\in\mathcal{N}}k_{ij}/(k_i k_j)\tilde{\Upsilon}_{ji}=\sum_{i,j\in\mathcal{N}}k_{ij}/(k_i k_j)\tilde{\Upsilon}_{ij}$. Therefore, $\sum_{i,j\in\mathcal{N}}k_{ij}/(k_i k_j)(\tilde{\Upsilon}_{ij}+\tilde{\Upsilon}_{ji})=2\sum_{i,j\in\mathcal{N}}k_{ij}/(k_i k_j)\tilde{\Upsilon}_{ij}$. As a result, Eq.~(\ref{eq_coopcondi_2_BD}) can be further simplified as 
\begin{equation}\label{eq_coopcondi_3_BD}
	r>\frac{\sum_{i,j\in\mathcal{N}}\dfrac{k_{ij}}{k_i k_j} \tilde{\tau}_{ij}}{\sum_{i,j\in\mathcal{N}}\dfrac{k_{ij}}{k_i k_j}\tilde{\Upsilon}_{ij}}.
\end{equation}
The right-hand side is the $r^\star$ value under the BD rule. Furthermore, let $\tilde{\tau}^{(1)}=\sum_{i,j\in\mathcal{N}}k_{ij}/(k_i k_j)\tilde{\tau}_{ij}$, $\tilde{\Upsilon}^{(1)}=\sum_{i,j\in\mathcal{N}}k_{ij}/(k_i k_j)\tilde{\Upsilon}_{ij}$, then the $r^\star$ value under the BD rule can also be expressed in shorthand as
\begin{equation}
	r^\star=\frac{\tilde{\tau}^{(1)}}{\tilde{\Upsilon}^{(1)}}.
\end{equation}

\subsection{Variation of the model: accumulated payoff}
In the previous deduction, we assume that an agent $i$'s actual payoff is averaged over the $1+k_i$ games organized by itself and its neighbors, ensuring the consistency of payoff scales across different numbers of neighbors. Another approach is to take the accumulated payoff from these games. On homogeneous graphs, there is no difference between these two approaches in the weak selection limit, but on heterogeneous graphs, agents with more neighbors tend to have higher \& lower payoffs because they participate in more games. From a physical perspective, the accumulated payoff is also a more intuitive model detail in real-world systems. We are thus interested in examining the conditions for the success of cooperation when using accumulated payoffs.

\subsubsection{Modified payoff calculation}
We do not normalize agent $i$'s actual payoff by dividing $1+k_i$ but take the accumulated payoff directly. Similar to Eq.~(\ref{eq_payoff}), the calculation of agent $i$'s actual payoff $f_i(\mathbf{x})$ follows Eq.~(\ref{eq_payoff_accu}).
\begin{align}\label{eq_payoff_accu}
	f_i(\mathbf{x})
	&=\sum_{l\in\mathcal{G}_i}\left(\frac{r\sum_{\ell\in\mathcal{G}_l} x_\ell c}{G_l}-x_i c\right)
	\nonumber\\
	&=\left(\frac{r(x_i+\sum_{l\in\mathcal{N}_i} x_l)c}{k_i+1}-x_i c\right)
	+\sum_{l\in\mathcal{N}_i} \left(\frac{r(x_l+\sum_{\ell\in\mathcal{N}_l} x_\ell)c}{k_l+1}-x_i c\right)
	\nonumber\\
	&=\left(\frac{rc}{k_i+1}-(k_i+1)c\right)x_i
	+rc\sum_{l\in\mathcal{N}_i}\left(\frac{1}{k_i+1}+\frac{1}{k_l+1}\right)x_l
	+rc\sum_{l\in\mathcal{N}_i}\frac{1}{k_l+1}\sum_{\ell\in\mathcal{N}_l}x_\ell.
\end{align}

The dynamics of strategy evolution remain the same under neutral drift. Only the quantity $\mathbb{E}_\text{RMC}^\circ [(x_i-x_j)(f_i(\mathbf{x})-f_j(\mathbf{x}))]$ is influenced by the modified payoff calculation $f_i(\mathbf{x})$ ($f_j(\mathbf{x})$) and is recalculated as follows.
\begin{align}\label{eq_coopcondi_1_PC_accu}
	&~\mathbb{E}_\text{RMC}^\circ [(x_i-x_j)(f_i(\mathbf{x})-f_j(\mathbf{x}))]
	\nonumber\\
	=&~\mathbb{E}_\text{RMC}^\circ \Bigg[
	\left(\frac{rc}{k_i+1}-(k_i+1)c\right)(x_i^2-x_i x_j)
	+rc\sum_{l\in\mathcal{N}_i}\left(\frac{1}{k_i+1}+\frac{1}{k_l+1}\right)(x_i x_l-x_j x_l)
	\nonumber\\
	&+rc\sum_{l\in\mathcal{N}_i}\frac{1}{k_l+1}\sum_{\ell\in\mathcal{N}_l}(x_i x_\ell-x_j x_\ell)
	-\left(\frac{rc}{k_j+1}-(k_j+1)c\right)(x_i x_j-x_j^2)
	\nonumber\\
	&-rc\sum_{l\in\mathcal{N}_j}\left(\frac{1}{k_j+1}+\frac{1}{k_l+1}\right)(x_i x_l-x_j x_l)-rc\sum_{l\in\mathcal{N}_j}\frac{1}{k_l+1}\sum_{\ell\in\mathcal{N}_l}(x_i x_\ell-x_j x_\ell)
	\Bigg]
	\nonumber\\
	=&~\left(\frac{rc}{k_i+1}-(k_i+1)c\right) \left(\mathbb{E}_\text{RMC}^\circ[x_i^2]-\mathbb{E}_\text{RMC}^\circ[x_i x_j]\right)
	+rc\sum_{l\in\mathcal{N}_i}\left(\frac{1}{k_i+1}+\frac{1}{k_l+1}\right) \left(\mathbb{E}_\text{RMC}^\circ[x_i x_l]-\mathbb{E}_\text{RMC}^\circ[x_j x_l]\right)
	\nonumber\\
	&+rc\sum_{l\in\mathcal{N}_i}\frac{1}{k_l+1}\sum_{\ell\in\mathcal{N}_l}\left(\mathbb{E}_\text{RMC}^\circ[x_i x_\ell]-\mathbb{E}_\text{RMC}^\circ[x_j x_\ell]\right)
	-\left(\frac{rc}{k_j+1}-(k_j+1)c\right)\left(\mathbb{E}_\text{RMC}^\circ[x_i x_j]-\mathbb{E}_\text{RMC}^\circ[x_j^2]\right)
	\nonumber\\
	&-rc\sum_{l\in\mathcal{N}_j}\left(\frac{1}{k_j+1}+\frac{1}{k_l+1}\right)\left(\mathbb{E}_\text{RMC}^\circ[x_i x_l]-\mathbb{E}_\text{RMC}^\circ[x_j x_l]\right)
	-rc\sum_{l\in\mathcal{N}_j}\frac{1}{k_l+1}\sum_{\ell\in\mathcal{N}_l}\left(\mathbb{E}_\text{RMC}^\circ[x_i x_\ell]-\mathbb{E}_\text{RMC}^\circ[x_j x_\ell]\right).
\end{align}

\subsubsection{Pairwise comparison}
The cooperation condition under the PC rule is still Eq.~(\ref{eq_coopcondi_PC}), because the condition was obtained under neutral drift and thus remains independent of the later introduced marginal effect of games. Using the result of Eq.~(\ref{eq_coopcondi_1_PC_accu}) and $\tau_{ij}=(1/2-\mathbb{E}_\text{RMC}^\circ[x_i x_j])/(K/4)$ as defined by Eq.~(\ref{eq_tauijERMC_PC}), we calculate 
\begin{align}\label{eq_coopcondi_2_PC_accu}
	&~\frac{1}{4N^2\langle k\rangle} \sum_{i,j\in\mathcal{N}}k_i p_{ij}\mathbb{E}_\text{RMC}^\circ [(x_i-x_j)(f_i(\mathbf{x})-f_j(\mathbf{x}))]>0 \nonumber\\
	\Leftrightarrow
	&~\sum_{i,j\in\mathcal{N}}k_i p_{ij}\Bigg\{\left(\frac{rc}{k_i+1}-(k_i+1)c\right) \tau_{ij}
	+rc\sum_{l\in\mathcal{N}_i}\left(\frac{1}{k_i+1}+\frac{1}{k_l+1}\right) \left(-\tau_{il}+\tau_{jl}\right)
	\nonumber\\
	&+rc\sum_{l\in\mathcal{N}_i}\frac{1}{k_l+1}\sum_{\ell\in\mathcal{N}_l}\left(-\tau_{i\ell}+\tau_{j\ell}\right)
	-\left(\frac{rc}{k_j+1}-(k_j+1)c\right)\left(-\tau_{ij}\right)
	\nonumber\\
	&-rc\sum_{l\in\mathcal{N}_j}\left(\frac{1}{k_j+1}+\frac{1}{k_l+1}\right)\left(-\tau_{il}+\tau_{jl}\right)
	-rc\sum_{l\in\mathcal{N}_j}\frac{1}{k_l+1}\sum_{\ell\in\mathcal{N}_l}\left(-\tau_{i\ell}+\tau_{j\ell}\right)\Bigg\}
	>0 \nonumber\\
	\Leftrightarrow
	&~r>\frac{\sum_{i,j\in\mathcal{N}}k_i p_{ij}(k_i+k_j+2)\tau_{ij}}{\sum_{i,j\in\mathcal{N}}k_i p_{ij}[(k_i+1)\Upsilon_{ij}+(k_j+1)\Upsilon_{ji}]}.
\end{align}
Further simplifying Eq.~(\ref{eq_coopcondi_2_PC_accu}) (using Eq.~(\ref{eq_tau01234_1})) leads to 
\begin{equation}\label{eq_coopcondi_3_PC_accu}
	r>\frac{\sum_{i,j\in\mathcal{N}}k_i (k_i+1)p_{ij}\tau_{ij}}{\sum_{i,j\in\mathcal{N}}k_i (k_i+1)p_{ij}\Upsilon_{ij}}.
\end{equation}
The right-hand side is the $r^\star$ value for the success of cooperation when using accumulated payoffs under the PC rule. The $\tau_{ij}$ values are still obtained by solving Eqs.~(\ref{eq_eqtauij_PC}) on the given network, and the $\Upsilon_{ij}$ values are still obtained by Eq.~(\ref{eq_Upsilonij_PC}).

\subsubsection{Death-birth}
The cooperation condition under the DB rule is Eq.~(\ref{eq_coopcondi_DB}). Using the result of Eq.~(\ref{eq_coopcondi_1_PC_accu}) and $\tau_{ij}=(1/2-\mathbb{E}_\text{RMC}^\circ[x_i x_j])/(K/4)$ as defined by Eq.~(\ref{eq_tauijERMC_DB}), we calculate 
\begin{align}\label{eq_coopcondi_2_DB_accu}
	&~\frac{1}{2N^2\langle k\rangle} \sum_{i,j\in\mathcal{N}}k_i p^{(2)}_{ij}\mathbb{E}_\text{RMC}^\circ [(x_i-x_j)(f_i(\mathbf{x})-f_j(\mathbf{x}))]>0 \nonumber\\
	\Leftrightarrow
	&~\sum_{i,j\in\mathcal{N}}k_i p^{(2)}_{ij}\Bigg\{\left(\frac{rc}{k_i+1}-(k_i+1)c\right) \tau_{ij}
	+rc\sum_{l\in\mathcal{N}_i}\left(\frac{1}{k_i+1}+\frac{1}{k_l+1}\right) \left(-\tau_{il}+\tau_{jl}\right)
	\nonumber\\
	&+rc\sum_{l\in\mathcal{N}_i}\frac{1}{k_l+1}\sum_{\ell\in\mathcal{N}_l}\left(-\tau_{i\ell}+\tau_{j\ell}\right)
	-\left(\frac{rc}{k_j+1}-(k_j+1)c\right)\left(-\tau_{ij}\right)
	\nonumber\\
	&-rc\sum_{l\in\mathcal{N}_j}\left(\frac{1}{k_j+1}+\frac{1}{k_l+1}\right)\left(-\tau_{il}+\tau_{jl}\right)
	-rc\sum_{l\in\mathcal{N}_j}\frac{1}{k_l+1}\sum_{\ell\in\mathcal{N}_l}\left(-\tau_{i\ell}+\tau_{j\ell}\right)\Bigg\}
	>0 \nonumber\\
	\Leftrightarrow
	&~r>\frac{\sum_{i,j\in\mathcal{N}}k_i p^{(2)}_{ij}(k_i+k_j+2)\tau_{ij}}{\sum_{i,j\in\mathcal{N}}k_i p^{(2)}_{ij}[(k_i+1)\Upsilon_{ij}+(k_j+1)\Upsilon_{ji}]}.
\end{align}
Further simplifying Eq.~(\ref{eq_coopcondi_2_DB_accu}) leads to 
\begin{equation}\label{eq_coopcondi_3_DB_accu}
	r>\frac{\sum_{i,j\in\mathcal{N}}k_i (k_i+1)p^{(2)}_{ij}\tau_{ij}}{\sum_{i,j\in\mathcal{N}}k_i (k_i+1)p^{(2)}_{ij}\Upsilon_{ij}}.
\end{equation}
The right-hand side is the $r^\star$ value for the success of cooperation when using accumulated payoffs under the DB rule. The $\tau_{ij}$ values are still obtained by solving Eqs.~(\ref{eq_eqtauij_PC}) on the given network, and the $\Upsilon_{ij}$ values are still obtained by Eq.~(\ref{eq_Upsilonij_PC}).

\subsubsection{Birth-death}
The cooperation condition under the BD rule is Eq.~(\ref{eq_coopcondi_BD}). Using the result of Eq.~(\ref{eq_coopcondi_1_PC_accu}) and $\tau_{ij}=(1/2-\mathbb{E}_\text{RMC}^\circ[x_i x_j])/(K/2)$ as defined by Eq.~(\ref{eq_tauijERMC_BD}), we calculate 
\begin{align}\label{eq_coopcondi_2_BD_accu}
	&~\frac{1}{2N^2\langle k^{-1}\rangle} \sum_{i,j\in\mathcal{N}}\frac{k_{ij}}{k_i k_j}\mathbb{E}_\text{RMC}^\circ [(x_i-x_j)(f_i(\mathbf{x})-f_j(\mathbf{x}))]>0 \nonumber\\
	\Leftrightarrow
	&~\sum_{i,j\in\mathcal{N}}\frac{k_{ij}}{k_i k_j}\Bigg\{\left(\frac{rc}{k_i+1}-(k_i+1)c\right) \tilde{\tau}_{ij}
	+rc\sum_{l\in\mathcal{N}_i}\left(\frac{1}{k_i+1}+\frac{1}{k_l+1}\right) \left(-\tilde{\tau}_{il}+\tilde{\tau}_{jl}\right)
	\nonumber\\
	&+rc\sum_{l\in\mathcal{N}_i}\frac{1}{k_l+1}\sum_{\ell\in\mathcal{N}_l}\left(-\tilde{\tau}_{i\ell}+\tilde{\tau}_{j\ell}\right)
	-\left(\frac{rc}{k_j+1}-(k_j+1)c\right)\left(-\tilde{\tau}_{ij}\right)
	\nonumber\\
	&-rc\sum_{l\in\mathcal{N}_j}\left(\frac{1}{k_j+1}+\frac{1}{k_l+1}\right)\left(-\tilde{\tau}_{il}+\tilde{\tau}_{jl}\right)
	-rc\sum_{l\in\mathcal{N}_j}\frac{1}{k_l+1}\sum_{\ell\in\mathcal{N}_l}\left(-\tilde{\tau}_{i\ell}+\tilde{\tau}_{j\ell}\right)\Bigg\}
	>0 \nonumber\\
	\Leftrightarrow
	&~r>\frac{\sum_{i,j\in\mathcal{N}}\dfrac{k_{ij}}{k_i k_j}(k_i+k_j+2)\tilde{\tau}_{ij}}{\sum_{i,j\in\mathcal{N}}\dfrac{k_{ij}}{k_i k_j}[(k_i+1)\tilde{\Upsilon}_{ij}+(k_j+1)\tilde{\Upsilon}_{ji}]}. 
\end{align}
Further simplifying Eq.~(\ref{eq_coopcondi_2_DB_accu}) leads to 
\begin{equation}\label{eq_coopcondi_3_BD_accu}
	r>\frac{\sum_{i,j\in\mathcal{N}}\dfrac{k_{ij}}{k_i k_j}(k_i+1)\tilde{\tau}_{ij}}{\sum_{i,j\in\mathcal{N}}\dfrac{k_{ij}}{k_i k_j}(k_i+1)\tilde{\Upsilon}_{ij}}.
\end{equation}
The right-hand side is the $r^\star$ value for the success of cooperation when using accumulated payoffs under the BD rule. The $\tilde{\tau}_{ij}$ values are obtained by solving Eqs.~(\ref{eq_eqtauij_BD}) on the given network, and the $\tilde{\Upsilon}_{ij}$ values are obtained by Eq.~(\ref{eq_Upsilonij_BD}).

\section{: Applications to specific networks}\label{sec_appl}
With the cooperation conditions obtained in \ref{sec_theory}, we can calculate the critical synergy factor for the success of cooperation in PGGs on any given network. Here, we present the calculation process for five examples: regular graphs, star graphs, hub-to-hub star graphs, $m$-hub star graphs, and fans.

\subsection{Regular graphs}
The theoretical results of PGG on regular graphs have been previously obtained~\cite{su2019spatial}, and our framework on any network can reproduce these results when applied to regular networks. On a regular graph, all nodes have the same number of neighbors, $k_i\equiv k$ for $i\in\mathcal{N}$, so $p_{ij}=p_{ji}\equiv 1/k$ for $i,j\in\mathcal{N}$, $i\neq j$. 

The calculation for regular graphs is, in fact, the least intuitive compared to other heterogeneous networks when using this framework. We cannot directly solve the system of linear equations for $\tau_{ij}$ but instead need to construct intermediate quantities and special recurrence relations for regular graphs~\cite{allen2017evolutionary}.

We first define
\begin{equation}
	\tau_i=1+\sum_{j\in\mathcal{N}}p_{ij}\tau_{ij},
\end{equation}
with which we calculate the recurrence relation of $\tau^{(n)}$ defined in the main text:
\begin{align}\label{eq_taun}
	\tau^{(n)}
	&=\sum_{i,j\in\mathcal{N}} k_i p_{ij}^{(n)}\tau_{ij} \nonumber\\
	&=\sum_{\substack{i,j\in\mathcal{N}\\i\neq j}}k_i p_{ij}^{(n)}\left(1+\frac{1}{2}\sum_{l\in\mathcal{N}}p_{il}\tau_{jl}+\frac{1}{2}\sum_{l\in\mathcal{N}}p_{jl}\tau_{il}\right) \nonumber\\
	&=\sum_{i,j\in\mathcal{N}}k_i p_{ij}^{(n)}\left(1+\frac{1}{2}\sum_{l\in\mathcal{N}}p_{il}\tau_{jl}+\frac{1}{2}\sum_{l\in\mathcal{N}}p_{jl}\tau_{il}\right)
	-\sum_{i\in\mathcal{N}}k_i p_{ii}^{(n)}\left(1+\sum_{l\in\mathcal{N}}p_{il}\tau_{il}\right) \nonumber\\
	&=\sum_{i,j\in\mathcal{N}}k_i p_{ij}^{(n)}
	+\frac{1}{2}\sum_{i,j,l\in\mathcal{N}}k_j p_{ji}^{(n)}p_{il}\tau_{jl}+\frac{1}{2}\sum_{i,j,l\in\mathcal{N}}k_i p_{ij}^{(n)}p_{jl}\tau_{il}
	-\sum_{i\in\mathcal{N}}k_i p_{ii}^{(n)}\tau_i \nonumber\\
	&=N\langle k\rangle
	+\frac{1}{2}\sum_{j,l\in\mathcal{N}}k_j p_{jl}^{(n+1)}\tau_{jl}+\frac{1}{2}\sum_{i,l\in\mathcal{N}}k_i p_{il}^{(n+1)}\tau_{il}
	-\sum_{i\in\mathcal{N}}k_i p_{ii}^{(n)}\tau_i \nonumber\\
	&=N\langle k\rangle
	+\tau^{(n+1)}
	-\sum_{i\in\mathcal{N}}k_i p_{ii}^{(n)}\tau_i.
\end{align}
Therefore, we have the following recurrence relation:
\begin{equation}\label{eq_taun+1}
	\tau^{(n+1)}=\tau^{(n)}+\sum_{i\in\mathcal{N}}k_i p_{ii}^{(n)}\tau_i-N\langle k\rangle.
\end{equation}

The fourth line in Eq.~(\ref{eq_taun}) used the following fact:
\begin{equation}\label{eq_kipijreverse}
	k_i p_{ij}^{(n)}
	=\sum_{\ell_1\in\mathcal{N}}k_{i\ell_1}p_{\ell_1 j}^{(n-1)}
	=\sum_{\ell_1,\ell_2,\dots,\ell_{n-1}\in\mathcal{N}}k_{i\ell_1}\frac{k_{\ell_1 \ell_2}\cdots k_{\ell_{n-1} j}}{k_{\ell_1}\cdots k_{\ell_{n-1}}}
	=\sum_{\ell_1,\ell_2,\dots,\ell_{n-1}\in\mathcal{N}}k_{j\ell_{n-1}}\frac{k_{\ell_{n-1} \ell_{n-2}}\cdots k_{\ell_{1} i}}{k_{\ell_{n-1}}\cdots k_{\ell_{1}}}=k_j p_{ji}^{(n)}.
\end{equation}

With the help of the recurrence relation Eq.~(\ref{eq_taun+1}), we can start from $\tau^{(0)}$ and obtain all $\tau^{(n)}$ values step by step. For $i=j$, we have $\tau_{ij}=0$. Moreover, one stays in the original position if not walking, so $p^{(0)}_{ij}=1$ if $i=j$ and $p^{(0)}_{ij}=0$ if $i\neq j$. Therefore, $\tau^{(0)}=\sum_{i,j\in\mathcal{N}} k_i p_{ij}^{(0)}\tau_{ij}=0$. Substituting these to Eq.~(\ref{eq_taun+1}), we have the following results:
\begin{subequations}\label{eq_tau01234_1}
	\begin{align}
		\tau^{(0)}&=0, \\
		\tau^{(1)}&=\sum_{i\in\mathcal{N}}k_i \tau_i-N\langle k\rangle, \\
		\tau^{(2)}&=\sum_{i\in\mathcal{N}}k_i \tau_i (1+p_{ii}^{(1)})-2N\langle k\rangle, \\
		\tau^{(3)}&=\sum_{i\in\mathcal{N}}k_i \tau_i (1+p_{ii}^{(1)}+p_{ii}^{(2)})-3N\langle k\rangle, \\
		\tau^{(4)}&=\sum_{i\in\mathcal{N}}k_i \tau_i (1+p_{ii}^{(1)}+p_{ii}^{(2)}+p_{ii}^{(3)})-4N\langle k\rangle.
	\end{align}
\end{subequations}

As declared by Ref.~\cite{allen2017evolutionary}, there is a relation: $\lim_{n\to\infty}p_{ii}^{{(n)}}=k_i/(N\langle k\rangle)$. Then, taking $n\to\infty$ in the recurrence relation Eq.~(\ref{eq_taun+1}), we have 
\begin{equation}\label{eq_tauinfty}
	\tau^{(\infty)}=\tau^{(\infty)}+\sum_{i\in\mathcal{N}}k_i p_{ii}^{(\infty)}\tau_i-N\langle k\rangle
	\Leftrightarrow
	\sum_{i\in\mathcal{N}}k^2_i \tau_i=N^2\langle k\rangle^2.
\end{equation}

Supposing all nodes on the regular graph are transitive, we denote $p^{(n)}_{ii}\equiv p^{(n)}$ in shorthand. Obviously, $p^{(1)}=0$, because we assumed no self-loops on the network, and one cannot leave and return to the same node within a single step; $p^{(2)}=1/k$, since for each possible first step, the probability of returning to the original node in the second step is $1/k$. The average degree of the network is $\langle k\rangle=k$, so we have $\sum_{i\in\mathcal{N}}k_i \tau_i\equiv N^2 k$ according to Eq.~(\ref{eq_tauinfty}). To summarize, Eqs.~(\ref{eq_tau01234_1}) can be calculated as
\begin{subequations}\label{eq_tau01234_2}
	\begin{align}
		\tau^{(0)}&=0, \\
		\tau^{(1)}&=\sum_{i\in\mathcal{N}}k_i \tau_i-Nk
		=(N-1)Nk, \\
		\tau^{(2)}&=\sum_{i\in\mathcal{N}}k_i \tau_i-2Nk
		=(N-2)Nk, \\
		\tau^{(3)}&=\sum_{i\in\mathcal{N}}k_i \tau_i\left(1+\frac{1}{k}\right)-3Nk
		=\left[N\left(1+\frac{1}{k}\right)-3\right]Nk, \\
		\tau^{(4)}&=\sum_{i\in\mathcal{N}}k_i \tau_i\left(1+\frac{1}{k}+p_{ii}^{(3)}\right)-4Nk
		=\left[N\left(1+\frac{1}{k}+p^{(3)}\right)-4\right]Nk.
	\end{align}
\end{subequations}

Since $k_i\equiv k$ on a regular graph, $\Upsilon_{ij}$ in Eq.~(\ref{eq_Upsilonij_PC}) can be simplified as
\begin{equation}\label{eq_Upsilonij_regular}
	\Upsilon_{ij}=
	\frac{1}{(k+1)^2}\left(
	\tau_{ij}+2k\sum_{l\in\mathcal{N}}p_{il} (\tau_{jl}-\tau_{il})+
	k^2 \sum_{\ell\in\mathcal{N}}p^{(2)}_{i\ell}(\tau_{j\ell}-\tau_{i\ell})
	\right).
\end{equation}

Now, we can calculate the critical synergy factor for the success of cooperation in PGGs. The general approach is to calculate the numerator $\sum_{i,j\in\mathcal{N}}k_i p^{(n)}_{ij}\tau_{ij}$ and denominator $\sum_{i,j\in\mathcal{N}}k_i p^{(n)}_{ij}\Upsilon_{ij}$ separately, expressing them by $\tau^{(0)}$, $\tau^{(1)}$, $\tau^{(2)}$, etc., and applying the results of Eqs.~(\ref{eq_tau01234_2}).

For the PC rule, the numerator of $r^\star$ is 
\begin{equation}\label{eq_nume_regular_PC}
	\sum_{i,j\in\mathcal{N}}k_i p_{ij}\tau_{ij}
	=\tau^{(1)},
\end{equation}
and by Eq.~(\ref{eq_Upsilonij_regular}), the denominator is
\begin{align}\label{eq_deno_regular_PC}
	\sum_{i,j\in\mathcal{N}}k_i p_{ij}\Upsilon_{ij}
	=&~\frac{1}{(k+1)^2}\Bigg(
	\sum_{i,j\in\mathcal{N}}k_i p_{ij}\tau_{ij}+2k\sum_{i,j,l\in\mathcal{N}}k_ip_{ij}p_{il}\tau_{jl}-2k\sum_{i,j,l\in\mathcal{N}}k_ip_{ij}p_{il}\tau_{il} \nonumber\\
	&+k^2 \sum_{i,j,\ell\in\mathcal{N}}k_ip_{ij}p^{(2)}_{i\ell}\tau_{j\ell}-k^2 \sum_{i,j,\ell\in\mathcal{N}}k_ip_{ij}p^{(2)}_{i\ell}\tau_{i\ell}
	\Bigg) \nonumber\\
	=&~\frac{1}{(k+1)^2}\Bigg(\sum_{i,j\in\mathcal{N}}k_i p_{ij}\tau_{ij}+2k\sum_{j,l\in\mathcal{N}}k_jp^{(2)}_{jl}\tau_{jl}-2k\sum_{i,l\in\mathcal{N}}k_ip_{il}\tau_{il} \nonumber\\
	&+k^2 \sum_{j,\ell\in\mathcal{N}}k_jp^{(3)}_{j\ell}\tau_{j\ell}-k^2 \sum_{i,\ell\in\mathcal{N}}k_ip^{(2)}_{i\ell}\tau_{i\ell}
	\Bigg) \nonumber\\
	=&~\frac{\tau^{(1)}+2k(\tau^{(2)}-\tau^{(1)})+k^2 (\tau^{(3)}-\tau^{(2)})}{(k+1)^2}.
\end{align}
Assembling Eq.~(\ref{eq_nume_regular_PC}) and Eq.~(\ref{eq_deno_regular_PC}) and inserting the results of Eqs.~(\ref{eq_tau01234_2}), we have 
\begin{align}\label{eq_appenr_regular_PC}
	r^\star
	&=\frac{\sum_{i,j\in\mathcal{N}}k_i p_{ij}\tau_{ij}}{\sum_{i,j\in\mathcal{N}}k_i p_{ij}\Upsilon_{ij}} \nonumber\\
	&=\frac{(k+1)^2 \tau^{(1)}}{\tau^{(1)}+2k(\tau^{(2)}-\tau^{(1)})+k^2 (\tau^{(3)}-\tau^{(2)})} \nonumber\\
	&=\frac{(N-1)G}{N-G}\xrightarrow{N\to \infty} G,
\end{align}
which is the critical synergy factor for the success of cooperation in PGGs on regular graphs under the PC rule, consistent with the previous research~\cite{su2019spatial,wang2023inertia}. Eq.~(\ref{eq_appenr_regular_PC}) has replaced the number of neighbors $k$ by the group size $G=k+1$ for intuitive understanding in PGGs.

For the DB rule, the calculation is similar. The numerator of $r^\star$ is 
\begin{equation}\label{eq_nume_regular_DB}
	\sum_{i,j\in\mathcal{N}}k_i p^{(2)}_{ij}\tau_{ij}=\tau^{(2)},
\end{equation}
and by Eq.~(\ref{eq_Upsilonij_regular}), the denominator is 
\begin{align}\label{eq_deno_regular_DB}
	\sum_{i,j\in\mathcal{N}}k_i p^{(2)}_{ij}\Upsilon_{ij}
	=&~\frac{1}{(k+1)^2}\Bigg(
	\sum_{i,j\in\mathcal{N}}k_i p^{(2)}_{ij}\tau_{ij}+2k\sum_{i,j,l\in\mathcal{N}}k_ip^{(2)}_{ij}p_{il}\tau_{jl}-2k\sum_{i,j,l\in\mathcal{N}}k_ip^{(2)}_{ij}p_{il}\tau_{il} \nonumber\\
	&+k^2 \sum_{i,j,\ell\in\mathcal{N}}k_ip^{(2)}_{ij}p^{(2)}_{i\ell}\tau_{j\ell}-k^2 \sum_{i,j,\ell\in\mathcal{N}}k_ip^{(2)}_{ij}p^{(2)}_{i\ell}\tau_{i\ell}
	\Bigg) \nonumber\\
	=&~\frac{1}{(k+1)^2}\Bigg(\sum_{i,j\in\mathcal{N}}k_i p^{(2)}_{ij}\tau_{ij}+2k\sum_{j,l\in\mathcal{N}}k_jp^{(3)}_{jl}\tau_{jl}-2k\sum_{i,l\in\mathcal{N}}k_ip_{il}\tau_{il} \nonumber\\
	&+k^2 \sum_{j,\ell\in\mathcal{N}}k_jp^{(4)}_{j\ell}\tau_{j\ell}-k^2 \sum_{i,\ell\in\mathcal{N}}k_ip^{(2)}_{i\ell}\tau_{i\ell}
	\Bigg) \nonumber\\
	=&~\frac{\tau^{(2)}+2k(\tau^{(3)}-\tau^{(1)})+k^2 (\tau^{(4)}-\tau^{(2)})}{(k+1)^2}.
\end{align}
Assembling Eq.~(\ref{eq_nume_regular_DB}) and Eq.~(\ref{eq_deno_regular_DB}) and inserting the results of Eq.~(\ref{eq_tau01234_2}), we have 
\begin{align}\label{eq_appenr_regular_DB}
	r^\star
	&=\frac{\sum_{i,j\in\mathcal{N}}k_i p^{(2)}_{ij}\tau_{ij}}{\sum_{i,j\in\mathcal{N}}k_i p^{(2)}_{ij}\Upsilon_{ij}} \nonumber\\
	&=\frac{(k+1)^2 \tau^{(2)}}{\tau^{(2)}+2k(\tau^{(3)}-\tau^{(1)})+k^2 (\tau^{(4)}-\tau^{(2)})} \nonumber\\
	&=\frac{(N-2)G^2}{N(G-1)^2 p^{(3)}+N(G+2)-2G^2} \nonumber\\
	&=\frac{(N-2)G^2}{N(G-2)\mathcal{C}+N(G+2)-2G^2}
	\xrightarrow{N\to \infty} \frac{G^2}{(G-2)\mathcal{C}+G+2}.
\end{align}
This is the critical synergy factor for the success of cooperation in PGGs on regular graphs under the DB rule, consistent with the previous research~\cite{su2019spatial,wang2023inertia,wang2023imitation}. Eq.~(\ref{eq_appenr_regular_DB}) has replaced the three-step random walk probability $p^{(3)}$ by clustering coefficient $\mathcal{C}=k^2 p^{(3)}/(k-1)$, which is an intuitive and commonly used concept in network science.

For the BD rule, the recurrence relation Eq.~(\ref{eq_tauij_BD_2}) reduces to the following form on a regular graph:
\begin{align}\label{eq_tauij_BD_regular}
	\begin{cases}
		\displaystyle{
			\tilde{\tau}_{ij}=\frac{1}{2}\left(1+\sum_{l\in\mathcal{N}_i}\tilde{\tau}_{jl}+\sum_{l\in\mathcal{N}_j}\tilde{\tau}_{il}\right)}, & \mbox{if $j\neq i$},
		\\[1em]
		\displaystyle{
			\tilde{\tau}_{ij}=0}, & \mbox{if $j=i$}.
	\end{cases}
\end{align}
For the critical synergy factor, $\tilde{\tau}_{ij}$ in the numerator and denominator are homogeneous (see Eq.~(\ref{eq_coopcondi_3_BD}) and Eq.~(\ref{eq_Upsilonij_BD})). Therefore, the critical synergy factor is invariant by replacing $\tilde{\tau}_{ij}\gets \tilde{\tau}_{ij}^*/2$, which makes Eq.~(\ref{eq_tauij_BD_regular}) the same recurrence relation for $\tilde{\tau}_{ij}^*$ as the one for $\tau_{ij}$ under the PC and DB rules. The solution for $\tilde{\tau}_{ij}^*$ is thus equal to $\tau_{ij}$. On the other hand, we have $k_{ij}/(k_i k_j)=k_i p_{ij}/k^2$ on regular graphs, where $k^2$ can be canceled simultaneously in the numerator and denominator of the critical synergy factor, making it (Eq.~(\ref{eq_coopcondi_3_BD})) the same as the one for the PC rule (Eq.~(\ref{eq_coopcondi_3_PC})). 

Therefore, on regular graphs, the condition for the success of cooperation in PGGs under the BD rule is equal to the one under the PC rule (i.e., Eq.~(\ref{eq_appenr_regular_PC})).

\begin{itemize}
	\item Accumulated payoff
\end{itemize}

On regular graphs, $k_i \equiv k$ for $i\in\mathcal{N}$. Therefore, $k_i+1$ can be canceled simultaneously in the numerator and denominator in Eq.~(\ref{eq_coopcondi_3_PC_accu}), Eq.~(\ref{eq_coopcondi_3_DB_accu}), and Eq.~(\ref{eq_coopcondi_3_BD_accu}). As a consequence, the critical synergy factors for accumulated payoff are equal to the ones for average payoff under the three update rules (Eq.~(\ref{eq_coopcondi_3_PC}), Eq.~(\ref{eq_coopcondi_3_DB}), and Eq.~(\ref{eq_coopcondi_3_BD})). For the results under the PC and BD rules, please refer to Eq.~(\ref{eq_appenr_regular_PC}), and for the result under the DB rule, please refer to Eq.~(\ref{eq_appenr_regular_DB}).

\subsection{Star graph}
For heterogeneous networks, we can calculate the critical synergy factor by solving the recurrence relation and assembling the resultant $\tau_{ij}$ values.

On a star graph, there is one hub ($H$) and $n$ leaves ($L$). The hub node has $k_H=n$ neighbors, and each leaf node has $k_L=1$ neighbor. The values are equal among $\tau_{ij}$ of the same type, and for the star graph, there are only two non-zero $\tau_{ij}$ types: $\tau_{HL}$, the relation between the hub and a leaf, and $\tau_{LL'}$, the relation between a leaf and another leaf. According to Eq.~(\ref{eq_tauij}) in the main text, we have the system of linear equations:
\begin{align}
	\begin{cases}
		\displaystyle{
			\tau_{HL}=1+\frac{n-1}{2n}\tau_{LL'}},\\[1em]
		\displaystyle{
			\tau_{LL'}=1+\frac{1}{2}\tau_{HL}+\frac{1}{2}\tau_{HL}}.
	\end{cases}
\end{align}
The solution is
\begin{align}\label{eq_tauij_star}
	\begin{cases}
		\displaystyle{
			\tau_{HL}=\frac{3n-1}{n+1}},\\[1em]
		\displaystyle{
			\tau_{LL'}=\frac{4n}{n+1}}.
	\end{cases}
\end{align}
Unlike $\tau_{ij}$, the values of $\Upsilon_{ij}$ are asymmetric with respect to $i$ and $j$. Therefore, we need to calculate three $\Upsilon_{ij}$ types: $\Upsilon_{HL}$, $\Upsilon_{LH}$, and $\Upsilon_{LL'}$. Inserting the $\tau_{ij}$ values of Eq.~(\ref{eq_tauij_star}) into Eq.~(\ref{eq_Upsilonij}) in the main text, we obtain the required $\Upsilon_{ij}$ values:
\begin{subequations}
	\begin{align}
		\Upsilon_{HL}
		=&~\frac{1}{k_H+1}\left(
		\frac{\tau_{HL}+\sum_{l\in\mathcal{N}_H}(\tau_{Ll}-\tau_{Hl})}{k_L+1}+
		\sum_{l\in\mathcal{N}_H}\frac{(\tau_{Ll}-\tau_{Hl})+ \sum_{\ell\in\mathcal{N}_l}(\tau_{L\ell}-\tau_{H\ell})}{k_l+1}
		\right) \nonumber\\
		=&~\frac{1}{k_H+1} \Bigg\{ \frac{\tau_{HL}+[(n-1)\tau_{LL'}-n\tau_{HL}]}{k_H+1} \nonumber\\
		&+\left[\frac{(\tau_{LL}-\tau_{HL})+(\tau_{HL}-\tau_{HH})}{k_L+1}+(n-1)\frac{(\tau_{LL'}-\tau_{HL})+(\tau_{HL}-\tau_{HH})}{k_L+1}\right] \Bigg) \nonumber\\
		=&~\frac{(n-1)[(n+3)\tau_{LL'}-2\tau_{HL}]}{2(n+1)^2}\nonumber\\
		=&~\frac{2n^2-n-1}{(n+1)^2}, \\
		\Upsilon_{LH}
		=&~\frac{1}{k_L+1} \left(\frac{\tau_{HL}+(\tau_{HH}-\tau_{HL} )}{k_L+1}+\frac{(\tau_{HH}-\tau_{HL} )+[(\tau_{HL}-\tau_{LL} )+(n-1)(\tau_{HL}-\tau_{LL'})]}{k_H+1}\right) \nonumber\\
		=&~\frac{(n-1)(\tau_{HL}-\tau_{LL'})}{2(n+1)} \nonumber\\
		=&-\frac{n-1}{2(n+1)}, \\
		\Upsilon_{LL'}
		=&~\frac{1}{k_L+1}\left(
		\frac{\tau_{LL'}+\sum_{l\in\mathcal{N}_L}(\tau_{L'l}-\tau_{Ll})}{k_L+1}+
		\sum_{l\in\mathcal{N}_L}\frac{(\tau_{L'l}-\tau_{Ll})+ \sum_{\ell\in\mathcal{N}_l}(\tau_{L'\ell}-\tau_{L\ell})}{k_l+1}
		\right) \nonumber\\
		=&~\frac{1}{k_L+1} \Bigg( \frac{\tau_{LL'}+(\tau_{L'H}-\tau_{LH})}{k_L+1} \nonumber\\
		&+\frac{(\tau_{L'H}-\tau_{LH})+[(\tau_{L'L}-\tau_{LL})+(\tau_{L'L'}-\tau_{LL'})+(n-2)(\tau_{L'L''}-\tau_{LL'})]}{k_H+1} \Bigg) \nonumber\\
		=&~\frac{n}{n+1}.
	\end{align}
\end{subequations}
Some informal variations of the symbols during the calculation are for intuitive understanding. For example, $L'$ refers to ``another leaf'' and thus $\tau_{L'L'}=\tau_{LL}$, $\tau_{L'L''}=\tau_{LL'}$.

Then, we can apply these $\tau_{ij}$ and $\Upsilon_{ij}$ values to calculate the critical synergy factor on the star graph. For the PC rule, the numerator is
\begin{align}
	\tau^{(1)} 
	&=\sum_{i,j\in\mathcal{N}} k_i p_{ij}\tau_{ij} \nonumber\\
	&=k_H (p_{HH}\tau_{HH}+n p_{HL}\tau_{HL})+n k_L (p_{LH}\tau_{LH}+p_{LL}\tau_{LL}+(n-1) p_{LL'}\tau_{LL'}) \nonumber\\
	&=2n\tau_{HL} \nonumber\\
	&=\frac{2n(3n-1)}{n+1},
\end{align}
where the $p_{ij}$ values are obtained by the network directly. For example, $p_{HH}=p_{LL}=0$ (no self-loops), $p_{LL'}=0$ (no edges between leaves), $p_{HL}=1/k_H=1/n$, $p_{LH}=1/k_L=1$. Similarly, the denominator is
\begin{align}
	\Upsilon^{(1)} 
	&=\sum_{i,j\in\mathcal{N}} k_i p_{ij}\Upsilon_{ij} \nonumber\\
	&=k_H (p_{HH}\Upsilon_{HH}+n p_{HL}\Upsilon_{HL})+n k_L (p_{LH}\Upsilon_{LH}+p_{LL}\Upsilon_{LL}+(n-1) p_{LL'}\Upsilon_{LL'}) \nonumber\\
	&=n(\Upsilon_{HL}+\Upsilon_{LH}) \nonumber\\
	&=\frac{n(3n^2-2n-1)}{2(n+1)^2}.
\end{align}
Therefore, the critical synergy factor on star graphs under the PC rule is
\begin{equation}
	r^\star=\frac{\tau^{(1)}}{\Upsilon^{(1)}}=\frac{4(3n-1)(n+1)}{3n^2-2n-1}
	\xrightarrow{n\to\infty}4.
\end{equation}

For the DB rule, the numerator is 
\begin{align}
	\tau^{(2)} 
	&=\sum_{i,j\in\mathcal{N}} k_i p_{ij}^{(2)}\tau_{ij} \nonumber\\
	&=k_H (p_{HH}^{(2)}\tau_{HH}+n p_{HL}^{(2)}\tau_{HL})+n k_L (p_{LH}^{(2)}\tau_{LH}+p_{LL}^{(2)}\tau_{LL}+(n-1) p_{LL'}^{(2)}\tau_{LL'}) \nonumber\\
	&=(n-1) \tau_{LL'} \nonumber\\
	&=\frac{4n(n-1)}{n+1},
\end{align}
where the $p^{(2)}_{ij}$ values are also directly obtained by the network: $p^{(2)}_{HH}=1$ (the first step must walk to one of the leaves and second step must walk to the hub), $p^{(2)}_{LL}=1/n$, $p^{(2)}_{LL'}=1/n$, $p^{(2)}_{HL}=0$, $p^{(2)}_{LH}=0$. Similarly, the denominator is
\begin{align}
	\Upsilon^{(2)} 
	&=\sum_{i,j\in\mathcal{N}} k_i p_{ij}^{(2)}\Upsilon_{ij} \nonumber\\
	&=k_H (p_{HH}^{(2)}\Upsilon_{HH}+n p_{HL}^{(2)}\Upsilon_{HL})+n k_L (p_{LH}^{(2)}\Upsilon_{LH}+p_{LL}^{(2)}\Upsilon_{LL}+(n-1) p_{LL'}^{(2)}\Upsilon_{LL'}) \nonumber\\
	&=(n-1) \Upsilon_{LL'} \nonumber\\
	&=\frac{n(n-1)}{n+1}.
\end{align}
Therefore, the critical synergy factor on star graphs under the DB rule is 
\begin{equation}
	r^\star=\frac{\tau^{(2)}}{\Upsilon^{(2)}}\equiv 4.
\end{equation}

For the BD rule, we list the system of linear equations according to Eq.~(\ref{eq_tauij_BD_2}):
\begin{align}
	\begin{cases}
		\displaystyle{
			\tilde{\tau}_{HL}=\frac{n}{n^2+1}+\frac{n(n-1)}{n^2+1}\tilde{\tau}_{LL'}},\\[1em]
		\displaystyle{
			\tilde{\tau}_{LL'}=\frac{n}{2}+\tilde{\tau}_{HL}}.
	\end{cases}
\end{align}
The solution is 
\begin{align}\label{eq_tauij_star_BD}
	\begin{cases}
		\displaystyle{
			\tilde{\tau}_{HL}=\frac{n(n^2-n+2)}{2(n+1)}},\\[1em]
		\displaystyle{
			\tilde{\tau}_{LL'}=\frac{n(n^2+3)}{2(n+1)}}.
	\end{cases}
\end{align}
Inserting these $\tilde{\tau}_{ij}$ values into Eq.~(\ref{eq_Upsilonij_BD}), we obtain the required $\tilde{\Upsilon}_{ij}$ values:
\begin{subequations}
	\begin{align}
		\tilde{\Upsilon}_{HL}
		=&~\frac{(n-1)[(n+3)\tilde{\tau}_{LL'}-2\tilde{\tau}_{HL}]}{2(n+1)^2}
		=\frac{n(n^3-n^2+5n-5)}{4(n+1)^2}, \\
		\tilde{\Upsilon}_{LH}
		=&~\frac{(n-1)(\tilde{\tau}_{HL}-\tilde{\tau}_{LL'})}{2(n+1)} 
		=\frac{n(n-1)}{4(n+1)}.
	\end{align}
\end{subequations}
Then, we apply these $\tilde{\tau}_{ij}$ and $\tilde{\Upsilon}_{ij}$ values to calculate the critical synergy factor on the star graph under the BD rule. The numerator is
\begin{equation}
	\tilde{\tau}^{(1)}=\sum_{i,j\in\mathcal{N}}\frac{k_{ij}}{k_i k_j} \tilde{\tau}_{ij}=2\tilde{\tau}_{HL}=\frac{n(n^2-n+2)}{n+1},
\end{equation}
and the denominator is
\begin{equation}
	\tilde{\Upsilon}^{(1)}=\sum_{i,j\in\mathcal{N}}\frac{k_{ij}}{k_i k_j} \tilde{\Upsilon}_{ij}=\tilde{\Upsilon}_{HL}+\tilde{\Upsilon}_{LH}=\frac{n(n^3-2n^2+5n-4)}{4(n+1)^2}.
\end{equation}
Therefore, the critical synergy factor on star graphs under the BD rule is 
\begin{equation}
	r^\star=\frac{\tilde{\tau}^{(1)}}{\tilde{\Upsilon}^{(1)}}=\frac{4n^3+4n+8}{n^3-2n^2+5n-4}
	\xrightarrow{n\to\infty}4.
\end{equation}

\begin{itemize}
	\item Accumulated payoff
\end{itemize}

When using accumulated payoffs, the values of $\tau_{ij}$, $\Upsilon_{ij}$, $\tilde{\tau}_{ij}$, and $\tilde{\Upsilon}_{ij}$ keep unchanged, but the formulas of the critical synergy factors are different.

For the PC rule, we follow Eq.~(\ref{eq_coopcondi_3_PC_accu}). The numerator is
\begin{equation}
	\sum_{i,j\in\mathcal{N}} k_i (k_{i}+1)p_{ij}\tau_{ij}
	=(n+3)\tau_{HL}
	=\frac{(3n-1)(n+3)}{n+1},
\end{equation}
and the denominator is 
\begin{equation}
	\sum_{i,j\in\mathcal{N}} k_i (k_i+1)p_{ij}\Upsilon_{ij}
	=(n+1)\Upsilon_{HL}+2\Upsilon_{LH}
	=\frac{2n(n-1)}{n+1}.
\end{equation}
The critical synergy factor on star graphs under the PC rule when using accumulated payoff is 
\begin{equation}
	r^\star_\text{accu}=\frac{(3n-1)(n+3)}{2n(n-1)}
	\xrightarrow{n\to\infty}\frac{3}{2}.
\end{equation}

For the DB rule, we follow Eq.~(\ref{eq_coopcondi_3_DB_accu}). The numerator is
\begin{equation}
	\sum_{i,j\in\mathcal{N}} k_i (k_i+1)p_{ij}^{(2)}\tau_{ij}=2(n-1)\tau_{LL'}
	=\frac{8n(n-1)}{n+1},
\end{equation}
and the denominator is 
\begin{equation}
	\sum_{i,j\in\mathcal{N}} k_i (k_i+1)p_{ij}^{(2)}\Upsilon_{ij}=2(n-1)\Upsilon_{LL'}
	=\frac{2n(n-1)}{n+1}.
\end{equation}
The critical synergy factor on star graphs under the DB rule when using accumulated payoff is 
\begin{equation}
	r^\star_\text{accu}\equiv 4.
\end{equation}

For the BD rule, we follow Eq.~(\ref{eq_coopcondi_3_BD_accu}). The numerator is
\begin{equation}
	\sum_{i,j\in\mathcal{N}}\frac{k_{ij}}{k_i k_j}(k_i+1) \tilde{\tau}_{ij}=(n+3)\tilde{\tau}_{HL}=\frac{n(n+3)(n^2-n+2)}{2(n+1)},
\end{equation}
and the denominator is 
\begin{equation}
	\sum_{i,j\in\mathcal{N}}\frac{k_{ij}}{k_i k_j}(k_i+1) \tilde{\Upsilon}_{ij}=(n+1)\tilde{\Upsilon}_{HL}+2\tilde{\Upsilon}_{LH}=\frac{n(n^3-n^2+3n-3)}{4(n+1)}.
\end{equation}
The critical synergy factor on star graphs under the BD rule when using accumulated payoff is 
\begin{equation}
	r^\star_\text{accu}=\frac{2(n+3)(n^2-n+2)}{n^3-n^2+3n-3}
	\xrightarrow{n\to\infty}2.
\end{equation}

\subsection{Hub-to-hub star}
On a hub-to-hub star graph, there are two hubs ($H$), each has $n$ leaves ($L$). A hub node has $k_H=n+1$ neighbors ($n$ leaves and the other hub), and each leaf node has $k_L=1$ neighbor. There are five non-zero $\tau_{ij}$ types: $\tau_{HH'}$, the relation between the two hubs; $\tau_{LL'}$, the relation between two leaves of the same hub; $\tau_{LL''}$, the relation between a leaf of one hub and another leaf of the other hub; $\tau_{HL}$, the relation between a hub and one of its leaves; $\tau_{HL'}$, the relation between a hub and a leaf of the other hub. According to Eq.~(\ref{eq_tauij}) in the main text, we have the system of linear equations:
\begin{align}
	\begin{cases}
		\displaystyle{
			\tau_{HH'}=1+\frac{n}{n+1}\tau_{HL'}},\\[1em]
		\displaystyle{
			\tau_{HL}=1+\frac{1}{2(n+1)}\tau_{HL'}+\frac{n-1}{2(n+1)}\tau_{LL'}},\\[1em]
		\displaystyle{
			\tau_{HL'}=1+\frac{1}{2(n+1)}\tau_{HL}+\frac{n}{2(n+1)}\tau_{LL''}+\frac{1}{2}\tau_{HH'}},\\[1em]
		\displaystyle{
			\tau_{LL'}=1+\tau_{HL}},\\[1em]
		\displaystyle{
			\tau_{LL''}=1+\tau_{HL'}}.
	\end{cases}
\end{align}
The solution is 
\begin{align}
	\begin{cases}
		\displaystyle{
			\tau_{HH'}=\frac{4n^3+20n^2+17n+5}{(2n+5)(n+1)}},\\[1em]
		\displaystyle{
			\tau_{HL}=\frac{5(2n+1)}{2n+5}},\\[1em]
		\displaystyle{
			\tau_{HL'}=\frac{2(2n^2+9n+5)}{2n+5}},\\[1em]
		\displaystyle{
			\tau_{LL'}=\frac{2(6n+5)}{2n+5}},\\[1em]
		\displaystyle{
			\tau_{LL''}=\frac{4n^2+20n+15}{2n+5}}.
	\end{cases}
\end{align}
Inserting these $\tau_{ij}$ values into Eq.~(\ref{eq_Upsilonij}) in the main text, we obtain the required $\Upsilon_{ij}$ values:
\begin{subequations}
	\begin{align}
		\Upsilon_{HH'}
		&=\frac{n}{2(n+2)}(\tau_{HH'}-\tau_{HL}+\tau_{HL'} ) \nonumber\\
		&=\frac{n(4n^3+16n^2+15n+5)}{2n^3+11n^2+19n+10}, \\
		\Upsilon_{HL}
		&=-\frac{2}{(n+2)^2}\tau_{HH'}-\frac{n-2}{(n+2)^2}\tau_{HL}-\frac{n-2}{(n+2)^2}\tau_{HL'}+\frac{n^2+3n-4}{2(n+2)^2}\tau_{LL'}+\frac{2}{(n+2)^2}\tau_{LL''} \nonumber\\
		&=\frac{n(6n^3+21n^2+30n+23)}{(n+2)^2 (2n^2+7n+5)}, \\
		\Upsilon_{LH}
		&=\frac{1}{2(n+2)}\tau_{HH'}+\frac{n-1}{2(n+2)}\tau_{HL}-\frac{1}{2(n+2)}\tau_{HL'}-\frac{n-1}{2(n+2)}\tau_{LL'} \nonumber\\
		&=-\frac{n(2n^2+7n+9)}{4n^3+22n^2+38n+20}, \\
		\Upsilon_{HL'}
		&=-\frac{2}{(n+2)^2}\tau_{HH'}-\frac{n^2+4n-4}{2(n+2)^2}\tau_{HL}+\frac{n^2+4}{2(n+2)^2}\tau_{HL'}+\frac{n-1}{(n+2)^2}\tau_{LL'}+\frac{n^2+4n}{2(n+2)^2}\tau_{LL''} \nonumber\\
		&=\frac{4n^5+26n^4+64n^3+84n^2+60n+10}{(n+2)^2 (2n^2+7n+5)}, \\
		\Upsilon_{LH'}
		&=\frac{n+4}{4(n+2)}\tau_{HH'}-\frac{n+4}{4(n+2)}\tau_{HL}+\frac{3n}{4(n+2)}\tau_{HL'}-\frac{n-1}{2(n+2)}\tau_{LL'} \nonumber\\
		&=\frac{8n^4+34n^3+53n^2+31n+10}{2(2n^3+11n^2+19n+10)}, \\
		\Upsilon_{LL'}
		&=\frac{\tau_{LL'}}{4} \nonumber\\
		&=\frac{6n+5}{2(2n+5)}.
	\end{align}
\end{subequations}

Then, we apply these $\tau_{ij}$ and $\Upsilon_{ij}$ values to calculate the critical synergy factor on the hub-to-hub star graph. For the PC rule, the numerator is
\begin{align}
	\tau^{(1)} 
	&=2\tau_{HH'}+4n\tau_{HL} \nonumber\\
	&=\frac{2(24n^3+50n^2+27n+5)}{2n^2+7n+5},
\end{align}
and the denominator is 
\begin{align}
	\Upsilon^{(1)} 
	&=2\Upsilon_{HH'}+2n\Upsilon_{HL}+2n\Upsilon_{LH} \nonumber\\
	&=\frac{n(18n^4+79n^3+131n^2+98n+20)}{(n+2)^2 (2n^2+7n+5)}.
\end{align}
Therefore, the critical synergy factor on hub-to-hub star graphs under the PC rule is
\begin{equation}
	r^\star=\frac{\tau^{(1)}}{\Upsilon^{(1)}}=\frac{2(n+2)^2 (24n^3+50n^2+27n+5)}{n(18n^4+79n^3+131n^2+98n+20)}
	\xrightarrow{n\to\infty}\frac{8}{3}.
\end{equation}

For the DB rule, the numerator is 
\begin{align}
	\tau^{(2)} 
	&=\sum_{i,j\in\mathcal{N}} k_i p_{ij}^{(2)}\tau_{ij} \nonumber\\
	&=\frac{2n}{n+1}\tau_{HL'}+\frac{2n(n-1)}{n+1}\tau_{LL'}+\frac{2n}{n+1}\tau_{HL'} \nonumber\\
	&=\frac{4n(10n^2+17n+5)}{2n^2+7n+5},
\end{align}
and the denominator is 
\begin{align}
	\Upsilon^{(2)} 
	&=\sum_{i,j\in\mathcal{N}} k_i p_{ij}^{(2)}\Upsilon_{ij} \nonumber\\
	&=\frac{2n}{n+1}\Upsilon_{HL'}+\frac{2n(n-1)}{n+1}\Upsilon_{LL'}+\frac{2n}{n+1}\Upsilon_{L'H} \nonumber\\
	&=\frac{n(22n^5+131n^4+287n^3+296n^2+148n+20)}{(2n+5)(n^2+3n+2)^2}.
\end{align}
Therefore, the critical synergy factor on hub-to-hub star graphs under the DB rule is 
\begin{equation}
	r^\star=
	\frac{4(n+2)^2 (10n^3+27n^2+22n+5)}{22n^5+131n^4+287n^3+296n^2+148n+20}
	\xrightarrow{n\to\infty}\frac{20}{11}.
\end{equation}

For the BD rule, we list the system of linear equations according to Eq.~(\ref{eq_tauij_BD_2}):
\begin{align}
	\begin{cases}
		\displaystyle{
			\tilde{\tau}_{HH'}=\frac{n+1}{2(n^2+n+1)}\left(1+2n\tilde{\tau}_{HL'} \right)},\\[1em]
		\displaystyle{
			\tilde{\tau}_{HL}=\frac{n+1}{n^2+n+2}\left(1+\frac{1}{n+1}\tilde{\tau}_{HL'}+(n-1)\tilde{\tau}_{LL'}\right)},\\[1em]
		\displaystyle{
			\tilde{\tau}_{HL'}=\frac{n+1}{n^2+n+2}\left(1+\frac{1}{n+1}\tilde{\tau}_{HH'}+\frac{1}{n+1}\tilde{\tau}_{HL}+n\tilde{\tau}_{LL''}\right)},\\[1em]
		\displaystyle{
			\tilde{\tau}_{LL'}=\frac{n+1}{2}+\tilde{\tau}_{HL}},\\[1em]
		\displaystyle{
			\tilde{\tau}_{LL''}=\frac{n+1}{2}+\tilde{\tau}_{HL'}}.
	\end{cases}
\end{align}
The solution is 
\begin{align}
	\begin{cases}
		\displaystyle{
			\tilde{\tau}_{HH'}=\frac{(n+1)(n^5+6n^4+10n^3+12n^2+9n+5)}{2(n^3+3n^2+4n+5)}},\\[1em]
		\displaystyle{
			\tilde{\tau}_{HL}=\frac{2n^5+5n^4+10n^3+11n^2+9n+5}{2(n^3+3n^2+4n+5)}},\\[1em]
		\displaystyle{
			\tilde{\tau}_{HL'}=\frac{n^6+7n^5+17n^4+28n^3+30n^2+23n+10}{2(n^3+3n^2+4n+5)}},\\[1em]
		\displaystyle{
			\tilde{\tau}_{LL'}=\frac{n^5+3n^4+7n^3+9n^2+9n+5}{n^3+3n^2+4n+5}},\\[1em]
		\displaystyle{
			\tilde{\tau}_{LL''}=\frac{n^6+7n^5+18n^4+32n^3+37n^2+32n+15}{2(n^3+3n^2+4n+5)}}.
	\end{cases}
\end{align}
Inserting these $\tilde{\tau}_{ij}$ values into Eq.~(\ref{eq_Upsilonij_BD}), we obtain the required $\tilde{\Upsilon}_{ij}$ values:
\begin{subequations}
	\begin{align}
		\tilde{\Upsilon}_{HH'}
		&=\frac{n}{2(n+2)}(\tilde{\tau}_{HH'}-\tilde{\tau}_{HL}+\tilde{\tau}_{HL'} ) \nonumber\\
		&=\frac{n(n^6+6n^5+14n^4+20n^3+20n^2+14n+5)}{2(n^4+5n^3+10n^2+13n+10)}, \\
		\tilde{\Upsilon}_{HL}
		&=-\frac{2}{(n+2)^2}\tilde{\tau}_{HH'}-\frac{n-2}{(n+2)^2}\tilde{\tau}_{HL}-\frac{n-2}{(n+2)^2}\tilde{\tau}_{HL'}+\frac{n^2+3n-4}{2(n+2)^2}\tilde{\tau}_{LL'}+\frac{2}{(n+2)^2}\tilde{\tau}_{LL''} \nonumber\\
		&=\frac{n(n^6+4n^5+12n^4+24n^3+36n^2+36n+15)}{2(n+2)^2 (n^3+3n^2+4n+5)}, \\
		\tilde{\Upsilon}_{LH}
		&=\frac{1}{2(n+2)}\tilde{\tau}_{HH'}+\frac{n-1}{2(n+2)}\tilde{\tau}_{HL}-\frac{1}{2(n+2)}\tilde{\tau}_{HL'}-\frac{n-1}{2(n+2)}\tilde{\tau}_{LL'} \nonumber\\
		&=-\frac{n(n^4+4n^3+9n^2+11n+5)}{4(n^4+5n^3+10n^2+13n+10)}.
	\end{align}
\end{subequations}
Then, we apply these $\tilde{\tau}_{ij}$ and $\tilde{\Upsilon}_{ij}$ values to calculate the critical synergy factor on the hub-to-hub star graph under the BD rule. The numerator is
\begin{equation}
	\tilde{\tau}^{(1)}=\frac{2}{(n+1)^2}\tilde{\tau}_{HH'}+\frac{4n}{n+1}\tilde{\tau}_{HL}=\frac{4n^6+11n^5+26n^4+32n^3+30n^2+19n+5}{n^4+4n^3+7n^2+9n+5},
\end{equation}
and the denominator is 
\begin{equation}
	\tilde{\Upsilon}^{(1)}=\frac{2}{(n+1)^2}\tilde{\Upsilon}_{HH'}+\frac{2n}{n+1}\tilde{\Upsilon}_{HL}+\frac{2n}{n+1}\tilde{\Upsilon}_{LH}=\frac{n(2n^7+9n^6+32n^5+69n^4+101n^3+107n^2+66n+20)}{2(n+2)^2 (n^4+4n^3+7n^2+9n+5)}.
\end{equation}
Therefore, the critical synergy factor on hub-to-hub star graphs under the BD rule is  
\begin{equation}
	r^\star=\frac{\tilde{\tau}^{(1)}}{\tilde{\Upsilon}^{(1)}}=\frac{2(n+2)^2 (4n^6+11n^5+26n^4+32n^3+30n^2+19n+5)}{n(2n^7+9n^6+32n^5+69n^4+101n^3+107n^2+66n+20)}
	\xrightarrow{n\to\infty}4.
\end{equation}

\begin{itemize}
\item Accumulated payoff
\end{itemize}

When using accumulated payoffs, we follow Eq.~(\ref{eq_coopcondi_3_PC_accu}) for the PC rule. The numerator is
\begin{align}
	\sum_{i,j\in\mathcal{N}} k_i (k_{i}+1)p_{ij}\tau_{ij}
	&=2(n+2)\tau_{HH'}+2n(n+4)\tau_{HL} \nonumber\\
	&=\frac{2(14n^4+83n^3+122n^2+59n+10)}{2n^2+7n+5},
\end{align}
and the denominator is 
\begin{align}
	\sum_{i,j\in\mathcal{N}} k_i (k_i+1)p_{ij}\Upsilon_{ij}
	&=2(n+2)\Upsilon_{HH'}+2n(n+2)\Upsilon_{HL}+4n\Upsilon_{LH} \nonumber\\
	&=\frac{2n(10n^4+43n^3+70n^2+49n+10)}{2n^3+11n^2+19n+10}.
\end{align}
The critical synergy factor on hub-to-hub star graphs under the PC rule when using accumulated payoff is 
\begin{equation}
	r^\star_\text{accu}=\frac{14n^5+111n^4+288n^3+303n^2+128n+20}{n(10n^4+43n^3+70n^2+49n+10)}
	\xrightarrow{n\to\infty}\frac{7}{5}.
\end{equation}

For the DB rule, we follow Eq.~(\ref{eq_coopcondi_3_DB_accu}). The numerator is 
\begin{align}
	\sum_{i,j\in\mathcal{N}} k_i (k_i+1)p_{ij}^{(2)}\tau_{ij}
	&=\frac{2n(n+4)}{n+1}\tau_{HL'}+\frac{4n(n-1)}{n+1}\tau_{LL'} \nonumber\\
	&=\frac{4n(2n^3+29n^2+39n+10)}{2n^2+7n+5},
\end{align}
and the denominator is 
\begin{align}
	\sum_{i,j\in\mathcal{N}} k_i (k_i+1)p_{ij}^{(2)}\Upsilon_{ij}
	&=\frac{2n(n+2)}{n+1}\Upsilon_{HL'}+\frac{4n}{n+1}\Upsilon_{LH'}+\frac{4n(n-1)}{n+1}\Upsilon_{LL'} \nonumber\\
	&=\frac{2n(4n^5+40n^4+115n^3+141n^2+74n+10)}{(n+1)^2 (2n^2+9n+10)}.
\end{align}
The critical synergy factor on hub-to-hub star graphs under the DB rule when using accumulated payoff is 
\begin{equation}\label{eq_r_superhubtohub}
	r^\star_\text{accu}=\frac{4n^5+70n^4+260n^3+370n^2+216n+40}{4n^5+40n^4+115n^3+141n^2+74n+10}
	\xrightarrow{n\to\infty}1.
\end{equation}
Eq.~(\ref{eq_r_superhubtohub}) is the result of the super structure for cooperation presented in the main text.

For the BD rule, we follow Eq.~(\ref{eq_coopcondi_3_BD_accu}). The numerator is 
\begin{align}
	\sum_{i,j\in\mathcal{N}}\frac{k_{ij}}{k_i k_j}(k_i+1) \tilde{\tau}_{ij}
	&=\frac{2(n+2)}{(n+1)^2}\tilde{\tau}_{HH'}+\frac{2n(n+4)}{n+1}\tilde{\tau}_{HL} \nonumber\\
	&=\frac{2n^7+14n^6+38n^5+73n^4+85n^3+74n^2+43n+10}{n^4+4n^3+7n^2+9n+5},
\end{align}
and the denominator is 
\begin{align}
	\sum_{i,j\in\mathcal{N}}\frac{k_{ij}}{k_i k_j}(k_i+1) \tilde{\Upsilon}_{ij}
	&=\frac{2(n+2)}{(n+1)^2}\tilde{\Upsilon}_{HH'}+\frac{2n(n+2)}{n+1}\tilde{\Upsilon}_{HL}+\frac{4n}{n+1}\tilde{\Upsilon}_{LH} \nonumber\\
	&=\frac{n(n^7+5n^6+18n^5+39n^4+56n^3+56n^2+33n+10)}{n^5+6n^4+15n^3+23n^2+23n+10}.
\end{align}
The critical synergy factor on hub-to-hub star graphs under the BD rule when using accumulated payoff is 
\begin{equation}
	r^\star_\text{accu}=\frac{2n^8+18n^7+66n^6+149n^5+231n^4+244n^3+191n^2+96n+20}{n(n^7+5n^6+18n^5+39n^4+56n^3+56n^2+33n+10)}
	\xrightarrow{n\to\infty}2.
\end{equation}

\subsection{$m$-hub star}
On an $m$-hub star graph, there are $m$ hubs ($H$), each has $n$ leaves ($L$). A hub node has $k_H=n+m-1$ neighbors ($n$ leaves and the remaining $m-1$ hubs), and each leaf node has $k_L=1$ neighbor. There are five non-zero $\tau_{ij}$ types: $\tau_{HH'}$, the relation between one hub and another hub; $\tau_{LL'}$, the relation between two leaves of the same hub; $\tau_{LL''}$, the relation between two leaves of different hubs; $\tau_{HL}$, the relation between a hub and one of its leaves; $\tau_{HL'}$, the relation between a hub and a leaf of another hub. The $m$-hub star reduces to a hub-to-hub star when $m=2$. According to Eq.~(\ref{eq_tauij}) in the main text, we have the system of linear equations:
\begin{align}
	\begin{cases}
		\displaystyle{
			\tau_{HH'}=1+\frac{m-2}{n+m-1}\tau_{HH'}+\frac{n}{n+m-1}\tau_{HL'}},\\[1em]
		\displaystyle{
			\tau_{HL}=1+\frac{m-1}{2(n+m-1)}\tau_{HL'}+\frac{n-1}{2(n+m-1)}\tau_{LL'}},\\[1em]
		\displaystyle{
			\tau_{HL'}=1+\frac{1}{2(n+m-1)}\tau_{HL}+\frac{m-2}{2(n+m-1)}\tau_{HL'}+\frac{n}{2(n+m-1)}\tau_{LL''}+\frac{1}{2}\tau_{HH'}},\\[1em]
		\displaystyle{
			\tau_{LL'}=1+\tau_{HL}},\\[1em]
		\displaystyle{
			\tau_{LL''}=1+\tau_{HL'}}.
	\end{cases}
\end{align}
The solution is 
\begin{align}
	\begin{cases}
		\displaystyle{
			\tau_{HH'}=\frac{2m^3+9m^2 n-4m^2+12mn^2-10mn+3m+4n^3-4n^2+n-1}{2m^2+4mn-2m+2n^2-n+1}},\\[1em]
		\displaystyle{
			\tau_{HL}=\frac{m^3+4m^2 n+m^2+4mn^2+2mn-4m+2n^2-5n+1}{2m^2+4mn-2m+2n^2-n+1}},\\[1em]
		\displaystyle{
			\tau_{HL'}=\frac{2m^3+9m^2 n-m^2+12mn^2-4mn+4n^3-2n^2-2}{2m^2+4mn-2m+2n^2-n+1}},\\[1em]
		\displaystyle{
			\tau_{LL'}=\frac{m^3+4m^2 n+3m^2+4mn^2+6mn-6m+4n^2-6n+2}{2m^2+4mn-2m+2n^2-n+1}},\\[1em]
		\displaystyle{
			\tau_{LL''}=\frac{2m^3+9m^2 n+m^2+12mn^2-2m+4n^3-n-1}{2m^2+4mn-2m+2n^2-n+1}}.
	\end{cases}
\end{align}
Inserting these $\tau_{ij}$ values into Eq.~(\ref{eq_Upsilonij}) in the main text, we obtain the required $\Upsilon_{ij}$ values: 
\begin{subequations}
	\begin{align}
		\Upsilon_{HH'}
		=&~\frac{n}{2(n+m)}(\tau_{HL'}-\tau_{HL}+\tau_{HH'}) \nonumber\\
		=&~\frac{n(3m^3+14m^2 n-6m^2+20mn^2-16mn+7m+8n^3-8n^2+6n-4}{2(2m^3+6m^2 n-2m^2+6mn^2-3mn+m+2n^3-n^2+n)},
		\\
		\Upsilon_{HL}
		=&-\frac{m(m-1)}{(m+n)^2}\tau_{HH'}+\frac{m-n}{(m+n)^2}\tau_{HL}+\frac{(m-n)(m-1)}{(m+n)^2}\tau_{HL'} \nonumber\\
		&+\frac{(n-1)(m+n+2)}{2(m+n)^2}\tau_{LL'}+\frac{n(m-1)}{(m+n)^2}\tau_{LL''} \nonumber\\
		=&~\Big\{m^4 n+7m^4+5m^3 n^2+22m^3 n-15m^3+8m^2 n^3+21m^2 n^2-37m^2 n-4m^2 \nonumber\\
		&+4mn^4+10mn^3-32mn^2-4mn+14m+4n^4-10n^3+10n-4\Big\}/
		\Big\{2(m+n)^2 \nonumber\\
		&\times (2m^2+4mn-2m+2n^2-n+1)\Big\},
		\\
		\Upsilon_{LH}
		=&~\frac{m-1}{2(m+n)}\tau_{HH'}+\frac{n-1}{2(m+n)}\tau_{HL}-\frac{m-1}{2(m+n)}\tau_{HL'}-\frac{n-1}{2(m+n)}\tau_{LL'} \nonumber\\
		=&~\frac{-3m^3-8m^2 n+8m^2-6mn^2+13mn-4m-2n^3+5n^2-3n}{2(2m^3+6m^2 n-2m^2+6mn^2-3mn+m+2n^3-n^2+n},
		\\
		\Upsilon_{HL'}
		=&-\frac{m(m-1)}{(m+n)^2}\tau_{HH'}+\frac{2m-2n-mn-n^2}{2(m+n)^2}\tau_{HL}+\frac{2m^2-mn-2m+n^2+2n}{2(m+n)^2}\tau_{HL'} \nonumber\\
		&+\frac{n-1}{(m+n)^2}\tau_{LL'}+\frac{3mn-2n+n^2}{2(m+n)^2}\tau_{LL''} \nonumber\\
		=&~\Big\{3m^4 n+8m^4+17m^3 n^2 +23m^3 n-12m^3+34m^2 n^3+13m^2 n^2-20m^2 n-10m^2 \nonumber\\
		&+28mn^4-6mn^3-4mn^2-22mn+16m+8n^5-4n^4+4n^3-12n^2+12n-4\Big\} \nonumber\\
		&~/\Big\{2(m+n)^2 (2m^2+4mn-2m+2n^2-n+1)\Big\},
		\\
		\Upsilon_{LH'}
		=&~\frac{3m+n-2}{4(m+n)}\tau_{HH'}-\frac{m+n+2}{4(m+n)}\tau_{HL}+\frac{3n-m+2}{4(m+n)}\tau_{HL'}-\frac{n-1}{2(m+n)}\tau_{LL'} \nonumber\\
		=&~\Big\{3m^4+19m^3 n-12m^3+44m^2 n^2-42m^2 n+23m^2+44mn^3-48mn^2+47mn \nonumber\\
		&-12m+16n^4-20n^3+26n^2-16n \Big\}/
		\Big\{4(2m^3+6m^2 n-2m^2+6mn^2-3mn \nonumber\\
		&+m+2n^3-n^2+n)\Big\},
		\\
		\Upsilon_{LL'}
		=&~\frac{\tau_{LL'}}{4} \nonumber\\
		=&~\frac{m^3+4m^2 n+3m^2+4mn^2+6mn-6m+4n^2-6n+2}{4(2m^2+4mn-2m+2n^2-n+1)}.
	\end{align}
\end{subequations}

Then, we apply these $\tau_{ij}$ and $\Upsilon_{ij}$ values to calculate the critical synergy factor on the $m$-hub star graph. For the PC rule, the numerator is 
\begin{align}
	\tau^{(1)} 
	=&~m(m-1)\tau_{HH'}+2mn\tau_{HL} \nonumber\\
	=&~\Big\{m(2m^4+11m^3 n-6m^3+20m^2 n^2-17m^2 n+7m^2+12mn^3-12mn^2+3mn \nonumber\\
	&-4m-6n^2+n+1)\Big\}/\Big\{2m^2+4mn-2m+2n^2-n+1\Big\},
\end{align}
and the denominator is 
\begin{align}
	\Upsilon^{(1)}
	=&~m(m-1)\Upsilon_{HH'}+mn\Upsilon_{HL}+mn\Upsilon_{LH} \nonumber\\
	=&~\Big\{mn(3m^5+18m^4 n-5m^4+39m^3 n^2-28m^3 n+6m^3+36m^2 n^3-51m^2 n^2 \nonumber\\
	&+19m^2 n-19m^2+12mn^4-34mn^3+16mn^2-28mn+18m-6n^4+3n^3-9n^2 \nonumber\\
	&+14n-4)\Big\}/\Big\{2(m+n)^2 (2m^2+4mn-2m+2n^2-n+1)\Big\}.
\end{align}
Therefore, the critical synergy factor on $m$-hub star graphs under the PC rule is 
\begin{equation}
	r^\star=
	\frac{\text{nume}}{\text{deno}}
	\xrightarrow{n\to \infty} \frac{4m}{2m-1}
	\xrightarrow{m=2} \frac{8}{3}
	\xrightarrow{m\to \infty} 2,
\end{equation}
where
\begin{align}
	\text{nume}=&~
	2(m+n)^2 (2m^4+11m^3 n-6m^3+20m^2 n^2-17m^2 n+7m^2+12mn^3 \nonumber\\
	&-12mn^2+3mn-4m-6n^2+n+1), \nonumber\\
	\text{deno}=&~
	n(3m^5+18m^4 n-5m^4+39m^3 n^2-28m^3 n+6m^3+36m^2 n^3-51m^2 n^2+19m^2 n \nonumber\\
	&-19m^2+12mn^4-34mn^3+16mn^2-28mn+18m-6n^4+3n^3-9n^2+14n-4).
\end{align}

For the DB rule, the numerator is 
\begin{align}
	\tau^{(2)} 
	=&~\frac{m(m-1)(m-2)}{n+m-1}\tau_{HH'}+\frac{2mn(m-1)}{n+m-1}\tau_{HL'}+\frac{mn(n-1)}{n+m-1}\tau_{LL'} \nonumber\\
	=&~\Big\{m(2m^4+11m^3 n-8m^3+20m^2 n^2-25m^2 n+11m^2+12mn^3-22mn^2+12mn \nonumber\\
	&-7m-4n^3-2n^2-2n+2)\Big\}/
	\Big\{2m^2+4mn-2m+2n^2-n+1\Big\},
\end{align}
and the denominator is 
\begin{align}
	\Upsilon^{(2)} 
	=&~\frac{m(m-1)(m-2)}{n+m-1}\Upsilon_{HH'}+\frac{mn(m-1)}{n+m-1}\Upsilon_{HL'}+\frac{mn(m-1)}{n+m-1}\Upsilon_{LH'}+\frac{mn(n-1)}{n+m-1}\Upsilon_{LL'} \nonumber\\
	=&~\Big\{mn(9m^6+63m^5 n-30m^5+171m^4 n^2-185m^4 n+54m^4+225m^3 n^3-414m^3 n^2 \nonumber\\
	&+246m^3 n-99m^3+144m^2 n^4-413m^2 n^3+384m^2 n^2-256m^2 n+114m^2+36mn^5 \nonumber\\
	&-182mn^4+242mn^3-215mn^2+168mn-56m-28n^5+50n^4-58n^3+62n^2 \nonumber\\
	&-40n+8)\Big\}/
	\Big\{4(m+n)^2 (2m^3+6m^2 n-4m^2+6mn^2-7mn+3m+2n^3-3n^2 \nonumber\\
	&+2n-1)\Big\}.
\end{align}
Therefore, the critical synergy factor on $m$-hub star graphs under the DB rule is 
\begin{equation}
	r^\star=
	\frac{\text{nume}}{\text{deno}}
	\xrightarrow{n\to \infty} \frac{12m-4}{9m-7}
	\xrightarrow{m=2} \frac{20}{11}
	\xrightarrow{m\to \infty} \frac{4}{3},
\end{equation}
where
\begin{align}
	\text{nume}=&~
	4(m+n)^2 (2m^3+6m^2 n-4m^2+6mn^2-7mn+3m+2n^3-3n^2+2n-1) \nonumber\\
	&\times (2m^4+11m^3 n-8m^3+20m^2 n^2-25m^2 n+11m^2+12mn^3-22mn^2+12mn \nonumber\\
	&-7m-4n^3-2n^2-2n+2), \nonumber\\
	\text{deno}=&~
	n(2m^2+4mn-2m+2n^2-n+1)(9m^6+63m^5 n-30m^5+171m^4 n^2-185m^4 n \nonumber\\
	&+54m^4+225m^3 n^3-414m^3 n^2+246m^3 n-99m^3+144m^2 n^4-413m^2 n^3+384m^2 n^2 \nonumber\\
	&-256m^2 n+114m^2+36mn^5-182mn^4+242mn^3-215mn^2+168mn-56m-28n^5 \nonumber\\
	&+50n^4-58n^3+62n^2-40n+8).
\end{align}

For the BD rule, we list the system of linear equations according to Eq.~(\ref{eq_tauij_BD_2}):
\begin{align}
	\begin{cases}
		\displaystyle{
			\tilde{\tau}_{HH'}=\frac{n+m-1}{2(n^2-n+mn+m-1)}\left(1+\frac{2(m-2)}{n+m-1}\tilde{\tau}_{HH'}+2n\tilde{\tau}_{HL'}\right)},\\[1em]
		\displaystyle{
			\tilde{\tau}_{HL}=\frac{n+m-1}{n^2-n+mn+m}\left(1+\frac{m-1}{n+m-1})\tilde{\tau}_{HL'}+(n-1)\tilde{\tau}_{LL'}\right)},\\[1em]
		\displaystyle{
			\tilde{\tau}_{HL'}=\frac{n+m-1}{n^2-n+mn+m}\left(1+\frac{1}{n+m-1}\tilde{\tau}_{HL}+\frac{m-2}{n+m-1}\tilde{\tau}_{HL'}+n\tilde{\tau}_{LL''}+\frac{1}{n+m-1}\tilde{\tau}_{HH'}\right)},\\[1em]
		\displaystyle{
			\tilde{\tau}_{LL'}=\frac{n+m-1}{2}+\tilde{\tau}_{HL}},\\[1em]
		\displaystyle{
			\tilde{\tau}_{LL''}=\frac{n+m-1}{2}+\tilde{\tau}_{HL}},
	\end{cases}
\end{align}
which can be solved, but the solution for $\tilde{\tau}_{ij}$ is too long and thus not presented here. Inserting these $\tilde{\tau}_{ij}$ values into Eq.~(\ref{eq_Upsilonij_BD}), we can obtain the required $\tilde{\Upsilon}_{ij}$ values. Finally, we can apply these $\tilde{\tau}_{ij}$ and $\tilde{\Upsilon}_{ij}$ values to calculate the critical synergy factor on the $m$-hub star graph under the BD rule. The numerator is
\begin{equation}
	\tilde{\tau}^{(1)}=\frac{m(m-1)}{(n+m-1)^2}\tilde{\tau}_{HH'}+\frac{2mn}{n+m-1}\tilde{\tau}_{HL},
\end{equation}
and the denominator is
\begin{equation}
	\tilde{\Upsilon}^{(1)}=\frac{m(m-1)}{(n+m-1)^2}\tilde{\Upsilon}_{HH'}+\frac{mn}{n+m-1}\tilde{\Upsilon}_{HL}+\frac{mn}{n+m-1}\tilde{\Upsilon}_{LH}.
\end{equation}
Therefore, the critical synergy factor on $m$-hub star graphs under the BD rule is 
\begin{equation}
	r^\star=\frac{\tilde{\tau}^{(1)}}{\tilde{\Upsilon}^{(1)}}=\frac{\text{nume}}{\text{deno}}
	\xrightarrow{n\to\infty}4,
\end{equation}
where
\begin{align}
	\text{nume}=&~
	2(m+n)^2 (2m^4 n^3+2m^4 n^2+6m^3 n^4-m^3 n^3-2m^3 n^2+3m^3 n+6m^2 n^5-8m^2 n^4+m^2 n^3 \nonumber\\
	&+6m^2 n^2-3m^2 n+3m^2+2mn^6-5mn^5+10mn^3-12mn^2+7mn-4m-3n^5+10n^4 \nonumber\\
	&-16n^3+14n^2-7n+1), \nonumber\\
	\text{deno}=&~
	n(m^5 n^3+4m^5 n^2+2m^5 n+4m^4 n^4+10m^4 n^3-7m^4 n^2+2m^4 n+2m^4+6m^3 n^5+6m^3 n^4 \nonumber\\
	&-23m^3 n^3+24m^3 n^2-7m^3 n+5m^3+4m^2 n^6-2m^2 n^5-20m^2 n^4+40m^2 n^3-35m^2 n^2 \nonumber\\
	&+17m^2 n-17m^2+mn^7-2mn^6-9mn^5+28mn^4-41mn^3+23mn^2-20mn+6m \nonumber\\
	&-3n^6+10n^5-19n^4+15n^3-7n^2-2n+4).
\end{align}

\begin{itemize}
	\item Accumulated payoff
\end{itemize}

When using accumulated payoffs, we follow Eq.~(\ref{eq_coopcondi_3_PC_accu}) for the PC rule. The numerator is
\begin{equation}
	\sum_{i,j\in\mathcal{N}} k_i (k_{i}+1)p_{ij}\tau_{ij}
	=m(m+n)(m-1)\tau_{HH'}+mn(m+n+2)\tau_{HL},
\end{equation}
and the denominator is 
\begin{equation}
	\sum_{i,j\in\mathcal{N}} k_i (k_i+1)p_{ij}\Upsilon_{ij}
	=m(m+n)(m-1)\Upsilon_{HH'}+mn(m+n)\Upsilon_{HL}+2mn\Upsilon_{LH}.
\end{equation}
The critical synergy factor on $m$-hub star graphs under the PC rule when using accumulated payoff is 
\begin{equation}
	r^\star_\text{accu}=\frac{\text{nume}}{\text{deno}}
	\xrightarrow{n\to\infty}\frac{4m-1}{3m-1}
	\xrightarrow{m\to\infty}\frac{4}{3},
\end{equation}
where 
\begin{align}
	\text{nume}=&~
	2(2m^3+6m^2 n-2m^2+6mn^2-3mn+m+2n^3-n^2+n)(2m^5+12m^4 n-6m^4 \nonumber\\
	&+26m^3 n^2-22m^3 n+7m^3+24m^2 n^3-24m^2 n^2+16m^2 n-4m^2+8mn^4-8mn^3 \nonumber\\
	&+10mn^2-12mn+m-2n^4+3n^3-10n^2+3n), \nonumber\\
	\text{deno}=&~
	n(2m^2+4mn-2m+2n^2-n+1)(3m^5+18m^4 n-2m^4+39m^3 n^2-17m^3 n-8m^3 \nonumber\\
	&+36m^2 n^3-37m^2 n^2-18m^2 n+m^2+12mn^4-26mn^3-14mn^2+5mn+10m \nonumber\\
	&-4n^4-6n^3+4n^2+8n-4).
\end{align}

For the DB rule, we follow Eq.~(\ref{eq_coopcondi_3_DB_accu}). The numerator is 
\begin{equation}
	\sum_{i,j\in\mathcal{N}} k_i (k_i+1)p_{ij}^{(2)}\tau_{ij}
	=\frac{m(m+n)(m-1)(m-2)}{m+n-1}\tau_{HH'}+\frac{mn(m-1)(m+n+2)}{m+n-1}\tau_{HL'}+\frac{2mn(n-1)}{m+n-1}\tau_{LL'},
\end{equation}
and the denominator is 
\begin{align}
	\sum_{i,j\in\mathcal{N}} k_i (k_i+1)p_{ij}^{(2)}\Upsilon_{ij}
	=&~\frac{m(m+n)(m-1)(m-2)}{m+n-1}\Upsilon_{HH'}+\frac{mn(m+n)(m-1)}{m+n-1}\Upsilon_{HL'}+\frac{2mn(n-1)}{m+n-1}\Upsilon_{LH'} \nonumber\\
	&+\frac{2mn(n-1)}{m+n-1}\Upsilon_{LL'}.
\end{align}
The critical synergy factor on $m$-hub star graphs under the DB rule when using accumulated payoff is 
\begin{equation}
	r^\star_\text{accu}=\frac{\text{nume}}{\text{deno}}
	\xrightarrow{n\to\infty}1,
\end{equation}
where 
\begin{align}
	\text{nume}=&~
	2(2m^4+8m^3 n-4m^3+12m^2 n^2-11m^2 n+3m^2+8mn^3-10mn^2+5mn-m+2n^4 \nonumber\\
	&-3n^3+2n^2-n)(2m^5+11m^4 n-8m^4+21m^3 n^2-27m^3 n+11m^3+16m^2 n^3-25m^2 n^2 \nonumber\\
	&+23m^2 n-7m^2+4mn^4-6mn^3+14mn^2-17mn+2m-4n^4+6n^3-18n^2+2n), \nonumber\\
	\text{deno}=&~
	n(2m^2+4mn-2m+2n^2-n+1)(3m^6+20m^5 n-4m^5+51m^4 n^2-33m^4 n-5m^4 \nonumber\\
	&+62m^3 n^3-81m^3 n^2+7m^3 n-3m^3+36m^2 n^4-80m^2 n^3+38m^2 n^2-15m^2 n+23m^2 \nonumber\\
	&+8mn^5-36mn^4+32mn^3-22mn^2+23mn-18m-8n^5+8n^4-10n^3+6n^2-6n+4).
\end{align}

For the BD rule, we follow Eq.~(\ref{eq_coopcondi_3_BD_accu}). The numerator is 
\begin{equation}
	\sum_{i,j\in\mathcal{N}}\frac{k_{ij}}{k_i k_j}(k_i+1) \tilde{\tau}_{ij}
	=\frac{m(m+n)(m-1)}{(n+m-1)^2}\tilde{\tau}_{HH'}+\frac{mn(m+n+2)}{n+m-1}\tilde{\tau}_{HL},
\end{equation}
and the denominator is 
\begin{equation}
	\sum_{i,j\in\mathcal{N}}\frac{k_{ij}}{k_i k_j}(k_i+1) \tilde{\Upsilon}_{ij}
	=\frac{m(m+n)(m-1)}{(n+m-1)^2}\tilde{\Upsilon}_{HH'}+\frac{mn(m+n)}{n+m-1}\tilde{\Upsilon}_{HL}+\frac{2mn}{n+m-1}\tilde{\Upsilon}_{LH}.
\end{equation}
The critical synergy factor on $m$-hub star graphs under the BD rule when using accumulated payoff is 
\begin{equation}
	r^\star_\text{accu}=\frac{\text{nume}}{\text{deno}}
	\xrightarrow{n\to\infty}2,
\end{equation}
where 
\begin{align}
	\text{nume}=&~
	2(m^4 n+4m^3 n^2-2m^3 n+3m^3+6m^2 n^3-6m^2 n^2+9m^2 n-4m^2+4mn^4-6mn^3+9mn^2 \nonumber\\
	&-7mn+m+n^5-2n^4+3n^3-3n^2+n)(m^5 n^3+2m^5 n^2+4m^4 n^4+6m^4 n^3-4m^4 n^2 \nonumber \\
	&+3m^4 n+6m^3 n^5+6m^3 n^4-15m^3 n^3+15m^3 n^2-4m^3 n+3m^3+4m^2 n^6+2m^2 n^5-20m^2 n^4 \nonumber\\
	&+29m^2 n^3-17m^2 n^2+10m^2 n-4m^2+mn^7-11mn^5+21mn^4-16mn^3+6mn^2-4mn \nonumber\\
	&+m-2n^6+4n^5-n^4-7n^3+10n^2-5n), \nonumber\\
	\text{deno}=&~
	n(m^3 n+3m^2 n^2-2m^2 n+3m^2+3mn^3-4mn^2+6mn-4m+n^4-2n^3+3n^2-3n+1) \nonumber\\
	&\times (m^5 n^3+4m^5 n^2+4m^5 n+4m^4 n^4+10m^4 n^3+m^4 n^2-9m^4 n+4m^4+6m^3 n^5+6m^3 n^4 \nonumber\\
	&-10m^3 n^3-11m^3 n^2+22m^3 n-5m^3+4m^2 n^6-2m^2 n^5-9m^2 n^4-3m^2 n^3+27m^2 n^2 \nonumber\\
	&-28m^2 n-m^2+mn^7-2mn^6-4mn^5+3mn^4+9mn^3-31mn^2+13mn-2m-2n^6 \nonumber\\
	&+4n^5-4n^4-6n^3+10n^2-8n+4).
\end{align}

\subsection{Ceiling fan}
On a ceiling fan graph, there is one hub ($H$) and $n$ leaves ($L$), each leaf consists of two connected nodes. The hub node has $k_H=2n$ neighbors, and each leaf node has $k_L=2$ neighbors (the hub and the other leaf node). There are three non-zero $\tau_{ij}$ types: $\tau_{HL}$, the relation between the hub and a leaf node; $\tau_{LL'}$, the relation between the two leaf nodes of the same leaf; $\tau_{LL''}$, the relation between two leaf nodes of two different leaves. According to Eq.~(\ref{eq_tauij}) in the main text, we have the system of linear equations: 
\begin{align}
	\begin{cases}
		\displaystyle{
			\tau_{HL}=1+\frac{1}{4n}(\tau_{LL'}+(2n-2)\tau_{LL''})+\frac{1}{4}\tau_{HL}},\\[1em]
		\displaystyle{
			\tau_{LL'}=1+\frac{1}{4}\tau_{HL}+\frac{1}{4}\tau_{HL}},\\[1em]
		\displaystyle{
			\tau_{LL''}=1+\frac{1}{4}(\tau_{HL}+\tau_{LL''})+\frac{1}{4}(\tau_{HL}+\tau_{LL''})}.
	\end{cases}
\end{align}
The solution is 
\begin{align}
	\begin{cases}
		\displaystyle{
			\tau_{HL}=\frac{2(8n-3)}{2n+3}},\\[1em]
		\displaystyle{
			\tau_{LL'}=\frac{10n}{2n+3}},\\[1em]
		\displaystyle{
			\tau_{LL''}=\frac{20n}{2n+3}}.
	\end{cases}
\end{align}
Inserting these $\tau_{ij}$ values into Eq.~(\ref{eq_Upsilonij}) in the main text, we calculate the required $\Upsilon_{ij}$ values:
\begin{subequations}
	\begin{align}
		\Upsilon_{HL}&=
		-\frac{4n^2+8n-3}{3(2n+1)^2}\tau_{HL}+\frac{4n+5}{3(2n+1)^2}\tau_{LL'}+\frac{8n^2+2n-10}{3(2n+1)^2}\tau_{LL''} \nonumber\\
		&=\frac{2(16n^3-4n^2-9n-3)}{(2n+1)^2 (2n+3)}, \\
		\Upsilon_{LH}&=
		\frac{10n-1}{9(2n+1)}\tau_{HL}-\frac{4n+5}{9(2n+1)}\tau_{LL'}-\frac{2n-2}{3(2n+1)}\tau_{LL''} \nonumber\\
		&=-\frac{2(n-1)}{3(4n^2+8n+3)}, \\
		\Upsilon_{LL'}&=
		0, \\
		\Upsilon_{LL''}&=
		-\frac{2}{9}\tau_{LL'}+\frac{4}{9}\tau_{LL''} 
		=\frac{20n}{6n+9}.
	\end{align}
\end{subequations}

Then, we apply these $\tau_{ij}$ and $\Upsilon_{ij}$ values to calculate the critical synergy factor on the ceiling fan graph. For the PC rule, the numerator is 
\begin{equation}
	\tau^{(1)}=4n\tau_{HL}+2n\tau_{LL'}
	=\frac{12n(7n-2)}{2n+3},
\end{equation}
and the denominator is 
\begin{equation}
	\Upsilon^{(1)}=2n\Upsilon_{HL}+2n\Upsilon_{LH}+2n\Upsilon_{LL'}
	=\frac{8n(24n^3-7n^2-13n-4)}{3(2n+1)^2 (2n+3)}.
\end{equation}
Therefore, the critical synergy factor on ceiling fan graphs under the PC rule is
\begin{equation}
	r^\star=\frac{9(2n+1)^2 (7n-2)}{2(24n^3-7n^2-13n-4)}
	\xrightarrow{n\to\infty}\frac{21}{4}.
\end{equation}

For the DB rule, the numerator is 
\begin{equation}
	\tau^{(2)}=2n\tau_{HL}+\tau_{LL'}+(2n-2)\tau_{LL''}
	=\frac{6n(12n-7)}{2n+3},
\end{equation}
and the denominator is 
\begin{equation}
	\Upsilon^{(2)}=n\Upsilon_{HL}+n\Upsilon_{LH}+\Upsilon_{LL'}+(2n-2)\Upsilon_{LL''}
	=\frac{4n(64n^3-7n^2-43n-14)}{3(2n+1)^2 (2n+3)}.
\end{equation}
Therefore, the critical synergy factor on ceiling fan graphs under the DB rule is
\begin{equation}
	r^\star=\frac{9(2n+1)^2 (12n-7)}{2(64n^3-7n^2-43n-14)}
	\xrightarrow{n\to\infty}\frac{27}{8}.
\end{equation}

For the BD rule, we list the system of linear equations according to Eq.~(\ref{eq_tauij_BD_2}):
\begin{align}
	\begin{cases}
		\displaystyle{
			\tilde{\tau}_{HL}=\frac{2n}{2n^2+n+1}\left(1+\frac{1}{2}\tilde{\tau}_{LL'}+(n-1)\tilde{\tau}_{LL''}+\frac{1}{2}\tilde{\tau}_{HL}\right)},\\[1em]
		\displaystyle{
			\tilde{\tau}_{LL'}=\frac{n}{n+1}\left(1+\frac{1}{n}\tilde{\tau}_{HL}\right)},\\[1em]
		\displaystyle{
			\tilde{\tau}_{LL''}=\frac{n}{n+1}\left(1+\frac{1}{n}\tilde{\tau}_{HL}+\tilde{\tau}_{LL''}\right)}.
	\end{cases}
\end{align}
The solution is 
\begin{align}
	\begin{cases}
		\displaystyle{
			\tilde{\tau}_{HL}=\frac{2n^4+n^2+2n}{2n^2+2n+1}},\\[1em]
		\displaystyle{
			\tilde{\tau}_{LL'}=\frac{n(2n^2+3)}{2n^2+2n+1}},\\[1em]
		\displaystyle{
			\tilde{\tau}_{LL''}=\frac{n(2n^2+3)(n+1)}{2n^2+2n+1}}.
	\end{cases}
\end{align}
Inserting these $\tilde{\tau}_{ij}$ values into Eq.~(\ref{eq_Upsilonij_BD}), we obtain the required $\tilde{\Upsilon}_{ij}$ values:
\begin{subequations}
	\begin{align}
		\tilde{\Upsilon}_{HL}&=
		-\frac{4n^2+8n-3}{3(2n+1)^2}\tilde{\tau}_{HL}+\frac{4n+5}{3(2n+1)^2}\tilde{\tau}_{LL'}+\frac{8n^2+2n-10}{3(2n+1)^2}\tilde{\tau}_{LL''} \nonumber\\
		&=\frac{n(8n^5+4n^4+18n^3+4n^2-25n-9)}{3(2n+1)^2 (2n^2+2n+1)}, \\
		\tilde{\Upsilon}_{LH}&=
		\frac{10n-1}{9(2n+1)}\tilde{\tau}_{HL}-\frac{4n+5}{9(2n+1)}\tilde{\tau}_{LL'}-\frac{2n-2}{3(2n+1)}\tilde{\tau}_{LL''} \nonumber\\
		&=\frac{n(8n^4-10n^3-6n^2+7n+1)}{9(4n^3+6n^2+4n+1)}, \\
		\tilde{\Upsilon}_{LL'}&=
		0.
	\end{align}
\end{subequations}
Then, we apply these $\tilde{\tau}_{ij}$ and $\tilde{\Upsilon}_{ij}$ values to calculate the critical synergy factor on the ceiling fan graph under the BD rule. The numerator is
\begin{equation}
	\tilde{\tau}^{(1)}=\tilde{\tau}_{HL}+\frac{n}{2}\tilde{\tau}_{LL'}=\frac{n(6n^3+5n+4)}{2(2n^2+2n+1)},
\end{equation}
and the denominator is 
\begin{equation}
	\tilde{\Upsilon}^{(1)}=\frac{1}{2}\tilde{\Upsilon}_{HL}+\frac{1}{2}\tilde{\Upsilon}_{LH}+\frac{n}{2}\tilde{\Upsilon}_{LL'}=\frac{n(20n^5+16n^3+10n^2-33n-13)}{9(2n+1)^2 (2n^2+2n+1)}.
\end{equation}
Therefore, the critical synergy factor on ceiling fan graphs under the BD rule is 
\begin{equation}
	r^\star=\frac{\tilde{\tau}^{(1)}}{\tilde{\Upsilon}^{(1)}}=\frac{9(2n+1)^2 (6n^3+5n+4)}{2(20n^5+16n^3+10n^2-33n-13)}
	\xrightarrow{n\to\infty}\frac{27}{5}.
\end{equation}

\begin{itemize}
	\item Accumulated payoff
\end{itemize}

When using accumulated payoffs, we follow Eq.~(\ref{eq_coopcondi_3_PC_accu}) for the PC rule. The numerator is
\begin{equation}
	\sum_{i,j\in\mathcal{N}} k_i (k_{i}+1)p_{ij}\tau_{ij}
	=4n(n+2)\tau_{HL}+6n\tau_{LL'}
	=\frac{4n(16n^2+41n-12)}{2n+3},
\end{equation}
and the denominator is 
\begin{equation}
	\sum_{i,j\in\mathcal{N}} k_i (k_i+1)p_{ij}\Upsilon_{ij}
	=2n(2n+1)\Upsilon_{HL}+6n\Upsilon_{LH}+6n\Upsilon_{LL'}
	=\frac{8n(4n^2-3n-1)}{2n+3}.
\end{equation}
The critical synergy factor on ceiling fan graphs under the PC rule when using accumulated payoff is 
\begin{equation}
	r^\star_\text{accu}=\frac{16n^2+41n-12}{2(4n^2-3n-1)}
	\xrightarrow{n\to\infty}2.
\end{equation}

For the DB rule, we follow Eq.~(\ref{eq_coopcondi_3_DB_accu}). The numerator is 
\begin{align}
	\sum_{i,j\in\mathcal{N}} k_i (k_i+1)p_{ij}^{(2)}\tau_{ij}
	&=2n(n+2)\tau_{HL}+3\tau_{LL'}+6(n-1)\tau_{LL''} \nonumber\\
	&=\frac{2n(16n^2+86n-57)}{2n+3},
\end{align}
and the denominator is 
\begin{align}
	\sum_{i,j\in\mathcal{N}} k_i (k_i+1)p_{ij}^{(2)}\Upsilon_{ij}
	&=n(2n+1)\Upsilon_{HL}+3n\Upsilon_{LH}+3\Upsilon_{LL'}+6(n-1)\Upsilon_{LL''} \nonumber\\
	&=\frac{4n(4n^2+7n-11)}{2n+3}.
\end{align}
The critical synergy factor on ceiling fan graphs under the DB rule when using accumulated payoff is 
\begin{equation}
	r^\star_\text{accu}=\frac{16n^2+86n-57}{2(4n^2+7n-11)}
	\xrightarrow{n\to\infty}2.
\end{equation}

For the BD rule, we follow Eq.~(\ref{eq_coopcondi_3_BD_accu}). The numerator is 
\begin{align}
	\sum_{i,j\in\mathcal{N}}\frac{k_{ij}}{k_i k_j}(k_i+1) \tilde{\tau}_{ij}
	&=(n+2)\tilde{\tau}_{HL}+\frac{3n}{2}\tilde{\tau}_{LL'} \nonumber\\
	&=\frac{n(4n^4+14n^3+2n^2+17n+8)}{2(2n^2+2n+1)},
\end{align}
and the denominator is 
\begin{align}
	\sum_{i,j\in\mathcal{N}}\frac{k_{ij}}{k_i k_j}(k_i+1) \tilde{\Upsilon}_{ij}
	&=\frac{2n+1}{2}\tilde{\Upsilon}_{HL}+\frac{3}{2}\tilde{\Upsilon}_{LH}+\frac{3n}{2}\tilde{\Upsilon}_{LL'} \nonumber\\
	&=\frac{n(2n^4+2n^3+n^2-n-4)}{3(2n^2+2n+1)}.
\end{align}
The critical synergy factor on ceiling fan graphs under the BD rule when using accumulated payoff is 
\begin{equation}
	r^\star_\text{accu}=\frac{3(4n^4+14n^3+2n^2+17n+8)}{2(2n^4+2n^3+n^2-n-4)}
	\xrightarrow{n\to\infty}3.
\end{equation}

By analogy, the critical synergy factor for PGGs on any other networks can be calculated using the same method in the future.

\section{: Some extensions to the donation game (DG)}
Here, we give the details of deducing cooperation conditions in pairwise donation games (DGs). These results can be derived through the techniques in previous literature~\cite{allen2017evolutionary}. However, only the results under the DB and BD updates using average payoff can be found in previous literature~\cite{allen2017evolutionary}. To compare our PGGs with pairwise DGs across different model details, we deduce the unpublished results of pairwise DGs and present them in our work.

The actual payoff of agent $i$ is averaged over $k_i$ DGs played with all neighbors $l\in\mathcal{N}_i$. In each DG, a cooperator pays $c$ and the other player receives $b$ ($b>c$), while a defector pays nothing and the other player receives nothing. Namely, the actual payoff $f_i(\mathbf{x})$ of agent $i$ is expressed as follows.
\begin{equation}\label{eq_payoff_dg}
	f_i(\mathbf{x})=\frac{1}{k_i}\sum_{l\in\mathcal{N}_i}(-x_i c+x_l b)
	=-x_i c+\frac{1}{k_i}\sum_{l\in\mathcal{N}_i}x_l b.
\end{equation}

The dynamics of strategy evolution under neutral drift remain the same. Only the quantity $\mathbb{E}_\text{RMC}^\circ [(x_i-x_j)(f_i(\mathbf{x})-f_j(\mathbf{x}))]$ is influenced by the payoff calculation in pairwise DGs. Applying the payoff in Eq.~(\ref{eq_payoff_dg}), we have 
\begin{align}\label{eq_DG_coopcondi_1}
	&~\mathbb{E}_\text{RMC}^\circ [(x_i-x_j)(f_i(\mathbf{x})-f_j(\mathbf{x}))]
	\nonumber\\
	=&~\mathbb{E}_\text{RMC}^\circ \Bigg[
	-(x_i^2-x_i x_j)c+\frac{1}{k_i}\sum_{l\in\mathcal{N}_i}(x_i x_l-x_j x_l)b
	+(x_i x_j-x_j^2)c-\frac{1}{k_j}\sum_{l\in\mathcal{N}_j}(x_i x_l-x_j x_l)b
	\Bigg]
	\nonumber\\
	=&-c\left(\mathbb{E}_\text{RMC}^\circ[x_i^2]-2\mathbb{E}_\text{RMC}^\circ[x_i x_j]+\mathbb{E}_\text{RMC}^\circ[x_j^2]\right)
	+\frac{b}{k_i}\sum_{l\in\mathcal{N}_i} \left(\mathbb{E}_\text{RMC}^\circ[x_i x_l]-\mathbb{E}_\text{RMC}^\circ[x_j x_l]\right)
	\nonumber\\
	&-\frac{b}{k_j}\sum_{l\in\mathcal{N}_j}
	\left(\mathbb{E}_\text{RMC}^\circ[x_i x_l]-\mathbb{E}_\text{RMC}^\circ[x_j x_l]\right).
\end{align}

\subsection{Pairwise comparison}
The cooperation condition under the PC rule is Eq.~(\ref{eq_coopcondi_PC}). Using the result of Eq.~(\ref{eq_DG_coopcondi_1}) and applying $\tau_{ij}=(1/2-\mathbb{E}_\text{RMC}^\circ[x_i x_j])/(K/4)$ defined by Eq.~(\ref{eq_tauijERMC_PC}), we calculate the cooperation condition under the PC rule as 
\begin{align}\label{eq_DG_coopcondi_2_PC}
	&~\frac{1}{4N^2\langle k\rangle} \sum_{i,j\in\mathcal{N}}k_i p_{ij}\mathbb{E}_\text{RMC}^\circ [(x_i-x_j)(f_i(\mathbf{x})-f_j(\mathbf{x}))]>0 \nonumber\\
	\Leftrightarrow
	&~\sum_{i,j\in\mathcal{N}}k_i p_{ij}\Bigg\{-2c\tau_{ij}
	+\frac{b}{k_i}\sum_{l\in\mathcal{N}_i} (-\tau_{il}+\tau_{jl})
	-\frac{b}{k_j}\sum_{l\in\mathcal{N}_j} (-\tau_{il}+\tau_{jl})\Bigg\}
	>0 \nonumber\\
	\Leftrightarrow
	&~\frac{b}{c}>\frac{2\sum_{i,j\in\mathcal{N}}k_i p_{ij}\tau_{ij}}{\sum_{i,j,l\in\mathcal{N}}k_i p_{ij}(p_{il}-p_{jl})(\tau_{jl}-\tau_{il})}.
\end{align}
Further simplifying Eq.~(\ref{eq_DG_coopcondi_2_PC}) (using Eq.~(\ref{eq_tau01234_1})) leads to 
\begin{equation}
	\frac{b}{c}>\frac{\sum_{i,j\in\mathcal{N}}k_i p_{ij}\tau_{ij}}{\sum_{i,j,l\in\mathcal{N}}k_i p_{ij}p_{il}(\tau_{jl}-\tau_{il})}
	=\frac{\tau^{(1)}}{\tau^{(2)}-\tau^{(1)}}.
\end{equation}
The right-hand side is the $(b/c)^\star$ value for the success of cooperation in pairwise DGs under the PC rule. The $\tau_{ij}$ values should be obtained by solving Eqs.~(\ref{eq_eqtauij_PC}) on a given network.

\subsection{Death-birth}
The critical $(b/c)^\star$ value under the DB update has been first obtained in Ref.~\cite{allen2017evolutionary}. Trivially, in our calculation, the cooperation condition under the DB rule is Eq.~(\ref{eq_coopcondi_DB}). Using the result of Eq.~(\ref{eq_DG_coopcondi_1}) and applying $\tau_{ij}=(1/2-\mathbb{E}_\text{RMC}^\circ[x_i x_j])/(K/4)$ defined by Eq.~(\ref{eq_tauijERMC_DB}), we calculate Eq.~(\ref{eq_coopcondi_DB}) as 
\begin{align}\label{eq_DG_coopcondi_2_DB}
	&~\frac{1}{2N^2\langle k\rangle} \sum_{i,j\in\mathcal{N}}k_i p^{(2)}_{ij}\mathbb{E}_\text{RMC}^\circ [(x_i-x_j)(f_i(\mathbf{x})-f_j(\mathbf{x}))]>0 \nonumber\\
	\Leftrightarrow
	&~\sum_{i,j\in\mathcal{N}}k_i p^{(2)}_{ij}\Bigg\{-2c\tau_{ij}
	+\frac{b}{k_i}\sum_{l\in\mathcal{N}_i} (-\tau_{il}+\tau_{jl})
	-\frac{b}{k_j}\sum_{l\in\mathcal{N}_j} (-\tau_{il}+\tau_{jl})\Bigg\}
	>0 \nonumber\\
	\Leftrightarrow
	&~\frac{b}{c}>\frac{2\sum_{i,j\in\mathcal{N}}k_i p^{(2)}_{ij}\tau_{ij}}{\sum_{i,j,l\in\mathcal{N}}k_i p^{(2)}_{ij}(p_{il}-p_{jl})(\tau_{jl}-\tau_{il})}.
\end{align}
Further simplifying Eq.~(\ref{eq_DG_coopcondi_2_PC}) leads to 
\begin{equation}
	\frac{b}{c}>\frac{\sum_{i,j\in\mathcal{N}}k_i p^{(2)}_{ij}\tau_{ij}}{\sum_{i,j,l\in\mathcal{N}}k_i p^{(2)}_{ij}p_{il}(\tau_{jl}-\tau_{il})}
	=\frac{\tau^{(2)}}{\tau^{(3)}-\tau^{(1)}}.
\end{equation}
The right-hand side is the $(b/c)^\star$ value for the success of cooperation in pairwise DGs under the DB rule, which is consistent with  ``$t_2/(t_3-t_1)$'' in the main text of Ref.~\cite{allen2017evolutionary}. The $\tau_{ij}$ values should be obtained by solving Eqs.~(\ref{eq_eqtauij_PC}) on a given network.

\subsection{Birth-death}
The critical $(b/c)^\star$ value under the BD update has also been mentioned in Ref.~\cite{allen2017evolutionary}. We examine their results here. The cooperation condition under the BD rule is Eq.~(\ref{eq_coopcondi_BD}). Using the result of Eq.~(\ref{eq_DG_coopcondi_1}) and applying $\tau_{ij}=(1/2-\mathbb{E}_\text{RMC}^\circ[x_i x_j])/(K/2)$ as defined by Eq.~(\ref{eq_tauijERMC_BD}), we calculate Eq.~(\ref{eq_coopcondi_BD}) follows.
\begin{align}\label{eq_DG_coopcondi_2_BD}
	&~\frac{1}{2N^2\langle k^{-1}\rangle} \sum_{i,j\in\mathcal{N}}\frac{k_{ij}}{k_i k_j}\mathbb{E}_\text{RMC}^\circ [(x_i-x_j)(f_i(\mathbf{x})-f_j(\mathbf{x}))]>0 \nonumber\\
	\Leftrightarrow
	&~\sum_{i,j\in\mathcal{N}}\frac{k_{ij}}{k_i k_j}\Bigg\{-2c\tilde{\tau}_{ij}
	+\frac{b}{k_i}\sum_{l\in\mathcal{N}_i} (-\tilde{\tau}_{il}+\tilde{\tau}_{jl})
	-\frac{b}{k_j}\sum_{l\in\mathcal{N}_j} (-\tilde{\tau}_{il}+\tilde{\tau}_{jl})\Bigg\}
	>0 \nonumber\\
	\Leftrightarrow
	&~\frac{b}{c}>\frac{2\sum_{i,j\in\mathcal{N}}\dfrac{k_{ij}}{k_i k_j}\tilde{\tau}_{ij}}{\sum_{i,j,l\in\mathcal{N}}\dfrac{k_{ij}}{k_i k_j}(p_{il}-p_{jl})(\tilde{\tau}_{jl}-\tilde{\tau}_{il})}
	=\frac{\sum_{i,j\in\mathcal{N}}\dfrac{k_{ij}}{k_i k_j}\tilde{\tau}_{ij}}{\sum_{i,j,l\in\mathcal{N}}\dfrac{k_{ij} k_{il}}{k_i^2 k_j}(\tilde{\tau}_{jl}-\tilde{\tau}_{il})}.
\end{align}
The right-hand side is the $(b/c)^\star$ value for the success of cooperation in pairwise DGs under the BD rule. The $\tilde{\tau}_{ij}$ values should be obtained by solving Eqs.~(\ref{eq_eqtauij_BD}) on a given network.

\subsection{Variation of the model: accumulated payoff}
When using accumulated payoffs, the actual payoff of agent $i$ is accumulated through the $k_i$ DGs played with neighbors. The actual payoff $f_i(\mathbf{x})$ of agent $i$ is 
\begin{equation}\label{eq_payoff_dg_accu}
	f_i(\mathbf{x})=\sum_{l\in\mathcal{N}_i}(-x_i c+x_l b)
	=-k_i x_i c+\sum_{l\in\mathcal{N}_i}x_l b.
\end{equation}

The dynamics of strategy evolution under neutral drift remain the same, no matter the payoff calculation is averaged, accumulated or other. Only the quantity $\mathbb{E}_\text{RMC}^\circ [(x_i-x_j)(f_i(\mathbf{x})-f_j(\mathbf{x}))]$ is influenced. Applying the accumulated payoff calculation in Eq.~(\ref{eq_payoff_dg_accu}), we have 
\begin{align}\label{eq_DG_coopcondi_1_PC_accu}
	&~\mathbb{E}_\text{RMC}^\circ [(x_i-x_j)(f_i(\mathbf{x})-f_j(\mathbf{x}))]
	\nonumber\\
	=&~\mathbb{E}_\text{RMC}^\circ \Bigg[
	-k_i (x_i^2-x_i x_j)c+\sum_{l\in\mathcal{N}_i}(x_i x_l-x_j x_l)b
	+k_j (x_i x_j-x_j^2)c-\sum_{l\in\mathcal{N}_j}(x_i x_l-x_j x_l)b
	\Bigg]
	\nonumber\\
	=&-c\left(k_i \mathbb{E}_\text{RMC}^\circ[x_i^2]-(k_i +k_j)\mathbb{E}_\text{RMC}^\circ[x_i x_j]+k_j \mathbb{E}_\text{RMC}^\circ[x_j^2]\right)
	+b\sum_{l\in\mathcal{N}_i} \left(\mathbb{E}_\text{RMC}^\circ[x_i x_l]-\mathbb{E}_\text{RMC}^\circ[x_j x_l]\right)
	\nonumber\\
	&-b\sum_{l\in\mathcal{N}_j}
	\left(\mathbb{E}_\text{RMC}^\circ[x_i x_l]-\mathbb{E}_\text{RMC}^\circ[x_j x_l]\right).
\end{align}

\subsubsection{Pairwise comparison}
The cooperation condition under the PC rule is still Eq.~(\ref{eq_coopcondi_PC}). Using the result of Eq.~(\ref{eq_DG_coopcondi_1}) and Eq.~(\ref{eq_tauijERMC_PC}), we calculate 
\begin{align}\label{eq_DG_coopcondi_2_PC_accu}
	&~\frac{1}{4N^2\langle k\rangle} \sum_{i,j\in\mathcal{N}}k_i p_{ij}\mathbb{E}_\text{RMC}^\circ [(x_i-x_j)(f_i(\mathbf{x})-f_j(\mathbf{x}))]>0 \nonumber\\
	\Leftrightarrow
	&~\sum_{i,j\in\mathcal{N}}k_i p_{ij}\Bigg\{-(k_i+k_j)c\tau_{ij}
	+b\sum_{l\in\mathcal{N}_i} (-\tau_{il}+\tau_{jl})
	-b\sum_{l\in\mathcal{N}_j} (-\tau_{il}+\tau_{jl})\Bigg\}
	>0 \nonumber\\
	\Leftrightarrow
	&~\frac{b}{c}>\frac{\sum_{i,j\in\mathcal{N}}k_i (k_i+k_j) p_{ij}\tau_{ij}}{\sum_{i,j,l\in\mathcal{N}}k_i p_{ij}(k_{il}-k_{jl})(\tau_{jl}-\tau_{il})}
	=\frac{\sum_{i,j\in\mathcal{N}}k_i^2 p_{ij}\tau_{ij}}{\sum_{i,j,l\in\mathcal{N}}k_i^2 p_{ij}p_{il}(\tau_{jl}-\tau_{il})}.
\end{align}
The right-hand side is the $(b/c)^\star_\text{accu}$ value in pairwise DGs using accumulated payoffs under the PC rule. The $\tau_{ij}$ values should be obtained by solving Eqs.~(\ref{eq_eqtauij_PC}) on a given network.

\subsubsection{Death-birth}
The cooperation condition under the DB rule is still Eq.~(\ref{eq_coopcondi_DB}). Using the result of Eq.~(\ref{eq_DG_coopcondi_1}) and Eq.~(\ref{eq_tauijERMC_DB}), we calculate 
\begin{align}\label{eq_DG_coopcondi_2_DB_accu}
	&~\frac{1}{2N^2\langle k\rangle} \sum_{i,j\in\mathcal{N}}k_i p^{(2)}_{ij}\mathbb{E}_\text{RMC}^\circ [(x_i-x_j)(f_i(\mathbf{x})-f_j(\mathbf{x}))]>0 \nonumber\\
	\Leftrightarrow
	&~\sum_{i,j\in\mathcal{N}}k_i p^{(2)}_{ij}\Bigg\{-(k_i+k_j)c\tau_{ij}
	+b\sum_{l\in\mathcal{N}_i} (-\tau_{il}+\tau_{jl})
	-b\sum_{l\in\mathcal{N}_j} (-\tau_{il}+\tau_{jl})\Bigg\}
	>0 \nonumber\\
	\Leftrightarrow
	&~\frac{b}{c}>\frac{2\sum_{i,j\in\mathcal{N}}k_i (k_i+k_j) p^{(2)}_{ij}\tau_{ij}}{\sum_{i,j,l\in\mathcal{N}}k_i p^{(2)}_{ij}(k_{il}-k_{jl})(\tau_{jl}-\tau_{il})}
	=\frac{\sum_{i,j\in\mathcal{N}}k_i^2 p_{ij}^{(2)}\tau_{ij}}{\sum_{i,j,l\in\mathcal{N}}k_i^2 p_{ij}^{(2)}p_{il}(\tau_{jl}-\tau_{il})}.
\end{align}
The right-hand side is the $(b/c)^\star_\text{accu}$ value in pairwise DGs using accumulated payoffs under the DB rule. The $\tau_{ij}$ values should be obtained by solving Eqs.~(\ref{eq_eqtauij_PC}) on a given network.

\subsubsection{Birth-death}
The cooperation condition under the BD rule is still Eq.~(\ref{eq_coopcondi_BD}). Using the result of Eq.~(\ref{eq_DG_coopcondi_1}) and Eq.~(\ref{eq_tauijERMC_BD}), we calculate 
\begin{align}\label{eq_DG_coopcondi_2_BD_accu}
	&~\frac{1}{2N^2\langle k^{-1}\rangle} \sum_{i,j\in\mathcal{N}}\frac{k_{ij}}{k_i k_j}\mathbb{E}_\text{RMC}^\circ [(x_i-x_j)(f_i(\mathbf{x})-f_j(\mathbf{x}))]>0 \nonumber\\
	\Leftrightarrow
	&~\sum_{i,j\in\mathcal{N}}\frac{k_{ij}}{k_i k_j}\Bigg\{-(k_i+k_j)c\tilde{\tau}_{ij}
	+b\sum_{l\in\mathcal{N}_i} (-\tilde{\tau}_{il}+\tilde{\tau}_{jl})
	-b\sum_{l\in\mathcal{N}_j} (-\tilde{\tau}_{il}+\tilde{\tau}_{jl})\Bigg\}
	>0 \nonumber\\
	\Leftrightarrow
	&~\frac{b}{c}>\frac{\sum_{i,j\in\mathcal{N}}\dfrac{k_{ij}}{k_i k_j}(k_i+k_j)\tilde{\tau}_{ij}}{\sum_{i,j,l\in\mathcal{N}}\dfrac{k_{ij}}{k_i k_j}(k_{il}-k_{jl})(\tilde{\tau}_{jl}-\tilde{\tau}_{il})}
	=\frac{\sum_{i,j\in\mathcal{N}}p_{ij}\tilde{\tau}_{ij}}{\sum_{i,j,l\in\mathcal{N}}p_{ji}p_{il}(\tilde{\tau}_{jl}-\tilde{\tau}_{il})}.
\end{align}
The right-hand side is the $(b/c)^\star_\text{accu}$ value in pairwise DGs using accumulated payoffs under the BD rule. The $\tilde{\tau}_{ij}$ values should be obtained by solving Eqs.~(\ref{eq_eqtauij_BD}) on a given network.

\begin{figure}
	\centering
	\includegraphics[width=\textwidth]{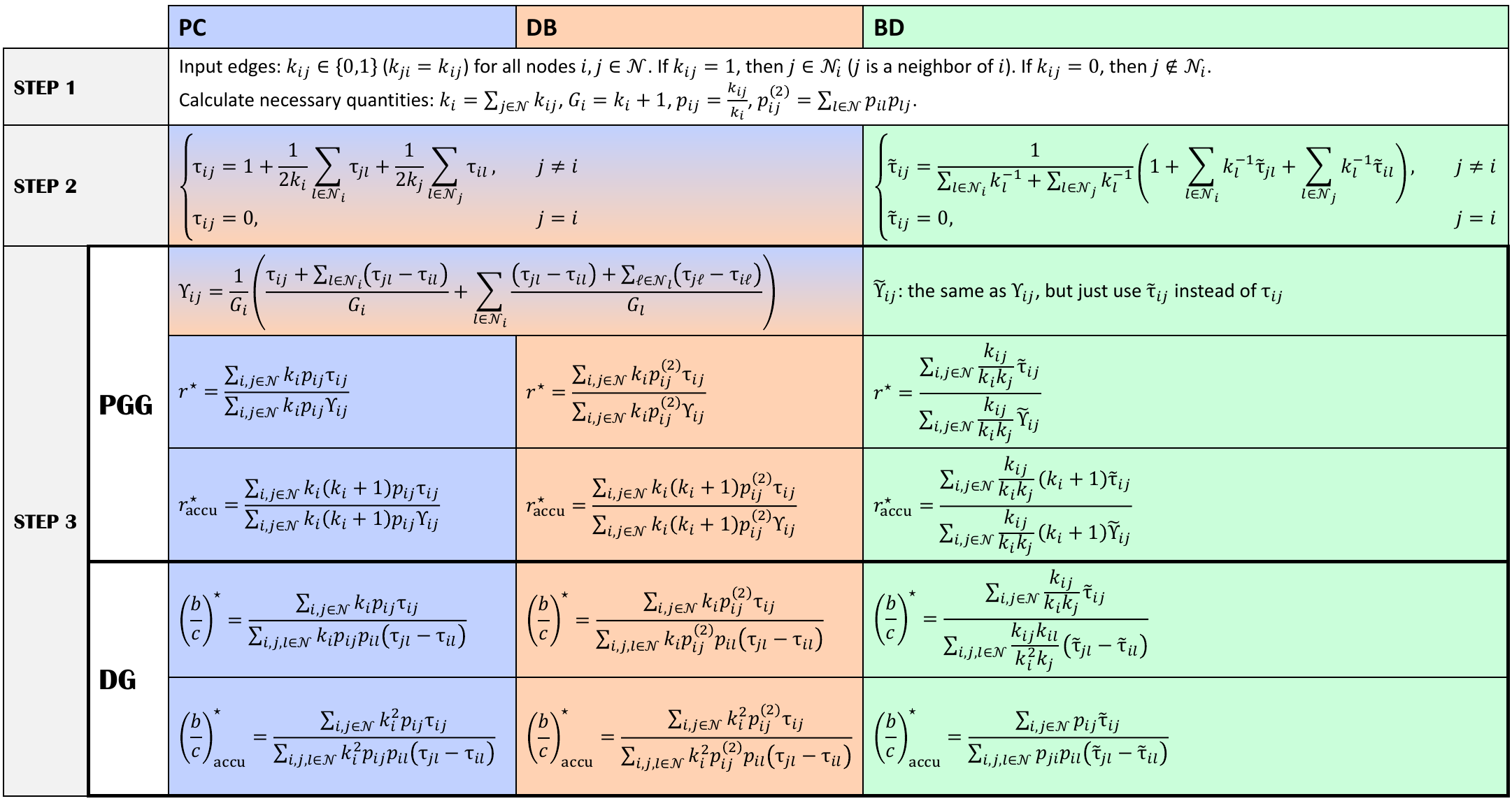}
	\caption{\textbf{Steps to calculate the theoretical conditions for the success of cooperation in both PGGs and DGs, including PC, DB, and BD update rules and average \& accumulated payoff calculations.} \textbf{Step 1}: Input edges $k_{ij}\in\{0,1\}$ between all nodes and calculate necessary quantities. \textbf{Step 2}: Solve for the linear equations to obtain $\tau_{ij}$ (for PC and DB updates) or $\tilde{\tau}_{ij}$ (for BD update). Note that $\tau_{ij}=\tau_{ji}$, $\tilde{\tau}_{ij}=\tilde{\tau}_{ji}$. \textbf{Step 3}: Insert the obtained values into the formulas for cooperation conditions. Note that usually $\Upsilon_{ij}\neq \Upsilon_{ji}$, $\tilde{\Upsilon}_{ij}\neq \tilde{\Upsilon}_{ji}$. $r^\star$ is the critical synergy factor using average payoff and $r^\star_\text{accu}$ is using accumulated payoff, and similar to $(b/c)^\star$ and $(b/c)^\star_\text{accu}$.}
	\label{fig_result_sup}
\end{figure}

The steps to calculate the cooperation conditions of both PGGs and DGs across all model details (PC, DB, and BD updates \& average and accumulated payoffs) are summarized in Fig.~\ref{fig_result_sup}.

\begin{figure}
	\centering
	\includegraphics[width=\textwidth]{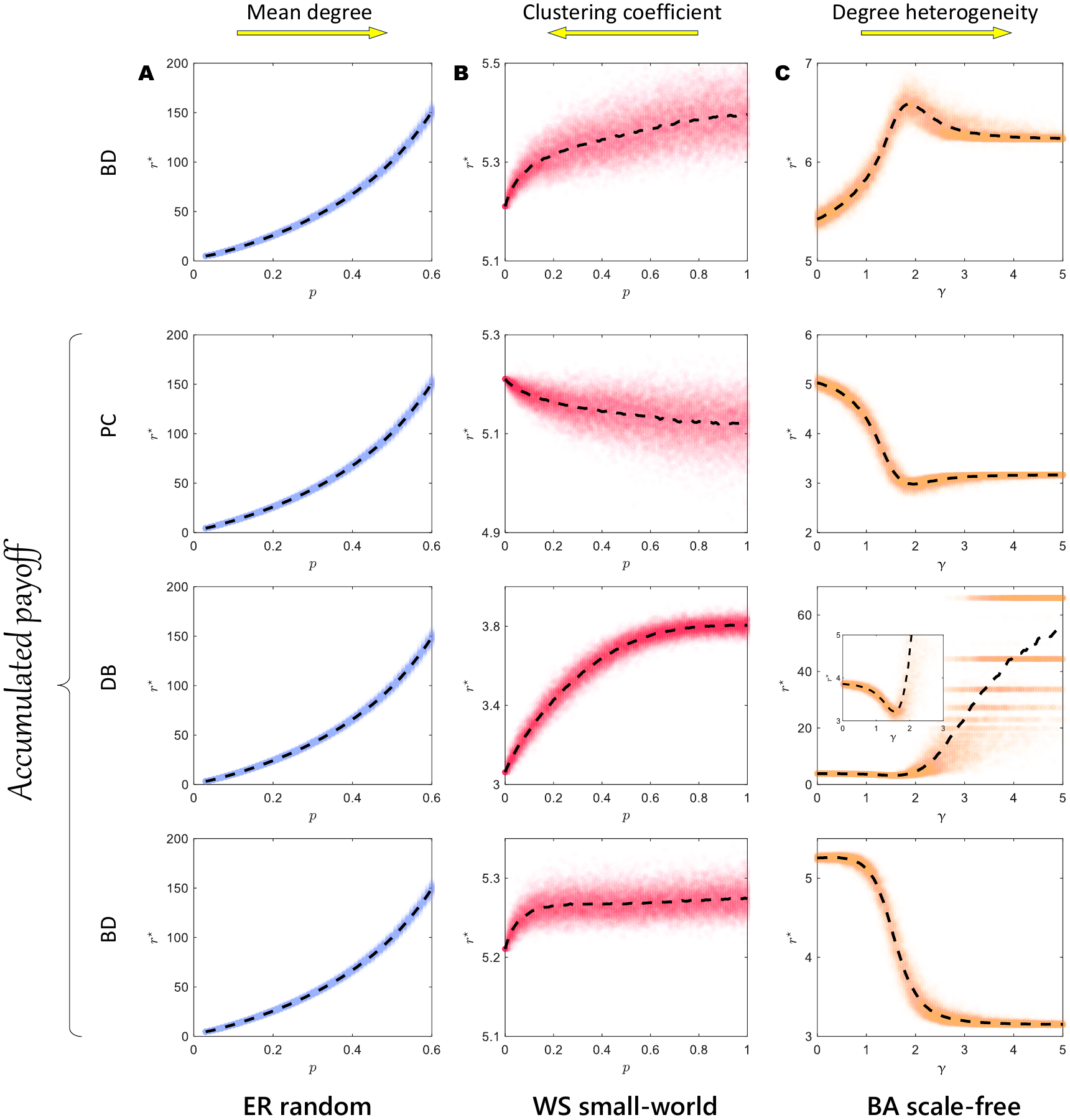}
	\caption{\textbf{Supplementary results across more model details (BD update and accumulated payoff) for effects of local structures on cooperation in PGGs.} (\textbf{A}) ER networks. The increasing average degree consistently inhibits cooperation. (\textbf{B}) WS networks. The increasing clustering coefficient promotes cooperation, but the PC rule using accumulated payoffs presents the opposite effect. (\textbf{C}) BA networks. The increasing degree heterogeneity initially promotes but ultimately inhibits cooperation under the PC and DB rules. However, this does not hold under the BD rule. The parameters are the same as the ones in Fig.~\ref{fig_randomgraph} in the main text.}
	\label{fig_randomgraph_sup}
\end{figure}

\begin{figure}
	\centering
	\includegraphics[width=.75\textwidth]{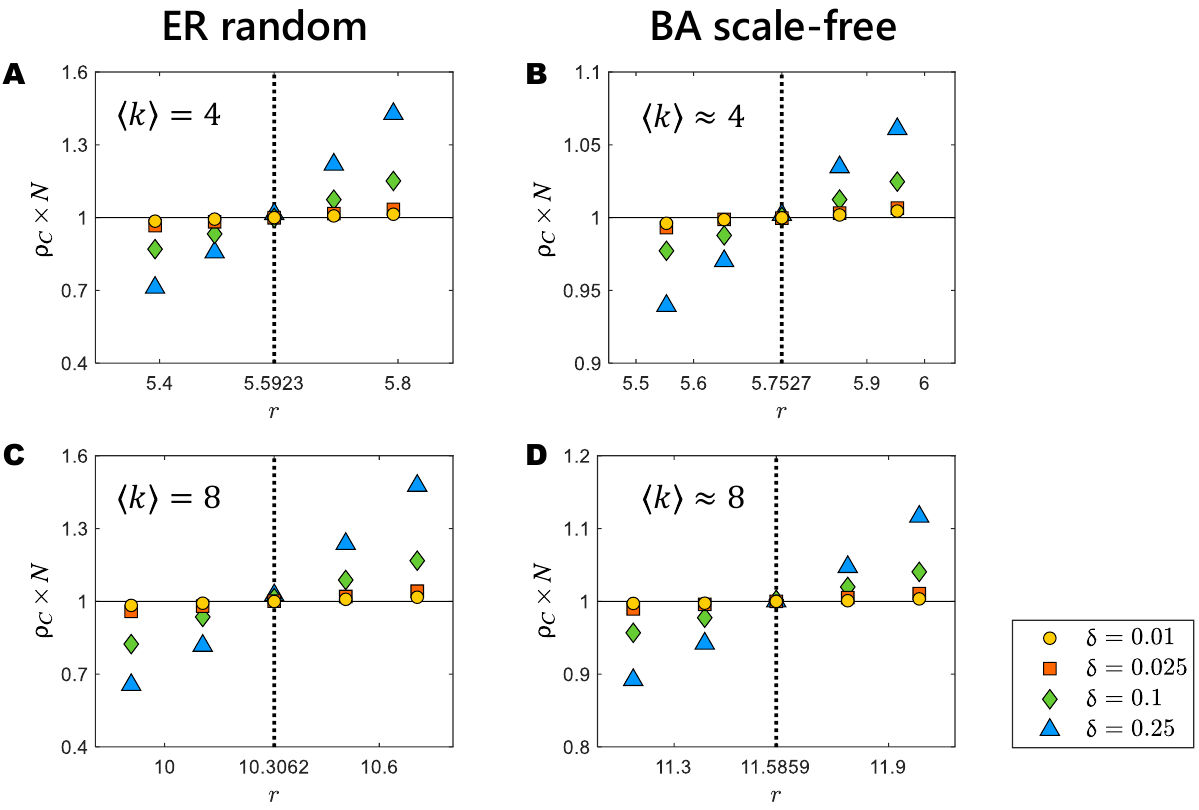}
	\caption{\textbf{Performance of our theory under non-marginal selection.} We examine the results on homogeneous (\textbf{A}, \textbf{C}) and heterogeneous networks (\textbf{B}, \textbf{D}) under selection strengths $\delta=0.01,0.025,0.1,0.25$ (PC rule). The critical threshold derived under weak selection remains accurate for $\delta = 0.1$. Under stronger selection, the theoretical threshold shows a modest deviation from simulation results (e.g., $\delta=0.25$ in \textbf{C}), but the qualitative conclusion remains robust: homogeneous networks support cooperation more effectively than heterogeneous ones ($5.5923 < 5.7527$ and $10.3062 < 11.5859$), consistent with our analytical predictions under weak selection. (\textbf{A}) ER network with $\langle k\rangle=4$ ($p=4/99$). (\textbf{B}) BA network with $\langle k\rangle\approx 4$ ($m=2$, $\gamma=2$). (\textbf{C}) ER network with $\langle k\rangle=8$ ($p=8/99$). (\textbf{D}) BA network with $\langle k\rangle\approx 8$ ($m=4$, $\gamma=2$). The population size is $N=100$ for all networks. }
	\label{fig_largeselection}
\end{figure}

\begin{figure}
	\centering
	\includegraphics[width=\textwidth]{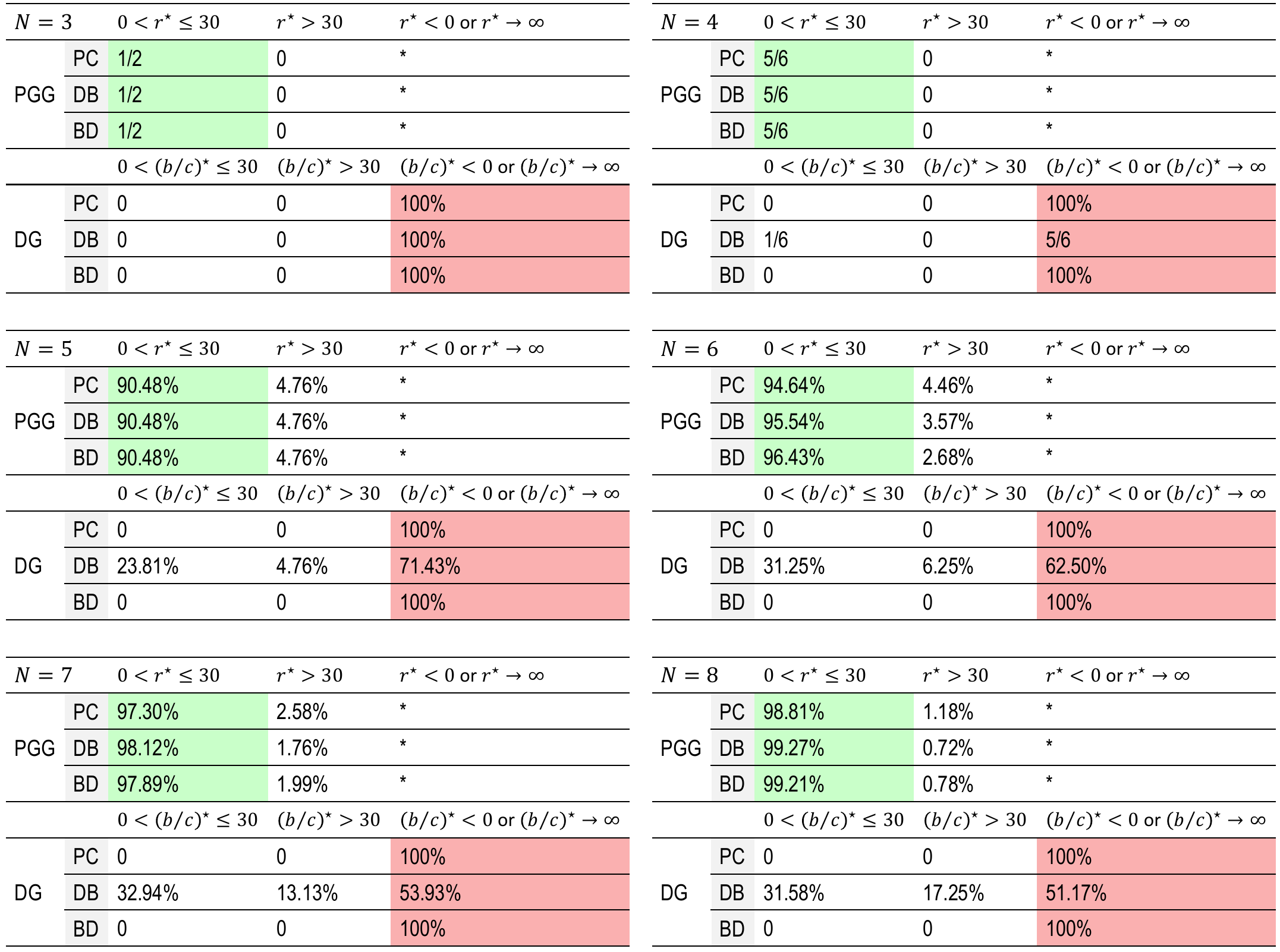}
	\caption{\textbf{Supplementary results for PGGs on all networks of different sizes $3\leq N\leq 8$.} The categories of networks classified by their critical synergy factors are presented. The symbol * means that the only structure that does not support cooperation is the fully connected network. The results are obtained using average payoffs.}
	\label{fig_allnetwork_sup}
\end{figure}

\begin{figure}
	\centering
	\includegraphics[width=\textwidth]{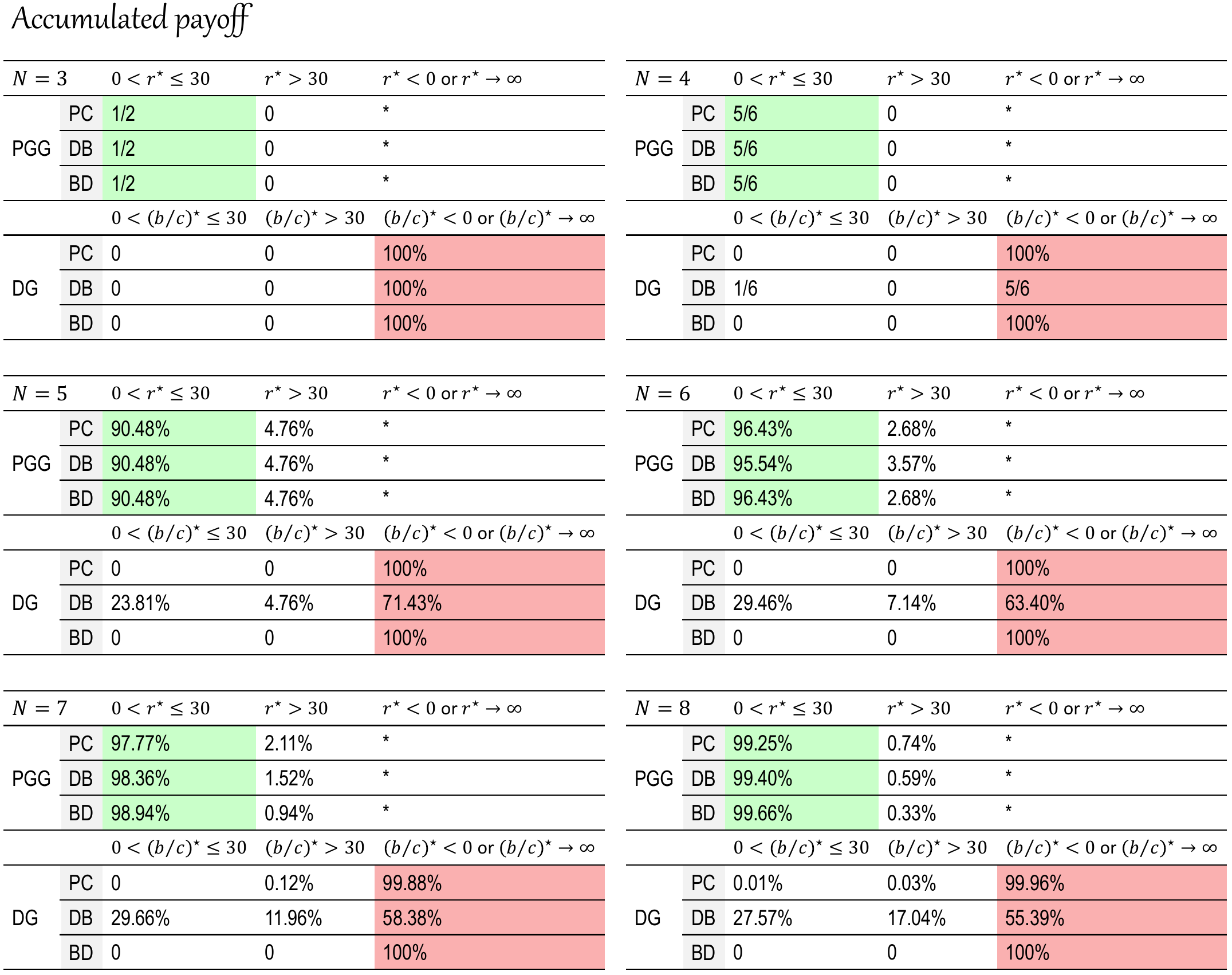}
	\caption{\textbf{Supplementary results to Fig.~\ref{fig_allnetwork_sup} for PGGs on all networks of different sizes $3\leq N\leq 8$.} The results are obtained using accumulated payoffs.}
	\label{fig_allnetwork_accu_sup}
\end{figure}

\begin{figure}
	\centering
	\includegraphics[width=\textwidth]{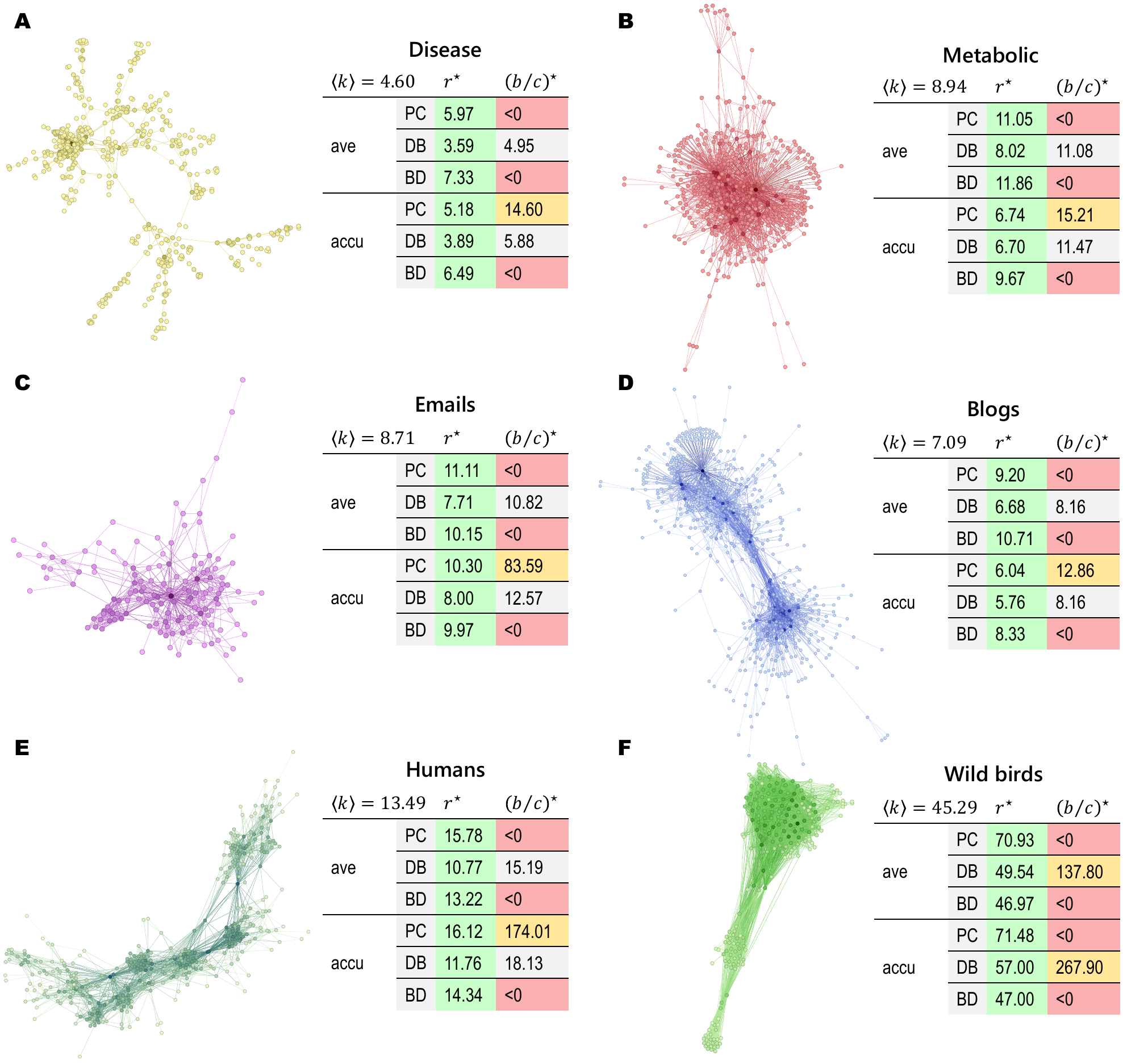}
	\caption{\textbf{Supplementary results for more empirical networks. The indicators $(b/c)^\star$ for cooperation in DGs vary depending on model details, whereas $r^\star$ in PGGs remain qualitatively consistent.} (\textbf{A}) Human disease network of size $N=516$~\cite{goh2007human,nr}. (\textbf{B}) {\it C. elegans} metabolic network of size $N=453$~\cite{jeong2000large,duch2005community,nr}. (\textbf{C}) Email network of size $N=143$~\cite{nr}. (\textbf{D}) Political blog network of size $N=643$~\cite{adamic2005political,nr}. (\textbf{E}) Human contact network of size $410$~\cite{infect,nr}. (\textbf{F}) Wild bird network of size $N=202$~\cite{nr}. }
	\label{fig_empirical_sup}
\end{figure}

\addtocontents{toc}{\protect\setcounter{tocdepth}{-1}}


\end{document}